\documentclass[12pt]{article}
\usepackage{fullpage}
\usepackage[T1]{fontenc}
\usepackage{newpxtext,newpxmath}
\usepackage{amsmath,mathrsfs}
\usepackage{graphicx}
\usepackage{caption}
\usepackage{subcaption}
\usepackage{cancel}
\usepackage[official]{eurosym}
\usepackage{natbib}
\usepackage{tikz,comment}
\usepackage{url}
\usepackage[font=footnotesize,labelfont=bf]{caption}
\usepackage{float}
\usepackage{ragged2e}
\usepackage{setspace}
\usepackage{changepage}
\usepackage{epstopdf}
\usepackage{indentfirst}
\usepackage[hang,flushmargin]{footmisc}
\usepackage{tikz}
\usepackage{breqn}
\usepackage{epigraph}
\usepackage{booktabs}
\usepackage{tcolorbox}
\usepackage[export]{adjustbox}

\usetikzlibrary{arrows,intersections,patterns}
\usetikzlibrary{shapes,arrows}
\usepackage[top=2.54cm,bottom=2.54cm,left=2.54cm,right=2.54cm]{geometry}
\usetikzlibrary{trees}
\usepackage{pgfplots}
\pgfplotsset{compat=1.18}

\newcommand{\bea}{\begin{eqnarray}}
\newcommand{\eea}{\end{eqnarray}}
\newcommand{\ben}{\begin{enumerate}}
\newcommand{\een}{\end{enumerate}}
\newcommand{\bi}{\begin{itemize}}
\newcommand{\ei}{\end{itemize}}

\usepackage{comment}
\usepackage{bbm}
\usepackage{bbold}
\usepackage{hyperref}
\hypersetup{
colorlinks = true,
linkcolor={red!80!black},
citecolor={blue!50!black},
urlcolor={blue!80!black}
}

 \newtcolorbox{mybox}[3][]
{
  colframe = #2!25,
  colback  = #2!10,
  coltitle = #2!20!black,  
  title    = {#3},
  #1,
}

\newcommand{\tablesize}{\fontsize{9.1pt}{11pt}\selectfont} 

\usepackage{minitoc}

\usepackage{authblk}

\title{\LARGE{\textbf{High-dimensional Four-Level Structured Functional Principal Component Analysis for Unbalanced Level Data}}\\ \textbf{\Large{with Application to Concussion Data}}}

\bigskip

\author[1]{Verena Werkmann}
\author[2]{Alok S. Shah}
\author[2]{Brian D. Stemper}
\author[1]{Jaroslaw Harezlak}

\affil{%
  \textsuperscript{1} Indiana University Bloomington
  \quad
  \textsuperscript{2} Medical College of Wisconsin
}

\date{}

\begin{document}
\doparttoc 
\faketableofcontents 
\maketitle
\maketitle\thispagestyle{empty}

\vspace{-0.5cm}
\begin{abstract}
\noindent
We implement the Structured Functional Principal Component Analysis (SFPCA) framework of \cite{shou2015structfpca} in a four-level nested setting with arbitrarily unbalanced group sizes at every level of the hierarchy. While the SFPCA framework can in principle accommodate deeper hierarchical structures, practical implementations remain challenging when the number of curves is large and the hierarchy is highly unbalanced. Our contributions are twofold. First, we operationalize SFPCA for four hierarchical levels with arbitrary unbalancedness across units. Second, we propose a memory-efficient computational strategy that makes the analysis of large collections of curves feasible on conventional hardware. In particular, we construct the four design-specific matrices required for the Method of Moments (MoM) estimators of the level-specific covariance operators while controlling their memory footprint, which would otherwise grow prohibitively with the number of curves. We illustrate the approach using three subsets of the NCAA-DoD Grand Alliance Concussion Assessment, Research, and Education (CARE) Consortium dataset (2014-2017).
 \end{abstract}


\thispagestyle{empty}
\bigskip\noindent \textbf{Classification}:

\newpage

\setcounter{page}{1}

\pagebreak

\section{Introduction}\label{sec:intro}

The framework of \cite{shou2015structfpca} provides a principled approach for multilevel functional principal component analysis (FPCA) of hierarchically structured functional data obtained from different sampling designs. In this framework, variability at each level of the hierarchy is captured through latent processes, whose covariance operators are decomposed into principal components. Structured Functional Principal Component Analysis (SFPCA) allows for a dimensionality reduction across the different levels of the data, but its practical implementation remains challenging in more complex hierarchical settings.

In this paper, we implement the Structured Functional Principal Component Analysis (SFPCA) framework of \cite{shou2015structfpca} in a four-level nested functional model with arbitrarily unbalanced group sizes at every level of the hierarchy. While the SFPCA framework in principle accommodates an arbitrary number of levels with unbalanced units, existing applications have been limited to at most three levels, with unbalancedness typically restricted to the finest level.
Our approach allows for unit-specific numbers of lower-level units throughout the hierarchy, from the top level down to individual curves. This flexibility is essential in many real-world applications, where hierarchical structures are inherently unbalanced. Enforcing balanced designs would require discarding or aggregating data, potentially leading to information loss and biased results.

This increased flexibility, however, introduces substantial computational challenges. In nested models, the number of observations at the finest level grows rapidly with each additional level or unit, so that even moderate hierarchical structures can result in datasets with thousands or tens of thousands of curves. For example, studies such as the Sleep Heart Health Study \citep[SHHS;][]{di2009multilevel} already feature multiple curves per subject, and additional hierarchical levels would further increase the total number of curves substantially.
In functional data analysis, high dimensionality is often associated with densely observed curves, where a fine grid leads to large covariance operators and expensive smoothing. For such settings, methods such as the Fast Covariance Estimation \citep[FACE;][]{xiao2016face} have been developed to reduce the computational burden. In contrast, the setting considered here is characterized by a large number of curves observed on relatively small grids. While smoothing is less problematic in this case, a different computational bottleneck arises: the estimation of level-specific covariance operators requires matrix multiplications involving dimensions equal to the total number of curves, which can quickly exceed available memory on conventional hardware. To address this challenge, we propose memory-efficient computational strategies based on matrix decomposition and block-wise operations, avoiding the explicit construction of large matrices whenever possible.

The proposed setting is further examined in a simulation study to disentangle the effects of multiple hierarchical levels, unbalancedness, observation noise, and a large number of curves. We additionally apply the method to real-world data, namely three subsets of the NCAA-DoD Grand Alliance Concussion Assessment, Research, and Education (CARE) Consortium, to assess the performance of the four-level SFPCA model under realistic conditions.

The remainder of the paper is organized as follows. Section \ref{sec:ext_sfpca} reviews the SFPCA procedure for the four-level nested structured functional model and outlines the proposed memory-efficient computational strategies. Section \ref{sec:simulation} presents the simulation study. Section \ref{sec:data_app} describes the data application and reports the results. Section \ref{sec:summary} concludes with a summary and discussion.


\section{Four-level Structured Functional Principal Component Analysis}\label{sec:ext_sfpca}

Following \citet{shou2015structfpca}, the four-way nested structured functional model is given by   
\bea\label{eq:4waynested}
Y_{ijkl}(t) &=& \mu(t) + X_i(t) + U_{ij}(t) + W_{ijk}(t) + V_{ijkl}(t) + \varepsilon_{ijkl}(t),\\ \nonumber i &=& 1,2,\dots,I, \, j = 1, 2, \dots, J_i,\, k = 1,2,\dots,K_{ij},\, l = 1, 2, \dots, L_{ijk},   
\eea
where $\varepsilon_{ijkl}(t) \sim \operatorfont{WN}(0, \mathnormal{\sigma}^2)$. $\mu(t)$ is the deterministic mean function and $X_i(t)$, $U_{ij}(t)$, $W_{ijk}(t)$ and $V_{ijkl}(t)$ are level-specific stochastic processes. These processes are assumed to have zero mean and to be uncorrelated which ensures identifiability. Their covariances are $K_X(t,s) = \operatorfont{cov}\mathnormal{\left(X_i(t), X_i(s)\right)}$, \dots, $K_V(t,s) = \operatorfont{cov}\mathnormal{\left(V_{ijkl}(t), V_{ijkl}(s)\right)}$. Due to the assumed uncorrelatedness of the level-specific random processes, the variability of $Y_{ijkl}$ is then represented as
\bea
K_Y(t,s) = \operatorfont{cov}\left(Y_i(t), Y_i(s)\right) = K_X(t,s) + K_U(t,s) + K_W(t,s) + K_V(t,s) + \operatorfont{I}_{(t=s)}\mathnormal{\sigma^2}.
\eea

Goals of SFPCA include efficient dimension reduction as well as the extraction of level-specific signals. As only the outcome function $Y(t)$ is observed, the latent processes $X(t)$, $U(t)$, $W(t)$ and $V(t)$ are modeled via the Karhunen-Lo\'{e}ve expansion as follows
\bea\label{eq:KL}
Y_{ijkl}(t) = \sum_{n_1=1}^{\infty} \phi_{n_1}^X(t) \xi_{i{n_1}} + \sum_{n_2=1}^{\infty} \phi_{n_2}^U(t) \theta_{j{n_2}} + \sum_{n_3=1}^{\infty} \phi_{n_3}^W(t) \zeta_{k{n_3}} + \sum_{n_4=1}^{\infty} \phi_{n_4}^V(t) \eta_{l{n_4}} + \varepsilon_{ijkl}(t)
\eea
with the eigenfunctions $\phi^X_{n_1}, \dots, \phi^V_{n_4}$ of the covariances $K_X, \dots, K_V$ and the scores $\xi_{i{n_1}}$, $\theta_{j{n_2}}$, $\zeta_{k{n_3}}$, and $\eta_{l{n_4}}$ which mean zero, mutually independent random variables with variances $\lambda_{n_1}^X$, $\lambda_{n_2}^U$, $\lambda_{n_3}^W$, and $\lambda_{n_4}^V$. These eigenvalues are assumed to be declining in magnitude for every $n_1$, $n_2$, $n_3$ and $n_4$, i.e., $\lambda_{n_1}^X \geq \lambda_{n_1 +1}^X$ etc. 

The first step in SFPCA is to obtain the estimators $\widehat{K}_X$, $\widehat{K}_U$, $\widehat{K}_W$ and $\widehat{K}_V$ to estimate $\left(\widehat{\phi^X}_{n_1},\xi_{i{n_1}}\right), \dots, \left(\widehat{\phi^V}_{n_4}, \eta_{l{n_4}}\right)$. The eigenfunctions and eigenvalues can be found via level-specific spectral decomposition which then allows for approximating model (\ref{eq:KL}) by the first $N_1, \dots, N_4$ level-specific principal components (PCs) of $X_i(t), \dots, V_{ijkl}(t)$
 \bea\label{eq:KLapprox}
Y_{ijkl}(t) = \sum_{n_1=1}^{N_1} \phi_{n_1}^X(t) \xi_{i{n_1}} + \sum_{n_2=1}^{N_2} \phi_{n_2}^U(t) \theta_{j{n_2}} + \sum_{n_3=1}^{N_3} \phi_{n_3}^W(t) \zeta_{k{n_3}} + \sum_{n_4=1}^{N_4} \phi_{n_4}^V(t) \eta_{l{n_4}} + \varepsilon_{ijkl}(t).
\eea
We collect $Y_{ijkl}(t)$ in a matrix $\operatorfont{Y}$ of dimension $p \times n$ where $p$ is the length of the discrete (and evenly spaced) sampling grid $\mathcal{T}$ and $n = \sum_{i}^I \sum_{j}^{J_i} \sum_{k}^{K_{ij}} L_{ijk}$ is the total number of curves in the model. We assume an unbalanced number of units in each of the levels. Next, we will provide the setup for the Methods of Moments (MoM) estimation of the covariances and give the estimator of the principal scores as best linear unbiased predictor (BLUP) of the mixed effects model (\ref{eq:KLapprox}) in matrix form. The following remarks are based on \citet{shou2015structfpca}'s original presentation for three levels in a purely balanced setup. 

The MoM estimator of each of the level-specific covariances $\widehat{K} \in \left\{\widehat{K}_X, \widehat{K}_U, \widehat{K}_W, \widehat{K}_V\right\}$ takes the "sandwich" form $\widehat{K} = \operatorfont{YGY'}$ where $G$ is the design-specific matrix of size $n\times n$ associated with each of the levels, $G \in  \left\{ G_X, G_U, G_W, G_V\right\}$.   To construct $G$, we us a bottom-up approach, i.e., starting with the finest level, to obtain $H_X$, $H_X$, $H_W$ and $H_V$: 
\bea\label{eq:hmatrices}
\nonumber
\widehat{H}_V &=& \frac{2}{k_1-n} \left(D_1 - E_1'E_1\right)\\
\nonumber
\widehat{H}_W &=& \frac{2}{k_2-k_1} \left(D_2 - E_2'E_2 - D_1 + E_1'E_1\right)\\
\widehat{H}_U &=& \frac{2}{k_3-k_2} \left(D_3 - E_3'E_3 - D_2 + E_2'E_2\right)\\
\nonumber
\widehat{H}_X &=& \frac{2}{k_4-k_3} \left(D_4 - E_4'E_4 - D_3 + E_3'E_3\right)\\
\nonumber
&=& \frac{2}{n^2-k_3} Y\left(nI - 11' - D_3 + E_3'E_3\right)Y'
\eea
with
\bea
\begin{tabular}{lll}
$D_1 = \operatorname{diag}\left(N_{1,1,1},\dots,N_{I,J,K} \right)$, & \, $N_{i,j,k} = n_{i,j,k}I_{n_{ijk}}$, &\\
$D_2 = \operatorname{diag}\left(N_{1,1},\dots,N_{I,J} \right)$, &\, $N_{i,j~} = n_{i,j,\cdot}I_{n_{ij\cdot}}$,& $n_{i,j,\cdot} = \sum_{k=1}^{K_{ij}}  n_{i,j,k}$\\
$D_3 = \operatorname{diag}\left(N_{1},\dots,N_{I} \right)$, &\, $N_{i~~} = n_{i,\cdot,\cdot}I_{n_{i\cdot\cdot}}$, & $n_{i,\cdot,\cdot} = \sum_{j=1}^{J_i}\sum_{k=1}^{K_{ij}}  n_{i,j,k}$\\
$D_4 = N$ & \, $N_{~~~} = n{I}_{n}$, & $n_{~~~} = \sum_{i=1}^{I}\sum_{j=1}^{J_i}\sum_{k=1}^{K_{ij}}  n_{i,j,k}$ 
\end{tabular}
\eea
and
\bea
\nonumber
{E}_1&=&\operatorname{diag}\left({1}_{\mathrm{n}_{1,1,1}}', \ldots, {1}_{\mathrm{n}_{I,J,K}}'\right),\quad {1}_{\mathrm{n}_{i,j,k}} = \left(1,1,\dots,1\right)'_{\#(n_{i,j,k})}\\
{E}_2&=&\operatorname{diag}\left({1}_{\mathrm{n}_{1,1,\cdot}}', \ldots, {1}_{\mathrm{n}_{I,J,\cdot }}'\right)~, \quad {1}_{\mathrm{n}_{i,j,\cdot}} = \left(1,1,\dots,1\right)'_{\#(n_{i,j,\cdot})}\\
\nonumber
{E}_3&=&\operatorname{diag}\left({1}_{\mathrm{n}_{1,\cdot,\cdot}}', \ldots, {1}_{\mathrm{n}_{I,\cdot,\cdot }}'\right)~~, \quad {1}_{\mathrm{n}_{i,\cdot,\cdot}} = \left(1,1,\dots,1\right)'_{\#(n_{i,\cdot,\cdot})}\\
\nonumber
{E}_4&=& 1_n1_n' \hspace{3.2cm}, \quad{1}_{\mathrm{n}_{~~~}} = \left(1,1,\dots,1\right)'_{\#(n)}
\eea
and
\bea
\begin{tabular}{llll}
$k_1 = \sum_{i=1}^I \sum_{j=1}^{J_i} \sum_{k=1}^{K_{ij}} {n_{i,j,k}^2}$,&
$k_2 = \sum_{i=1}^I \sum_{j=1}^{J_i} {n_{i,j,\cdot}^2}$, &
$k_3 = \sum_{i=1}^I  n_{i,\cdot,\cdot}^2$,& 
$k_4 = n^2$.
\end{tabular}
\eea
The $G-$matrices are given as
\bea\label{eq:gmatrices}
G_V = 0.5\widehat{H}_V, \; G_W = 0.5\left(\widehat{H}_W - \widehat{H}_V\right), \; G_U = 0.5\left(\widehat{H}_U - \widehat{H}_W\right), \; G_X = 0.5\left(\widehat{H}_X - \widehat{H}_U\right),
\eea 
and, using them in the MoM estimator, the level-specific covariances are obtained as
\bea\label{eq:covariances}
\widehat{K}_V = \operatorfont{Y}G_V\operatorfont{Y}',\; \widehat{K}_W = \operatorfont{Y}G_W\operatorfont{Y}',\; \widehat{K}_U = \operatorfont{Y}G_U\operatorfont{Y}',\; \widehat{K}_X = \operatorfont{Y}G_X\operatorfont{Y}'.
\eea

For at least the three- and four-level nested design, we can employ a notational short-cut for generating the level-specific $G-$matrices which proves particularly useful when level groups are unbalanced: Consider the differences $D - E'E$ as one matrix $DEE$ where each of these $DEE = \left\{DEE_1, DEE_2, DEE_3, DEE_4\right\}$ is of size $n \times n$. $DEE_1$, $DEE_2$ and $DEE_3$ are block-diagonal matrices and $DEE_4$ is a matrix containing only the constant $-1$ on the off-diagonal elements and the constant $n-1$ on the diagonal elements. The structure of $DEE_1$ to $DEE_4$ is given in Table (\ref{tab:structGmats}). The block sizes in matrix $DEE_1$ correspond to the group sizes of the $W-$level which are groups of the bottom level units (curves). For example, $n_{1,1,1}$ gives the size of the first group in the $W-$level and $n_{I,J,K}$ is the size of the last group in $W-$level. The block sizes in matrix $DEE_2$ are equal to the sum of all curves nested under each $U-$level group, e.g. the sum of $n_{1,1,1},\dots, n_{1,1,K}$ is the size of the first block (sum of all curves belonging to $i=j=1$) and the sum of $n_{I,J,1},\dots, n_{I,J,K}$ is the size of the last block (sum of all curves belonging to $i=I$, $j=J$). Lastly, the block sizes in matrix $DEE_3$ are equal to the sum of all curves nested under each $X-$level group.

In addition, the denominators of the scaling factors, $2/(n^2-k_3)$, $1/(k_3-k_2)$ and $1/(k_2-k_1)$, equal the traces of the matrix differences $\left(DEE_4-DEE_3\right)$, $\left(DEE_3-DEE_2\right)$, $\left(DEE_2-DEE_1\right)$ and $k_1-n = \operatorfont{tr}\left(DEE_1\right)$. Thus, the $H-$matrices can be considered trace-scaled.

The principal scores are estimated as the Best Linear Unbiased Predictor (BLUP) of the truncated mixed effects model given in (\ref{eq:KLapprox}) with estimated eigenvectors:
 \bea\label{eq:KLapprox_est}
\operatorfont{Y} = \operatorfont{\widehat{\Phi}}_{N_1}^X  \mathnormal{\xi}_{i} + \widehat{\operatorfont{\Phi}}_{N_2}^U \mathnormal{\theta}_{j} + \widehat{\operatorfont{\Phi}}_{N_3}^W \mathnormal{\zeta}_{k} + \widehat{\operatorfont{\Phi}}_{N_4}^V \mathnormal{\eta}_{l},
\eea
where the scores are normally distributed with zero mean and the covariances $\widehat{\Lambda}_{N_1}^X$, $\widehat{\Lambda}_{N_2}^U$, $\widehat{\Lambda}_{N_3}^W$ and $\widehat{\Lambda}_{N_4}^V$ which are diagonal matrices containing the first $N_1$, $N_2$, $N3$ and $N_4$ eigenvalues.  For simplicity, we assume a noise-free model ($\sigma^2 = 0$) for the following presentation of the scores. Analogously to \citet[][Web Appendix A]{shou2015structfpca}, we estimate ${\operatorfont{u}}_i = \left(\xi_i, \theta_i, \zeta_i, \eta_i\right)$ in
\bea\label{eq:scores_BLUP}
\widetilde{\operatorfont{Y}}_i &=& \operatorfont{B}\operatorfont{u}_i, \quad \widetilde{\operatorfont{Y}} = \operatorfont{vec}\left(\operatorfont{Y}\right), 
\eea
where $\xi_i = \left(\xi_{i1}, \dots, \xi_{iN_1}\right)$, $\theta_i = \left(\theta_{i11}, \dots, \theta_{i1N_2}, \dots, \theta_{iJ_i1}\dots, \theta_{iJ_1N_2}\right)$, $\zeta_i = \left(\zeta_{i111}, \dots, \zeta_{i11N_3},   \right.$\\ $\left. \dots,\zeta_{iJ_iK_{iJ_i}1}\dots,\zeta_{iJ_1N_3}\right)$ and $\eta_i = \left(\eta_{i1111}, \dots, \eta_{i111N_4}, \dots, \eta_{iJ_iK_{ij}1}\dots, \eta_{iJ_1K_{ij}N_4}\right)$.

The BLUP of $\operatorfont{u}_i$ is
\bea\label{eq:scoresestimation}
\widehat{\operatorfont{u}}_i &=& \left(\operatorfont{B'B}\right)^{-1} \operatorfont{B}'\widetilde{\operatorfont{Y}}_i
\eea
with
\bea
\nonumber
\operatorfont{B} &=& [\operatorfont{B}_X | \operatorfont{B}_U | \operatorfont{B}_W | \operatorfont{B}_V]\\
&=&\left(\begin{array}{ccccccccccccccc}
\operatorfont{\widehat{\Phi}}_X & \operatorfont{\widehat{\Phi}}_U &\dots &\dots & \operatorfont{\widehat{\Phi}}_W &\dots &\dots & \operatorfont{\widehat{\Phi}}_V &\dots &\dots & & \\
\vdots & \vdots & & & \ddots & & &\ddots \\
\operatorfont{\widehat{\Phi}}_X & \operatorfont{\widehat{\Phi}}_U &\dots &\dots & & \operatorfont{\widehat{\Phi}}_W & & & \operatorfont{\widehat{\Phi}}_V & & & \\
\vdots & & \ddots & & & \ddots & & & \ddots \\
\operatorfont{\widehat{\Phi}}_X & & & \operatorfont{\widehat{\Phi}}_U & & & \operatorfont{\widehat{\Phi}}_W & & &   \operatorfont{\widehat{\Phi}}_V & &  \\
\vdots & & & \vdots & & & \ddots & && \ddots &\\
\operatorfont{\widehat{\Phi}}_X & & & \operatorfont{\widehat{\Phi}}_U & & & & \operatorfont{\widehat{\Phi}}_W &&&\operatorfont{\widehat{\Phi}}_V &  
\end{array}\right).
\eea

The dimensions of the level-specific $\operatorfont{B}$ matrices displayed in Table \ref{tab:Bdimensions} reveal that, in many cases, $\operatorfont{B}$ is far too large to be materialized, much less to be squared or inverted as required for the BLUP in Equation (\ref{eq:scores_BLUP}), using common soft- and hardware. For illustration purposes, let's consider the following example scenario: The data matrix $\operatorfont{Y}_{\text{ex}}$ consists of $n=14,579$ curves on an evenly-spaced grid of $p=40$. The aggregated level dimensions are $\sum_{i=1}^I L_1(i) = I = 5$, $\sum_{i=1}^I L_2(i) = 20$,  $\sum_{i=1}^I L_2L_3(i) = 71$ and $\sum_{i=1}^I L_2L_3L_4(i) = n$. Vectorizing $\operatorfont{Y}_{\text{ex}}$ leads to a data vector of length $n_{\text{vec}} = 583,160 = np$. When we assume that each of the four levels' variability is sufficiently described by three principal components, $N_1 = N_2 = N_3 = N_4 = 3$, we will end up with 44,013 scores to be estimated overall: $N_1\cdot 5 = 15$ for the coarsest level, $N_2\cdot 20 = 60$ for the next finer level, $N_3\cdot 71 = 213$ for the second finest level and $N_4\cdot n = 43,737$ for the finest level. This means $\operatorfont{B}$ will be of size $583,160 \times 44,013$ which would need a little below 200GB of memory for mere storage of this matrix. This example shows that while, from a theoretical/methodological point of view, the application of a four- or ``more"-way nested model is straightforward, conventional soft- and hardware can rather quickly reach its limits when, for example, high curve dimensions are reached. With nested models, the number of curves increase many times over when one unit is added to the top level. Thus, as the example above  has made clear, even with modest level dimensions in the upper hierarchies, curve level dimensions can easily be in the ten thousands. In MATLAB, for example, each (double precision) number required 8 bytes which translates to $n^2\cdot 8$ bytes of memory for a dense $n\times n$ matrix \citep{mathworkshelp1}. 12GB is the conservatively estimated memory limit available to MATLAB on a conventional 16GB system given that background tasks need to be performed as well. It turns out that the four level-specific design matrices can have sizes around $44,000 \times 44,000$ as three of the $H-$matrices as well as three of the $G-$matrices (which are all $n\times n$) are block-diagonal and thus sparse to different degrees. The fourth $G-$matrix is diagonal and thus needs even less memory. However, additional matrix operations are prohibitively costly at $n$ that large and score estimation is completely ruled out without any further preprocessing of the matrices.   
 
\subsection{Memory-efficient methods}

To be able to work with curve dimensions of $n\geq 45,000$ in a memory-efficient way, we employ several matrix (processing) techniques and operations.\footnote{ChatGPT (OpenAI) was used to assist in identifying and implementing existing memory-efficient computational techniques for simulations and data analyses. All resulting code was reviewed and tested by the authors.} While the analyses are performed in R, we generate the $G-$matrices in MATLAB which allows for system buffering and, thus, can reach higher dimension limits than R. After generating the $G-$matrices, we use the \texttt{hodlr()} function from the hm-toolbox in MATLAB \citep[see][]{hmtoolbox} to separate the $G-$matrices into smaller blocks which can then be exported to the computer disk and then imported to R. HODLR stands for "hierarchically off-diagonal low rank" and is usually used for low-rank compression of matrices with recursive patterns across sub-matrices. However, we repurpose this function as a structured decomposition tool because it can automatically generate hierarchical tree representations of the $G-$matrices where level and index information are well-defined without altering the numerical content of the matrices. Thus, MATLAB can consistently identify and index the blocks which are then stored as HDF5 (Hierarchical Data Format version 5) chunks. HDF5 files can be considered containers that store complex data including metadata in one file. These files can be compressed to save memory and transferred across platforms (here: from MATLAB to R). As all $G-$matrix entries are numerically very small, all blocks compress efficiently. Each of these blocks or chunks can then be read separately into R allowing for block-wise matrix operations without the need for ever loading the full $G-$matrices into the R memory. In R, the consistent identifiers and indices of the chunks (generated by \texttt{hodlr()} and transferred through HDF5) facilitate block-wise matrix multiplication with the respective blocks of the data matrix $Y$ to obtain $GY'$. Subsequently, we use standard matrix multiplication $Y\times(GY')$ for constructing each level-specific covariance as given in Equation (\ref{eq:covariances}).\\

The latent scores are estimated jointly by solving the Tikhonov-regularized normal equations
\bea\label{eq:scoresestimation_ridge}
\widehat{\operatorfont{u}}_i &=& \left(\operatorfont{B'B}+\mathnormal{\lambda}\operatorfont{I}\right)^{-1} \operatorfont{B}'\widetilde{\operatorfont{Y}}_i.
\eea
To facilitate this estimation in the presence of high-dimensional $n$ and avoid exhausting memory, we never explicitly form $\operatorfont{B}$. Instead, we exploit its block-diagonal structure across $X-$level units and use only the associated block operators $B_i$. For each unit $i$, we define forward and adjoint linear operators $B_i x$ and $B_i'y$ which create the design mapping without ever forming the matrix $B_i$. We  define the Gram-matrix 
\bea\label{eq:grammat}
\operatorfont{G} = \operatorfont{B}'\operatorfont{B}
\eea
as the metric measuring inner products in the score space. During the estimation procedure, we evaluate $\operatorfont{G}$ implicitly via the block-structure:
\bea\label{eq:gramop}
\operatorfont{G}(x) = \sum_{i=1}^I \operatorfont{B}_i'\left(\operatorfont{B}_i x\right).
\eea
The evaluation of $G$ as a function (Gram-operator) saves us the memory and computational cost of saving $G$ as a matrix. The right-hand side of the normal equations is similarly constructed as
\bea
 b = \sum_{i=1}^I B_i'\widetilde{Y}_i.
\eea
Using a (preconditioned) conjugate gradient algorithm \citep[see][]{hestenes1952methods}, we obtain the global score vector $u$ by iteratively solving
\bea
\left(\operatorfont{G} +\mathnormal\lambda I\right)u = b.
\eea
All operations are performed through operator calls instead of explicit matrix multiplication. Therefore, the memory used scales linearly in $I$ and is not dependent on the full dimension of $\operatorfont{B}$. These features make the approach applicable to very large and unbalanced datasets.

\subsection{Model including noise}

When $\sigma^2 >0$ in model (\ref{eq:4waynested}), there are two alternatives to deal with this scenario: (a) Smoothing the entire data matrix or (b) smoothing the estimated level-specific covariance matrices \citep[see][]{shou2015structfpca}. The objections raised in \citet{shou2015structfpca} regarding the computational feasibility of option (b) do not apply in our scenarios as $p << n$. The off-diagonal surfaces of the four level-specific covariances of size $p \times p$ can be smoothed as in \citet{staniswalis1998nonparametric}. While option (a) is computationally slightly less costly, it is also less accurate in recovering the noise variance and the eigenfunctions. Therefore, option (b), smoothing the level-specific covariances, is more effective and is the approach we adopted here. In particular, we chose a generalized additive model (GAM) with a bivariate thin plate regression spline, with smoothing parameters estimated by REML \citep[see][]{wood2017generalized}.


\section{Simulation}\label{sec:simulation}

To evaluate the performance of the four-way nested structured functional model with unbalanced levels as well as the approach to handle high-dimensional data, we carry out simulation experiments. We use model (\ref{eq:KLapprox}),
 \bea\label{eq:KLapprox_sim}
 \nonumber
Y_{ijkl}(t) &=& \sum_{n_1=1}^{N_1} \phi_{n_1}^X(t) \xi_{i{n_1}} + \sum_{n_2=1}^{N_2} \phi_{n_2}^U(t) \theta_{ij{n_2}} + \sum_{n_3=1}^{N_3} \phi_{n_3}^W(t) \zeta_{ijk{n_3}} + \sum_{n_4=1}^{N_4} \phi_{n_4}^V(t) \eta_{ijkl{n_4}} + \varepsilon_{ijkl}(t)\\
\xi_{i{n_1}} &\stackrel{iid}{\sim}&\mathcal{N}(0,\lambda^X_{n_1}),\, \theta_{ij{n_2}} \stackrel{iid}{\sim}\mathcal{N}(0,\lambda^U_{n_2}),\, \zeta_{ijk{n_3}} \stackrel{iid}{\sim}\mathcal{N}(0,\lambda^W_{n_3}),\, \eta_{ijkl{n_2}} \stackrel{iid}{\sim}\mathcal{N}(0,\lambda^V_{n_4}),\\
\nonumber
\varepsilon_{ijkl}(t) &\stackrel{iid}{\sim}&\mathcal{N}(0,\sigma^2),
\eea
to generate level-specific unbalanced data, i.e. $i = 1,\dots, I$, $j=1, \dots, J_i$, $k = 1,\dots, K_{ijk}$, $l=1,\dots,L_{ijkl}$. Each level is made up of three principal components, $N_1 = N_2 = N_3 = N_4 = 3$, the eigenvalues are $\lambda^X_{n_1} = \lambda^U_{n_1} = \lambda^W_{n_1} = \lambda^V_{n_1} = 0.5^{{n_1}-1}$, $n_1 = 1,2,3$. The grid of time points is $p=40$ long: $t \in \mathcal{T}=\{1/p, 2/p, \dots, 1\}$. The $X-$level contains $I = 6 = \left(1, 1, 1, 1, 1, 1\right)$ units. All other dimensions are drawn from uniform distributions to make them randomly unbalanced within each level: $J_i \sim \operatorfont{U}\left(3,5\right)$, $K_{ij}\sim \operatorfont{U}\left(12,16\right)$, $L_{ijkl}\sim \operatorfont{U}\left(35,100\right)$. This results in the following structure: Each of the $I$ $X-$level units nests between three to five $U-$level units which again contain each between 12 and 16 $W-$level units. Eventually, each of the $W-$level units nests between 35 and 100 curves (bottom or $V-$level, finest). Taken together, the data matrix is $p \times n$ where $n=23,898$ is the total number of curves in the simulation model.  Following the FPCA literature \citep[see, e.g.,][]{di2009multilevel, crainiceanu2009generalized, shou2015structfpca}, we use the eigenfunctions:
\bea\label{eq:eigenfcts}
\nonumber
\phi^X_{1} =& 1,&\quad \phi^X_{2} =\sqrt{3}(2x-1),\quad \phi^X_{3} = \sqrt{5}(6x^2-6x+1)\\
\phi^U_{1} =& \sin\left({2\pi x}\right),&\quad \phi^U_{2} = \cos\left({2\pi x}\right),\hspace{0.9cm} \phi^U_{3} = \sin\left({4\pi x}\right),\\
\nonumber
\phi^W_{1} =& \cos\left({4\pi x}\right),&\quad \phi^W_{2} = \sin\left({6\pi x}\right),\hspace{0.9cm} \phi^W_{3} = \cos\left({6\pi x}\right),\\
\nonumber
\phi^V_{1} =& \sin\left({8\pi x}\right),&\quad \phi^V_{2} = \cos\left({8\pi x}\right),\hspace{0.9cm} \phi^V_{3} =\sin\left({10\pi x}\right).
\eea
We perform $M=100$ simulation repetitions. In the benchmark case, we let the standard deviation of the noise $\sigma = 0$ ("no noise" scenario) and furthermore investigate the cases $\sigma = 0.5$ and $\sigma= 1$. We treat the true number of PCs, $N_1$, $N_2$, $N_3$ and $N_4$, as given to be able to compare the accuracy of estimation under unbalanced levels only (without noise) and under unbalanced levels with two noise intensities.

\begin{figure}[H]
\begin{tabular}{ccc}
\multicolumn{3}{c}{\scriptsize{\textbf{$\mathbf{X-}$level}}}\\
\includegraphics[height=4.5cm, width=5cm]{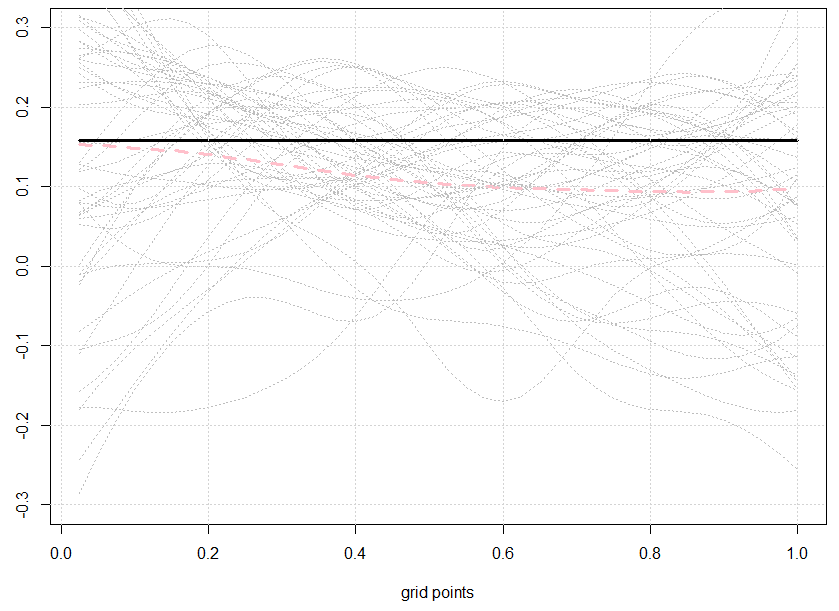}&\includegraphics[height=4.5cm, width=5cm]{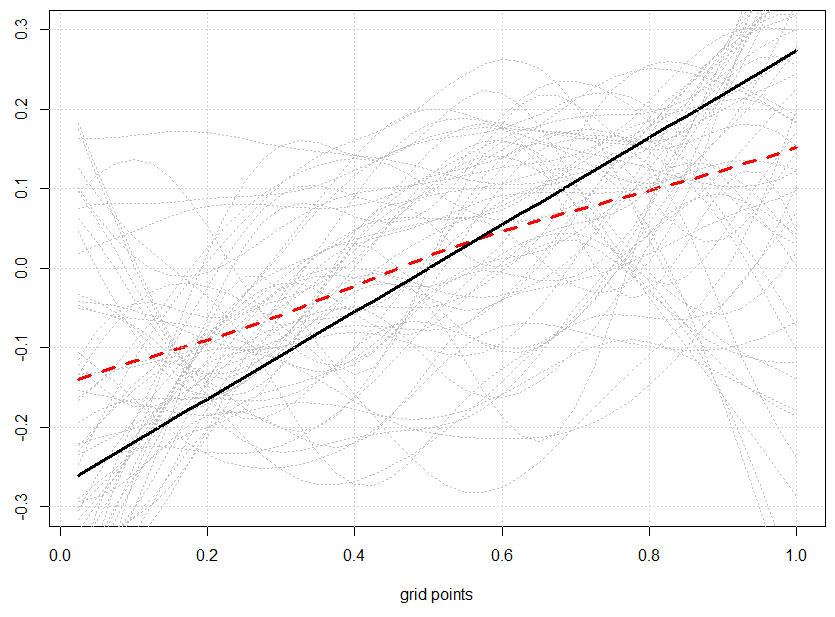}&
\includegraphics[height=4.5cm, width=5cm]{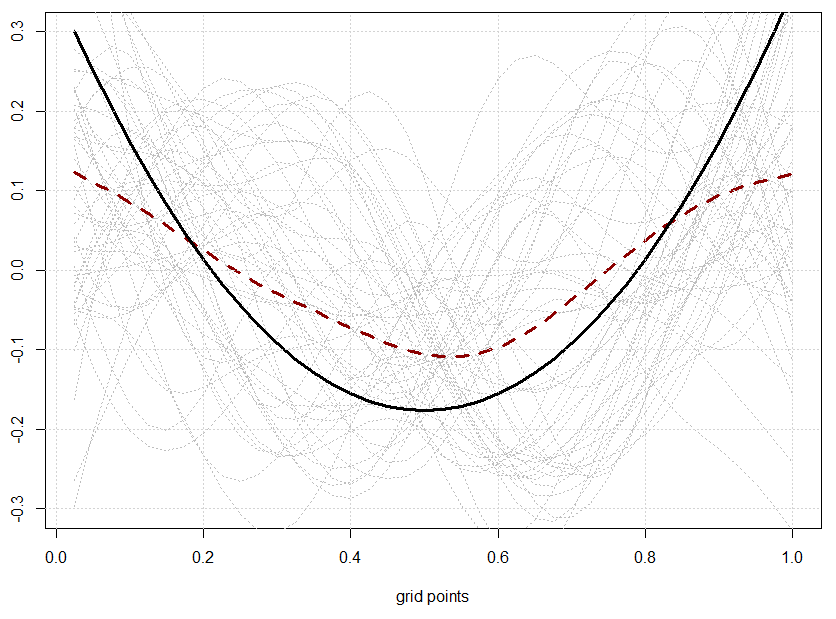}\\
\multicolumn{3}{c}{\scriptsize{\textbf{$\mathbf{U-}$level}}}\\
\includegraphics[height=4.5cm, width=5cm]{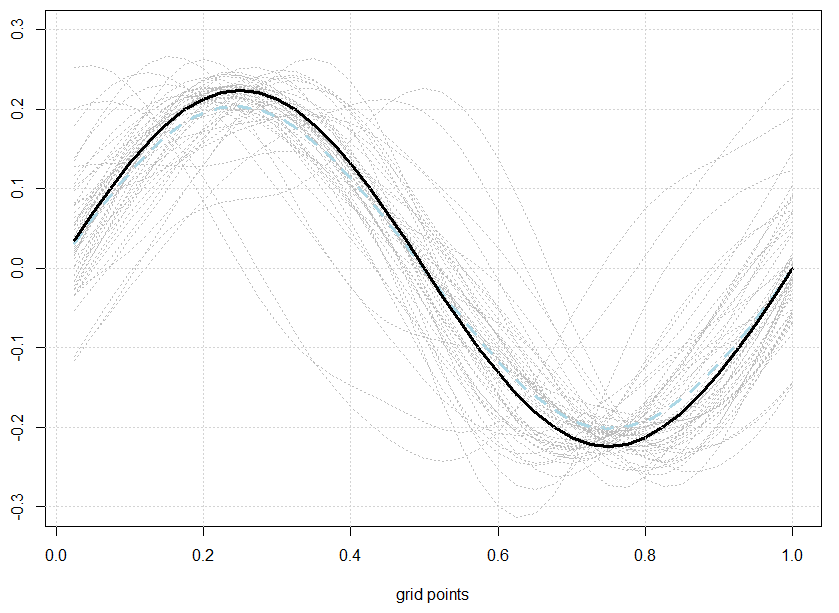}&\includegraphics[height=4.5cm, width=5cm]{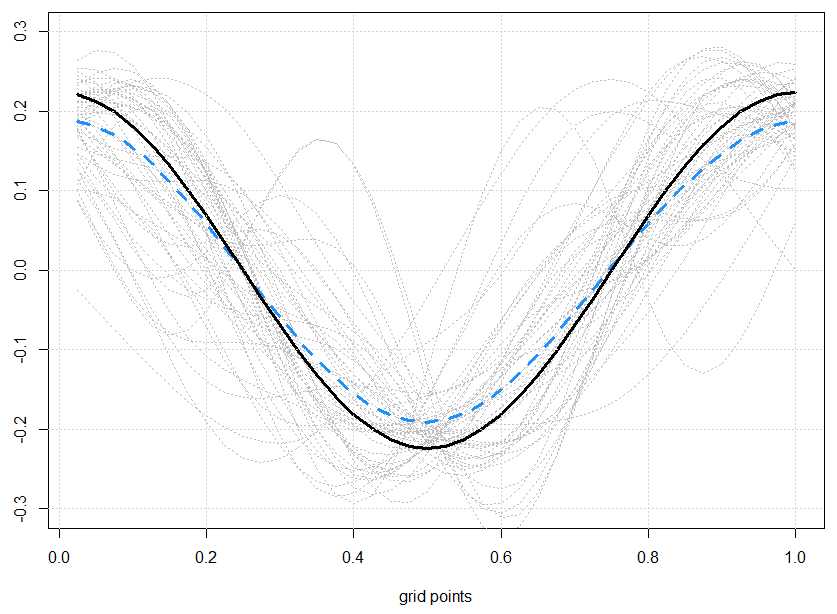}&
\includegraphics[height=4.5cm, width=5cm]{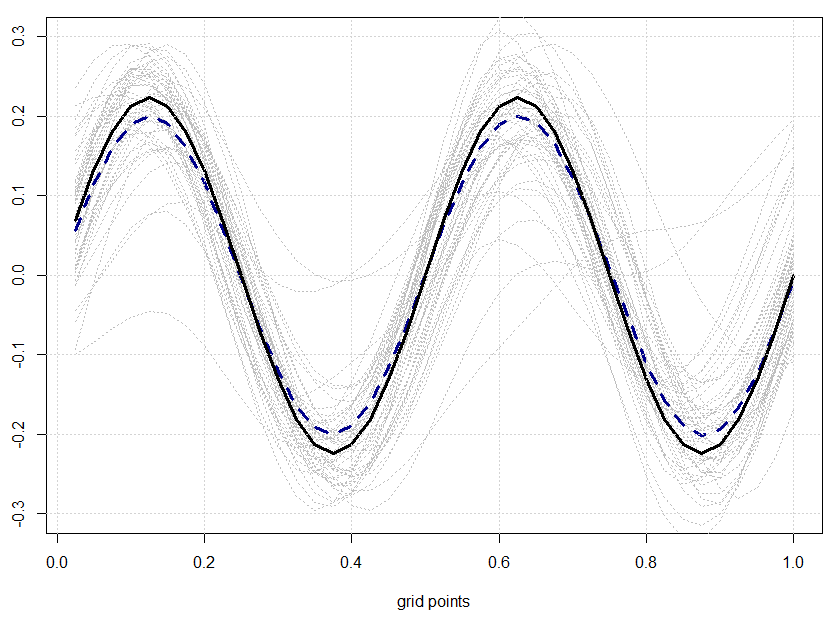}\\
\multicolumn{3}{c}{\scriptsize{\textbf{$\mathbf{W-}$level}}}\\
\includegraphics[height=4.5cm, width=5cm]{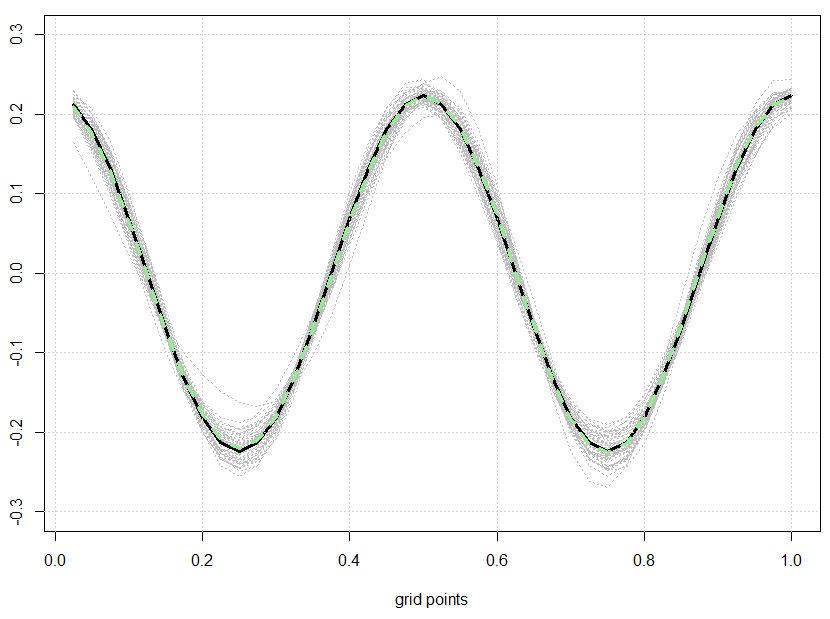}&\includegraphics[height=4.5cm, width=5cm]{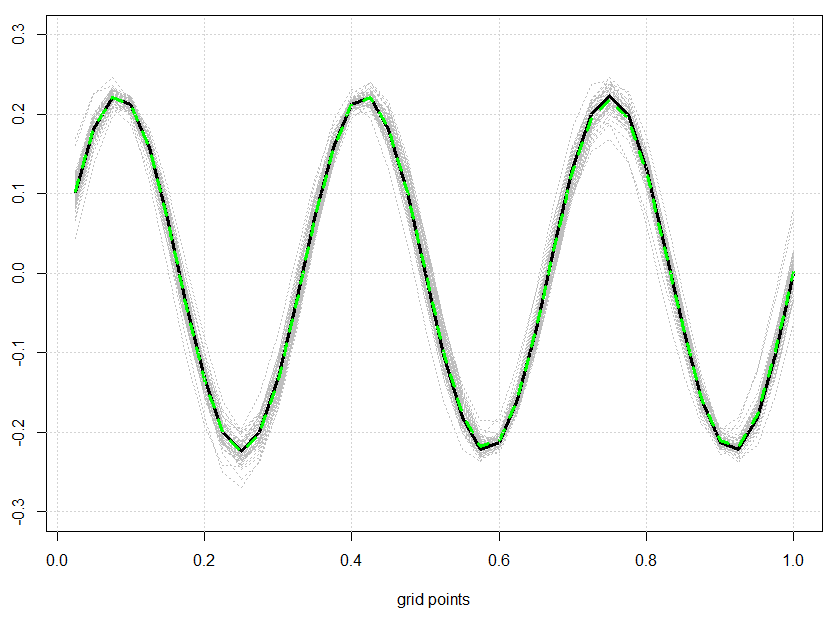}&
\includegraphics[height=4.5cm, width=5cm]{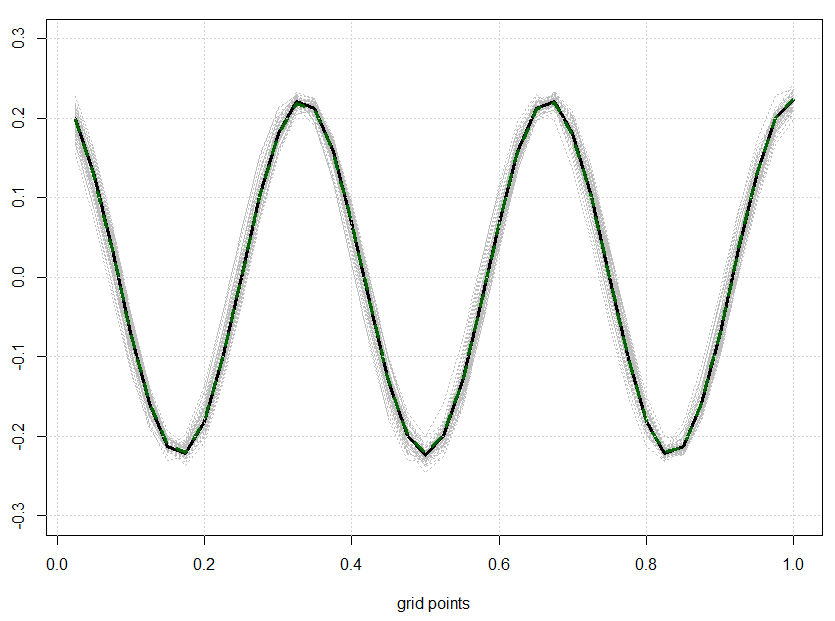}\\
\multicolumn{3}{c}{\scriptsize{\textbf{$\mathbf{V-}$level}}}\\
\includegraphics[height=4.5cm, width=5cm]{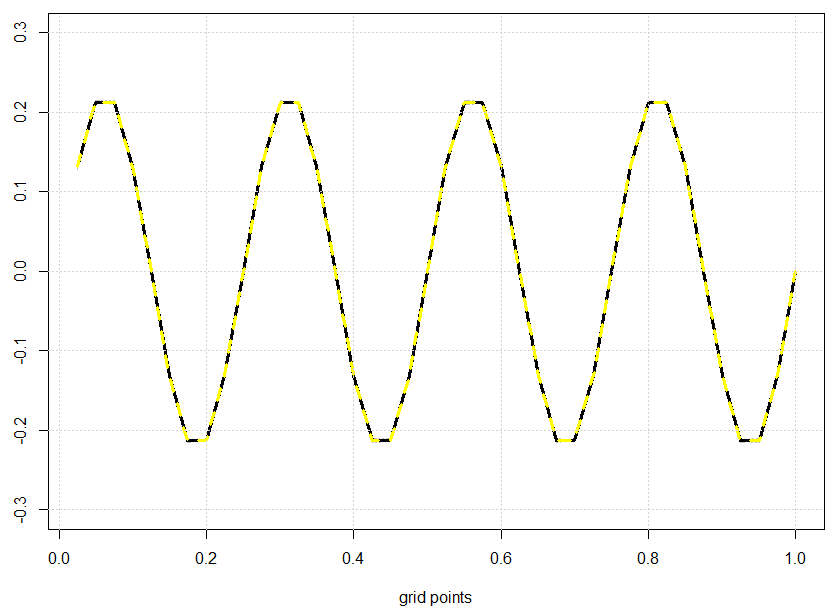}&\includegraphics[height=4.5cm, width=5cm]{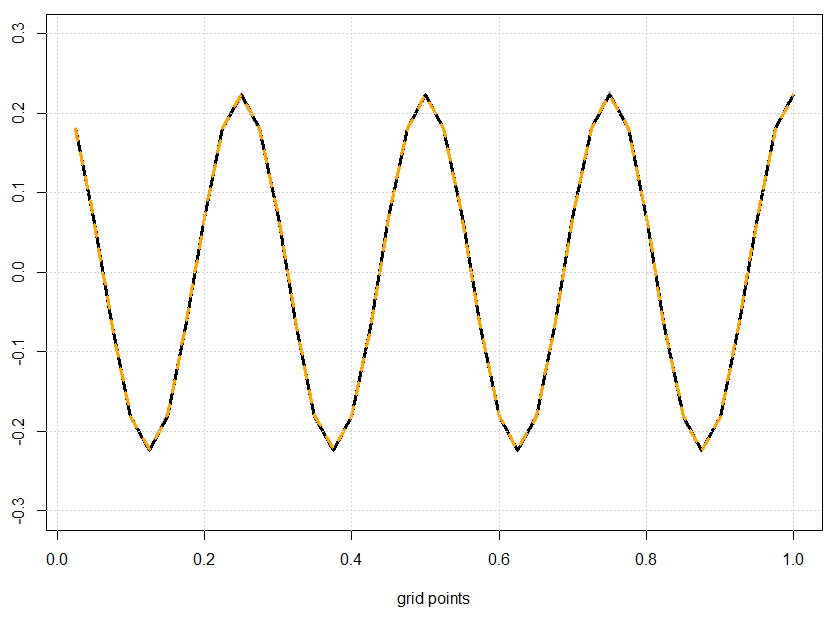}&
\includegraphics[height=4.5cm, width=5cm]{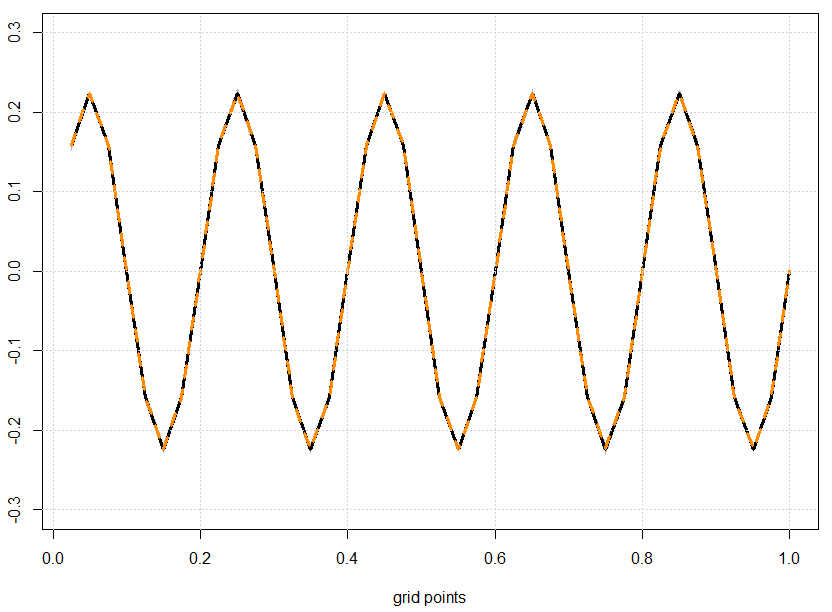}
\end{tabular}
\caption{\textbf{Eigenfunctions/principal components, $\boldsymbol{\sigma=0}$.} True eigenfunction (black line), mean of 100 estimated eigenfunctions (red line), 50 estimated eigenfunctions (gray lines).}\label{fig:eigfcts_sim_sig0}
\end{figure}

\begin{figure}[H]
\begin{tabular}{ccc}
\multicolumn{3}{c}{\scriptsize{\textbf{$\mathbf{X-}$level}}}\\
\includegraphics[height=4.5cm, width=5cm]{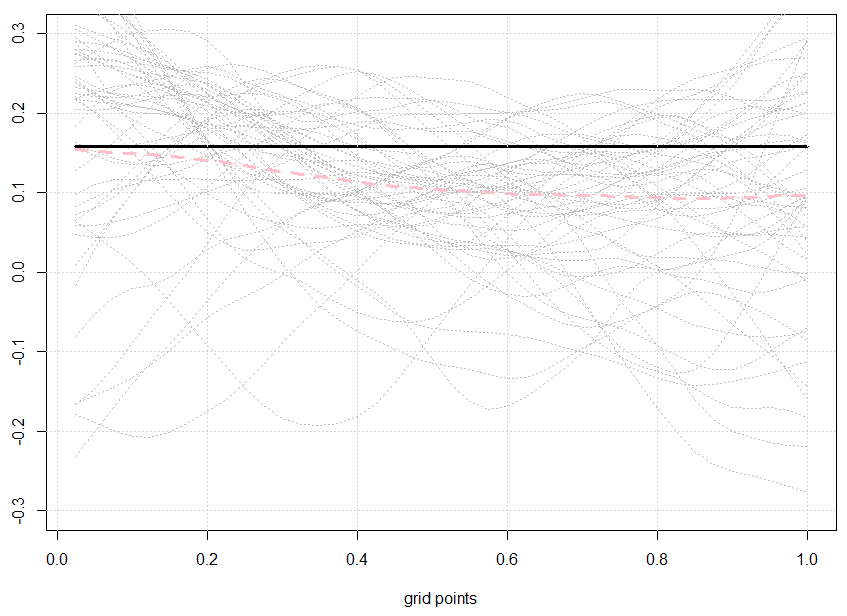}&\includegraphics[height=4.5cm, width=5cm]{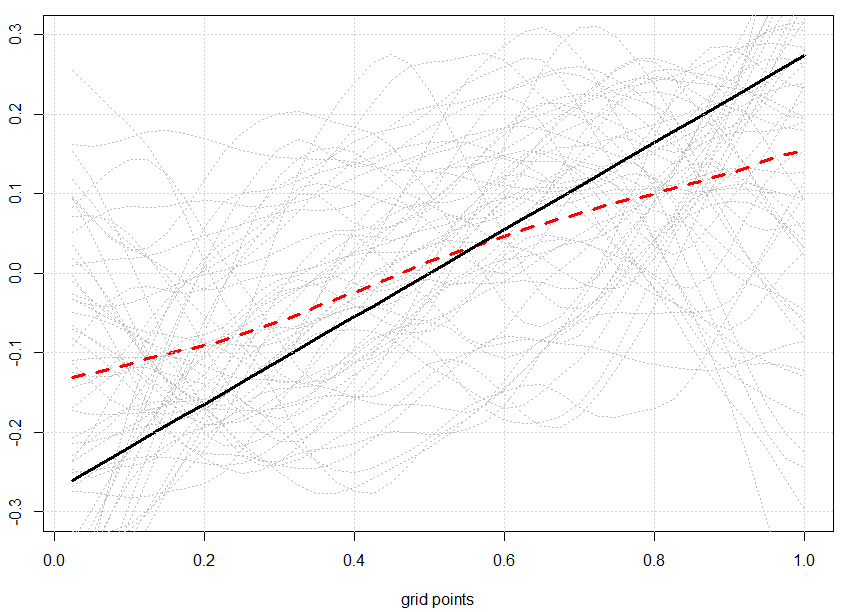}&
\includegraphics[height=4.5cm, width=5cm]{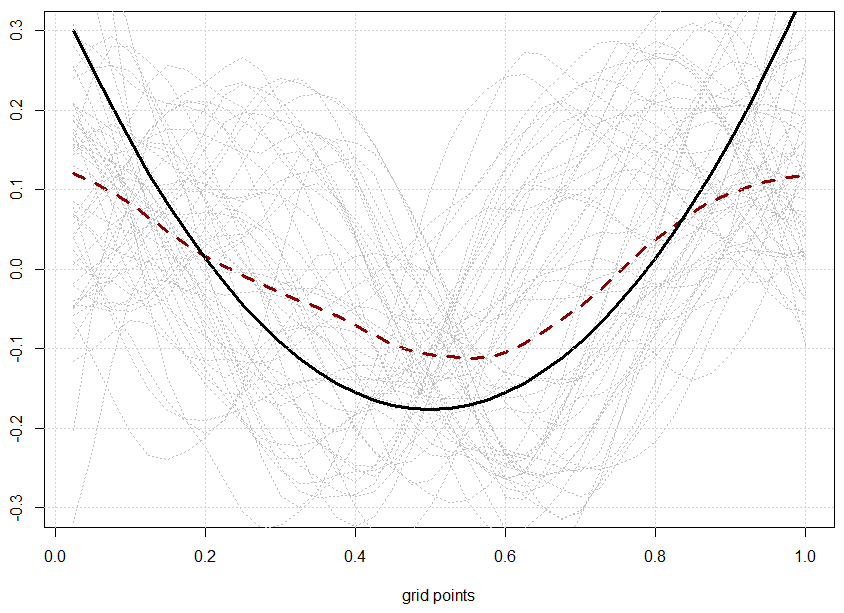}\\
\multicolumn{3}{c}{\scriptsize{\textbf{$\mathbf{U-}$level}}}\\
\includegraphics[height=4.5cm, width=5cm]{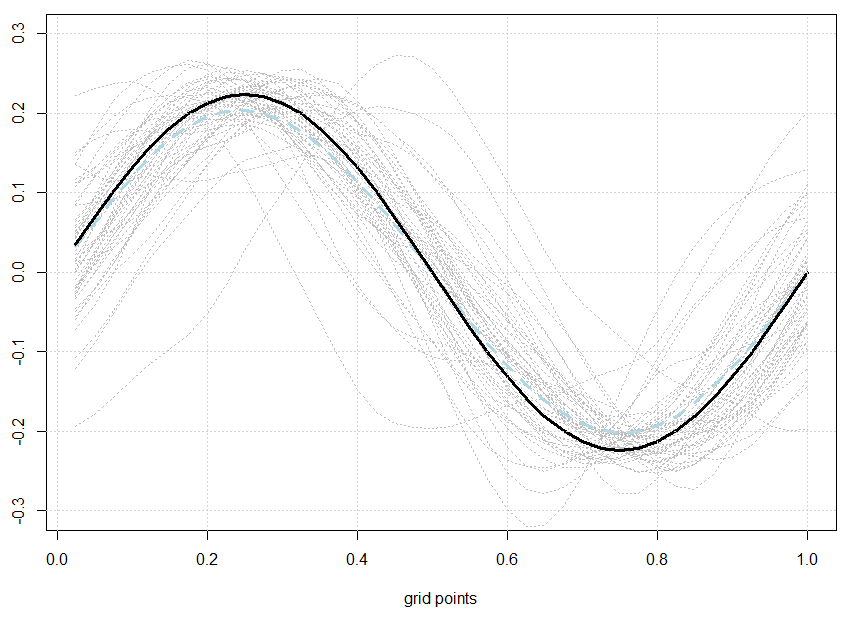}&\includegraphics[height=4.5cm, width=5cm]{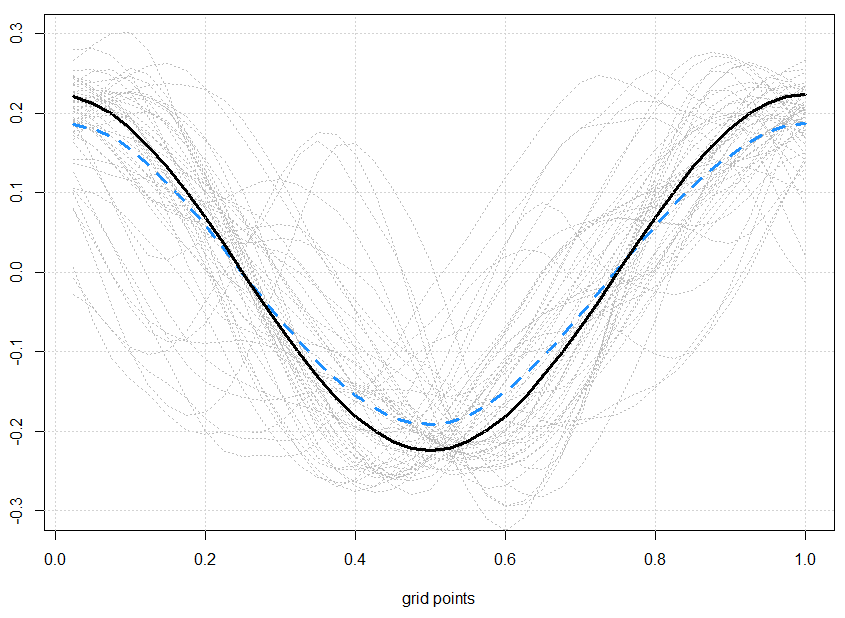}&
\includegraphics[height=4.5cm, width=5cm]{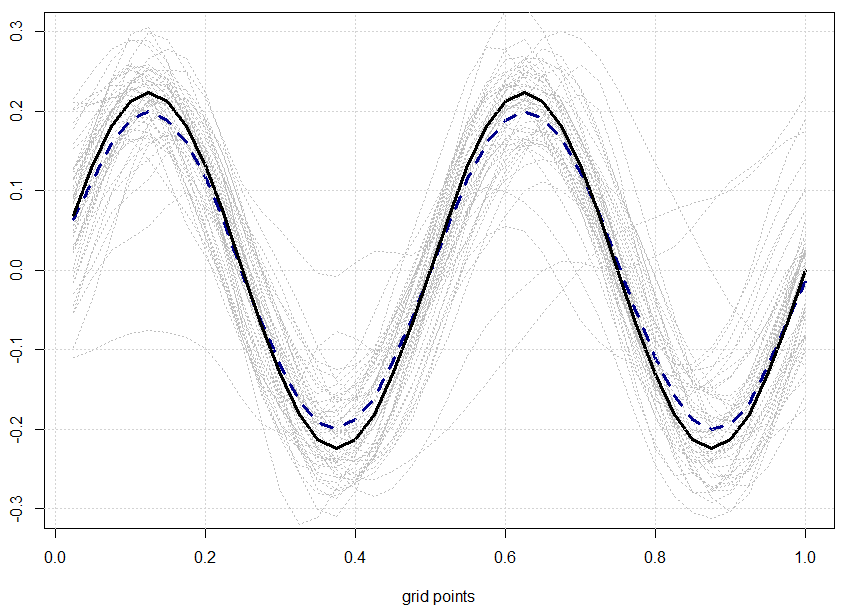}\\
\multicolumn{3}{c}{\scriptsize{\textbf{$\mathbf{W-}$level}}}\\
\includegraphics[height=4.5cm, width=5cm]{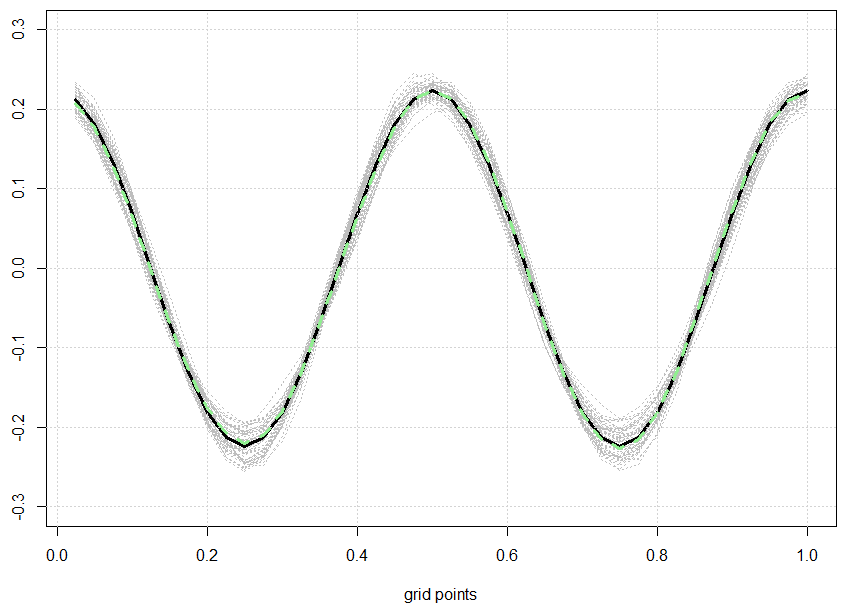}&\includegraphics[height=4.5cm, width=5cm]{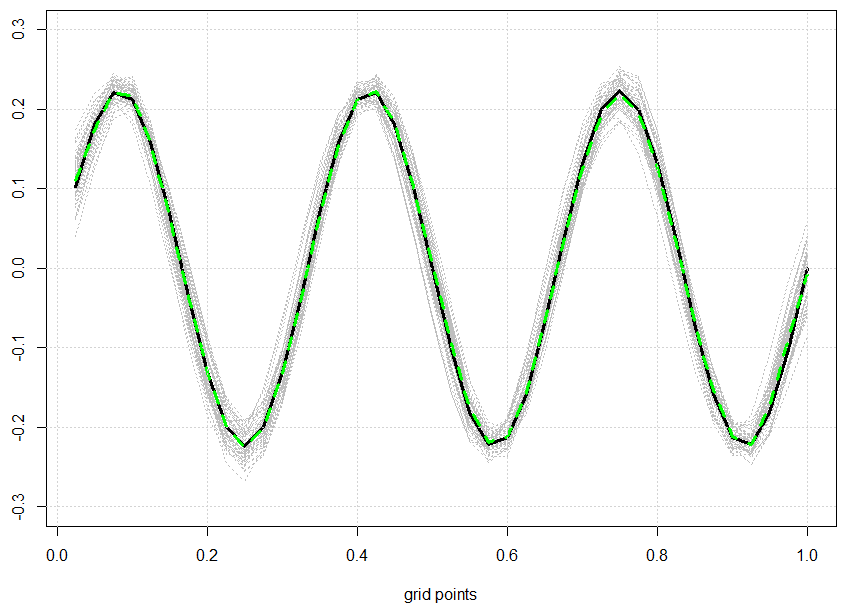}&
\includegraphics[height=4.5cm, width=5cm]{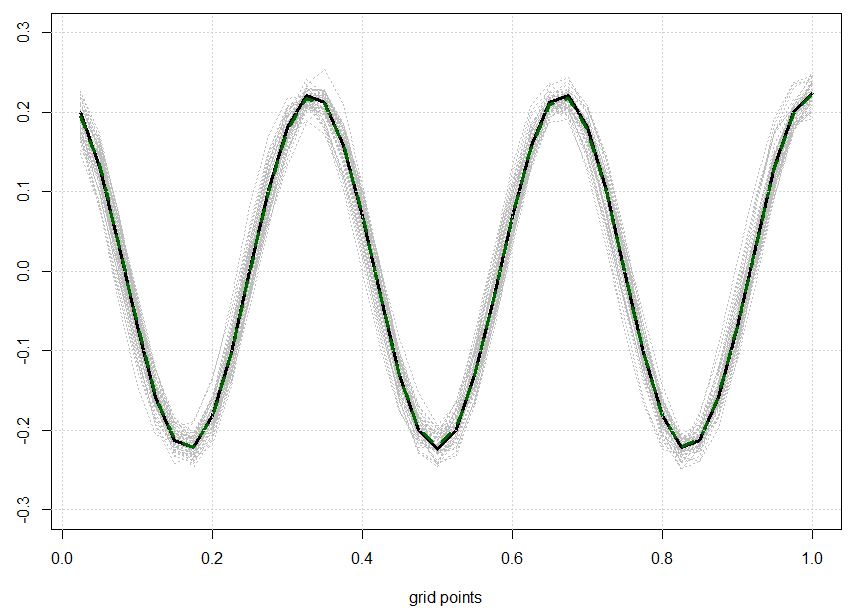}\\
\multicolumn{3}{c}{\scriptsize{\textbf{$\mathbf{V-}$level}}}\\
\includegraphics[height=4.5cm, width=5cm]{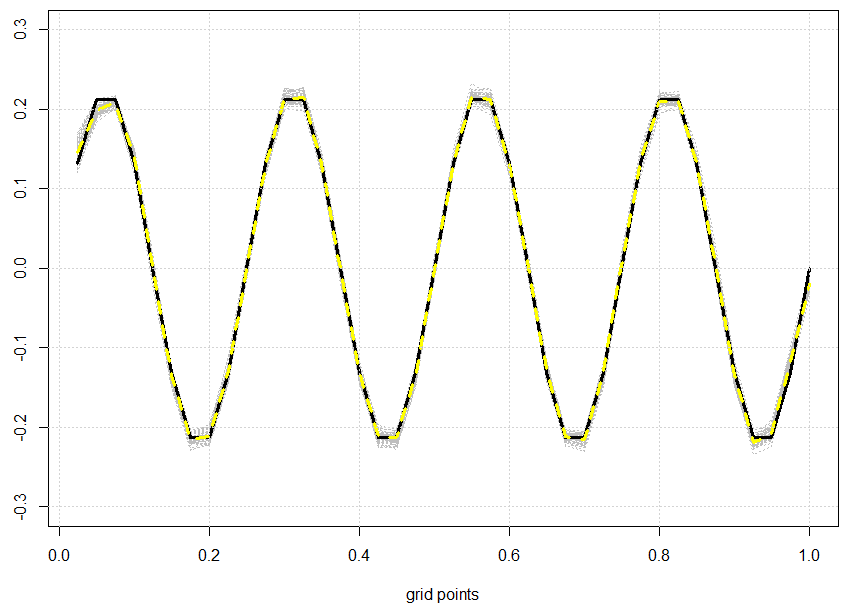}&\includegraphics[height=4.5cm, width=5cm]{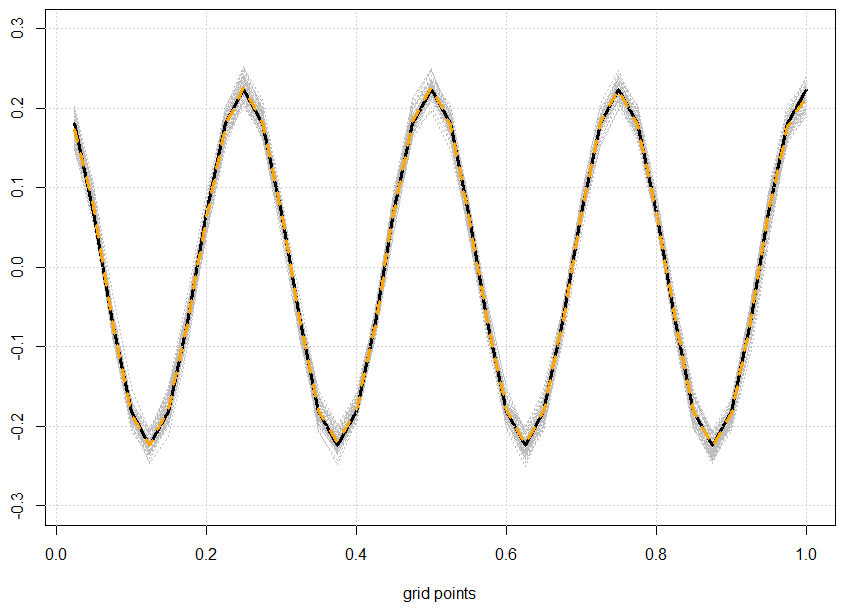}&
\includegraphics[height=4.5cm, width=5cm]{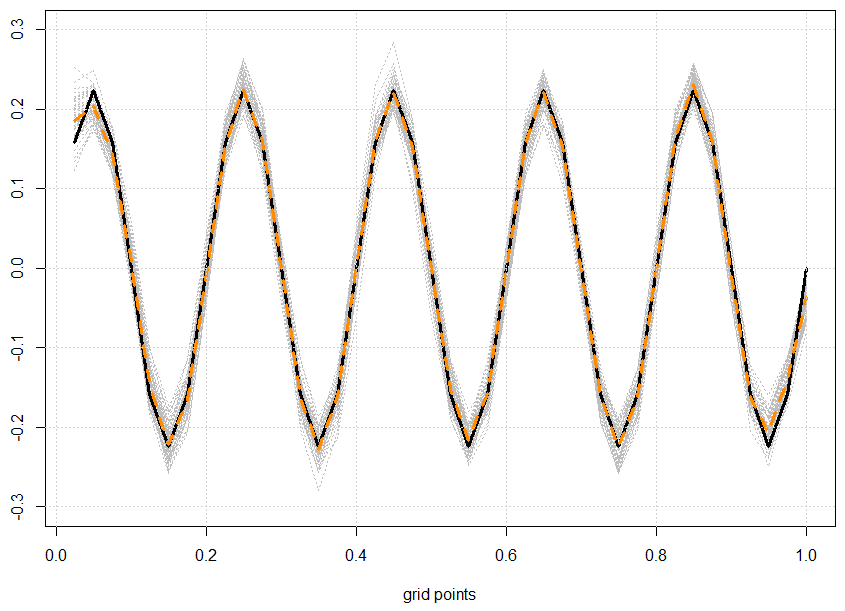}
\end{tabular}
\caption{\textbf{Eigenfunctions/principal components, $\boldsymbol{\sigma=1}$.} True eigenfunction (black line), mean of 100 estimated eigenfunctions (red line), 50 estimated eigenfunctions (gray lines).}\label{fig:eigfcts_sim_sig1}
\end{figure}

In Figures \ref{fig:eigfcts_sim_sig0} and \ref{fig:eigfcts_sim_sig1}, we display the estimated eigenfunctions for $\sigma=0$ and $\sigma=1$, respectively.  On average, the shape of all but the coarsest-level ($X-$level) eigenfunctions is well recovered when $\sigma=0$. With only six independent units, the $X-$level faces the issue of a very small sample size, which explains the large variation in estimated eigenfunctions in individual simulation repetitions. What is more, as the nested lower levels are unbalanced, the level contrasts collected in the $X-$level $G-$matrix $G_X$ (see Equation (\ref{eq:gmatrices})) are biased. To appreciate this issue more thoroughly, it is useful to examine the balanced case first. When the design is balanced, there is an equal number of observations in each group in level $n_\ell$ (nested in $X$) collected in the associated $H-$matrices. Thus, each group carries the same structural information which makes the resulting contrast or $G-$matrices and thus the estimated covariance representative. When the design is unbalanced, however, larger groups tend to dominate smaller groups in the $G-$matrices that are generated through successive subtraction of the $H-$matrices, see Equations (\ref{eq:gmatrices}). Consequently, the estimated covariances mostly reflect the structure of larger groups rather than that of the average group. $G_X$ is most sensitive to these over- and under-subtractions. The nested finer level $G-$matrices benefit from the fact that, in the hierarchically higher levels, part of this misassignment is mitigated by the next subtraction of $H-$matrices. However, $G_X$, belonging in the top level, inherits all the remaining deviations without the opportunity for further correction. Thus, although the $H-$matrices correctly encode unbalanced group sizes, unbalancedness induces non-uniform group-level information in the resulting covariance operator. As a result, this contrast-based approach for separating the levels leads to group-dependent over- and under-subtraction, which, in turn, becomes most apparent in the estimated eigenfunctions at coarser levels. Moving from the coarsest to the finer levels in Figure \ref{fig:eigfcts_sim_sig0}, we observe a rapid improvement in estimation accuracy due to the increasing level-specific sample size. A similar picture arises when $\sigma = 1$ in Figure \ref{fig:eigfcts_sim_sig1}. This result is also corroborated numerically in Table \ref{tab:comparison_true_vs_est}, top panel, where the median MSE of true and estimated eigenfunctions is shown for all signal-to-noise-ratios examined. The only notable difference comparing scenarios $\sigma=0$ and $\sigma=1$ is the variation in estimated $V-$level eigenfunctions in individual simulation repetitions. When $\sigma=1$, the variation is slightly larger, indicating that, in the simulation, the noise is introduced into the data through the finest level. However, the median MSE for comparing $\phi_{n_4}^V$ and $\widehat\phi_{n_4}^V$ shows only very small deviations from the case of $\sigma=0$, hinting that the covariance surface smoothing using penalized regression splines works sufficiently well.

The median MSE of the $X-$ and $U-$level scores is larger for $\sigma = 0$ than for $\sigma > 0$ scenarios where covariance smoothing is applied (see Table \ref{tab:comparison_true_vs_est}, middle panel). At first sight, this result seems counterintuitive as, without any noise, all the level components lie exactly in the same space. As there is no dispersion, the level spaces are strongly collinear, which leads to an almost perfect overlay of the level effects on the one hand, but to very little discriminatory power on the other. As a consequence, the $G-$matrices, acting as contrast operators, have difficulty separating these almost identical level spaces which becomes visible in the increased median MSE for $X-$ and $U-$level scores when we compare the no noise scenario to those scenarios featuring noise and smoothing. In the hierarchically lower levels, the score estimation improves overall as well as for $\sigma = 0$ relative to $\sigma > 0$ because these levels benefit of a larger sample size and are thus less dependent on a perfect separation. Generally, introducing noise to the model  has a similar effect as regularization and hence improves level separation in  the sense that it increases eigenvalue gaps, makes level spaces more stable and projections less ambiguous. This is particularly true for the coarser $X-$ and $U-$levels. Adding smoothing improves the estimation even more as the noise, which is orthogonal to the hierarchical structures in expectation, is removed but the smoothed, structured level effects are maintained. This effect also contributes to the smaller median MSEs for the $\sigma >0$ scenarios.

The estimation of the eigenvalues is largely unaffected by moderate noise (see Table \ref{tab:comparison_true_vs_est}, bottom panel); the MSE is almost constant across the noise scenarios. This is due to the fact that eigenvalues measure how much variance is collected in each level space, while scores measure how exactly each realization is projected onto this level space. Therefore, eigenvalue estimation is more stable when confronted with noise or imperfect separation.

Considering all estimates, the differences in magnitude of the (median) MSEs relative to a balanced setting in the spirit of  \cite{shou2015structfpca} (see Appendix Table \ref{tab:comparison_true_vs_est_balanced}) arise from the unequal group sizes in the lower hierarchical levels, which, as discussed above, induce a non-uniform contribution of group-level information in the level-specific covariance operators. Furthermore, \cite{shou2015structfpca}'s own simulation study employs a grid dimension of $p=50,000$ and $n=1,250$ curves while the study in this paper works with $p=40$ and $n = 23,898$. As mentioned before, estimation of eigenfunctions and scores on a coarser grid can be challenging because it implies a coarser discretization of the curves. Because, geometrically, eigenfunctions represent the shape of the curves and scores are the projections onto this shape, more grid points lead to better curve approximations and more stable projections. Eigenvalues, however, measure the entire area under each curve which is almost invariant to a reasonably large grid dimension.

Summing up, we find the following. \emph{Eigenvalue estimation} is largely insensitive to grid resolution and noise level, but becomes sensitive under more strongly unbalanced designs, as eigenvalues capture the total variance allocated to each hierarchical level, compare bottom panels of Tables \ref{tab:comparison_true_vs_est} (unbalanced design, coarse grid), \ref{tab:comparison_true_vs_est_unbalanced_p100} (unbalanced design, finer grid), \ref{tab:comparison_true_vs_est_balanced} (balanced design, coarse grid) and \ref{tab:comparison_true_vs_est_balanced_p100} (balanced design, finer grid). 
In contrast, \emph{eigenfunctions and scores} depend on a reliable separation of hierarchical levels and are therefore more sensitive to discretization, noise, and differing (unbalanced) group sizes within the levels. Introducing observation noise improves numerical identifiability by separating otherwise nearly collinear level-specific subspaces, while subsequent smoothing attenuates noise and stabilizes the underlying structure structure. Unbalanced designs further affect the magnitude and stability of higher-level estimates by inducing a non-uniform contribution of group-level information, leading to a larger estimation error compared to balanced settings, compare top and middle panels of Tables \ref{tab:comparison_true_vs_est} (unbalanced design, coarse grid), \ref{tab:comparison_true_vs_est_unbalanced_p100} (unbalanced design, finer grid), \ref{tab:comparison_true_vs_est_balanced} (balanced design, coarse grid) and \ref{tab:comparison_true_vs_est_balanced_p100} (balanced design, finer grid).


\section{Data application -- Concussion dataset}\label{sec:data_app}

\begin{center}
\begin{figure}[H]
\begin{tabular}{ccc}
\textbf{Raw}&\textbf{Shifted}&\textbf{Shifted and Mean-centered}\\
\vspace{-0.5cm}\\
\includegraphics[height=4.5cm, width=5cm]{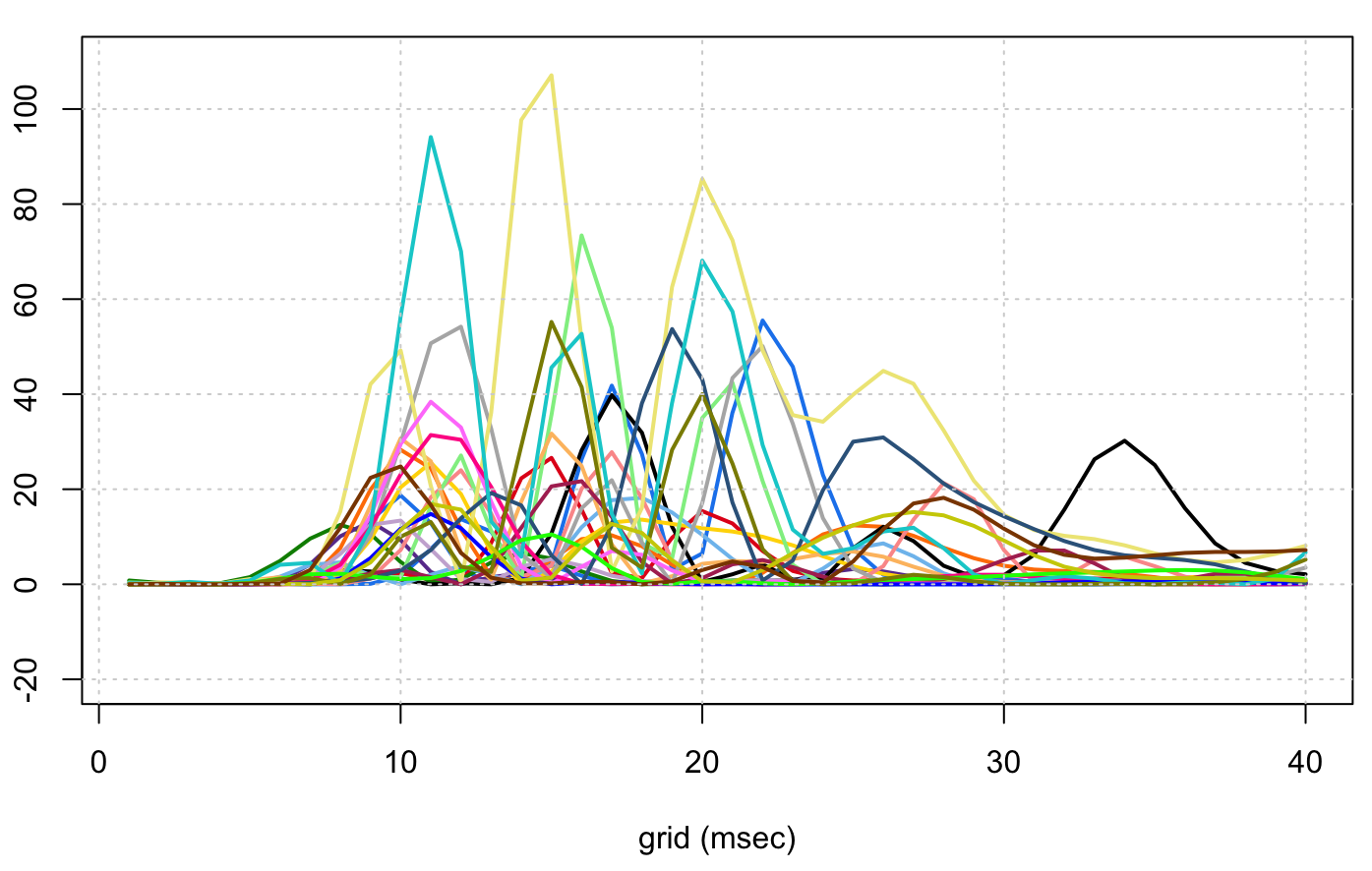}&\includegraphics[height=4.5cm, width=5cm]{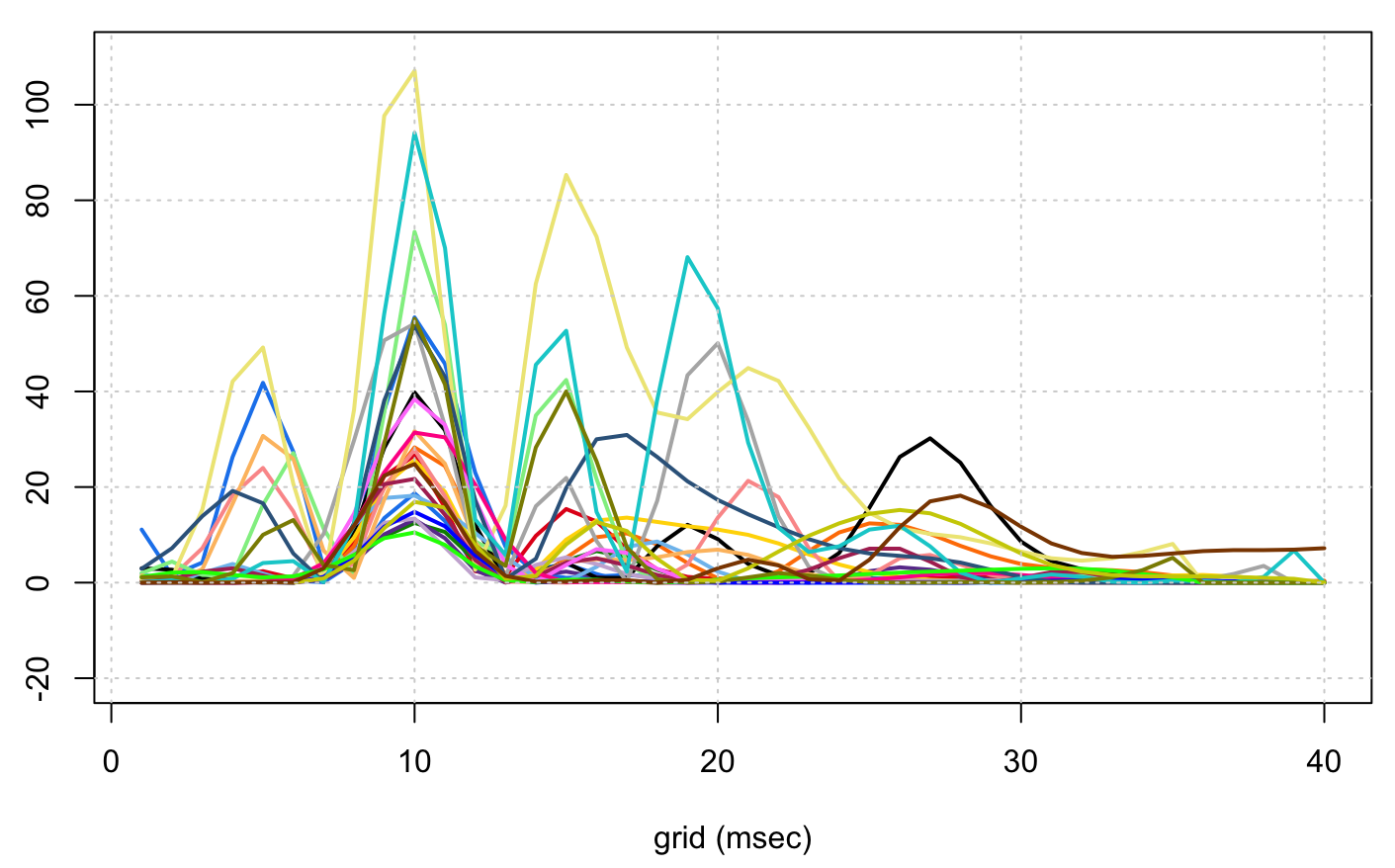}&\includegraphics[height=4.5cm, width=5cm]{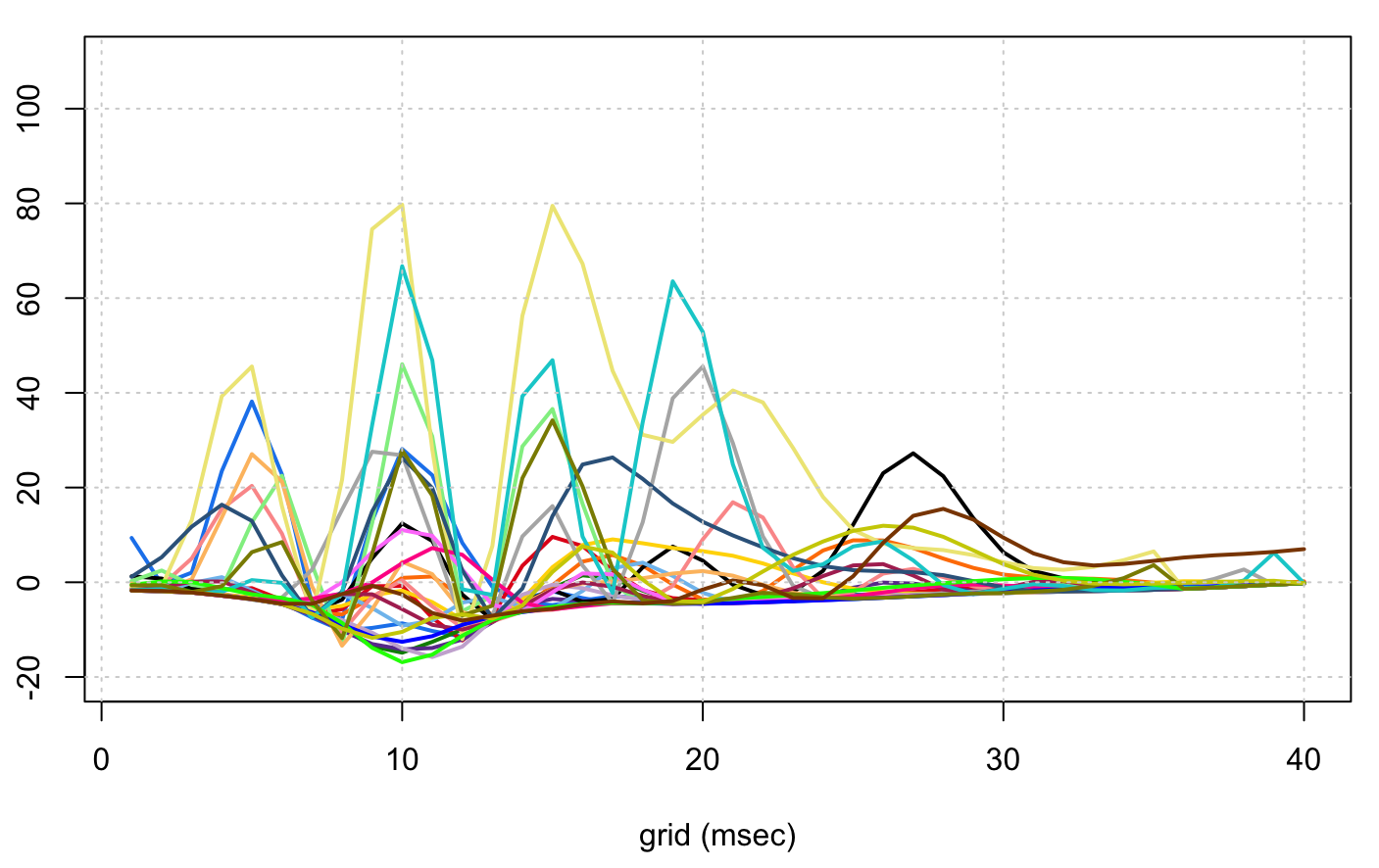}\\
\includegraphics[height=4.5cm, width=5cm]{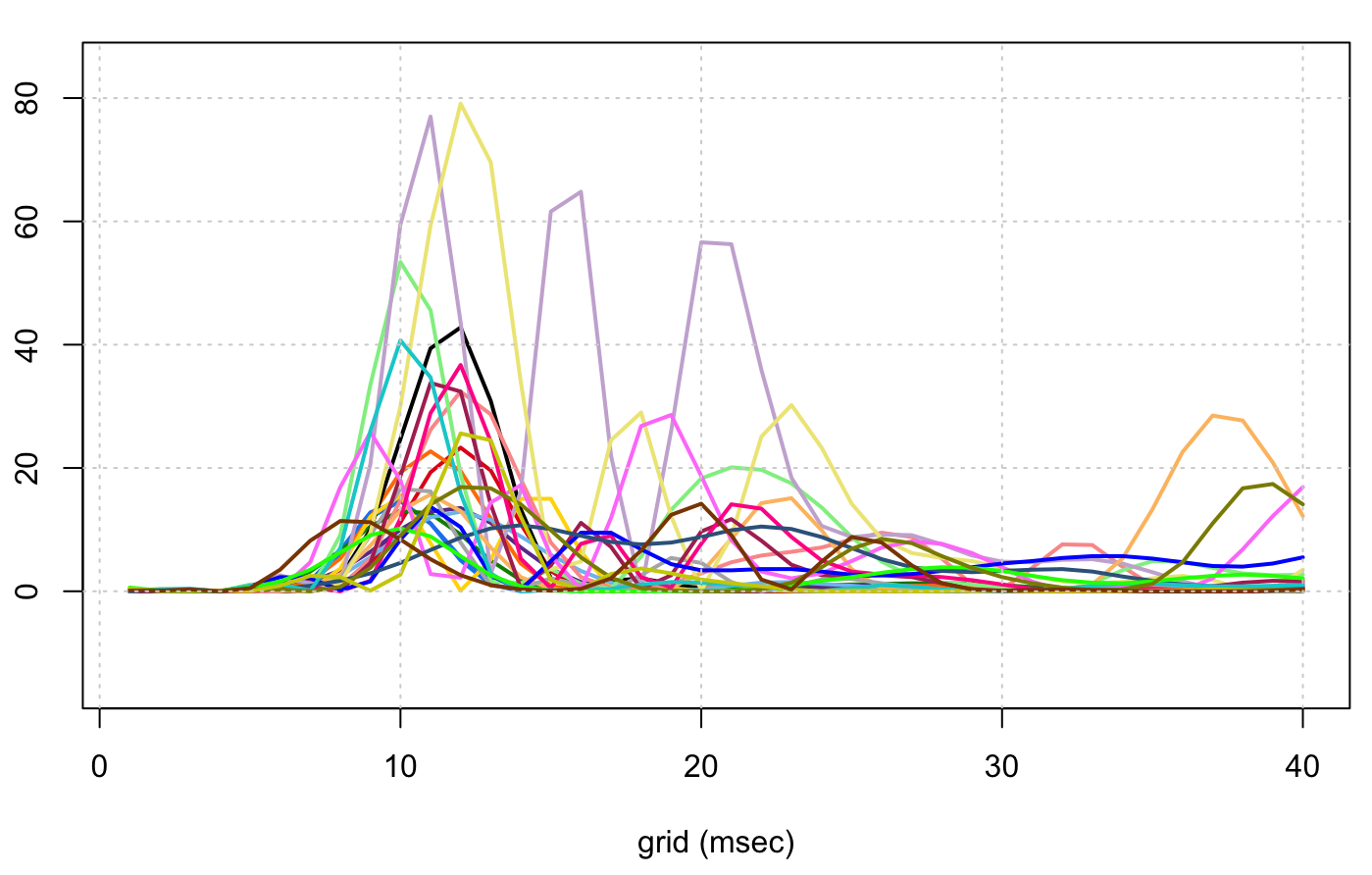}&\includegraphics[height=4.5cm, width=5cm]{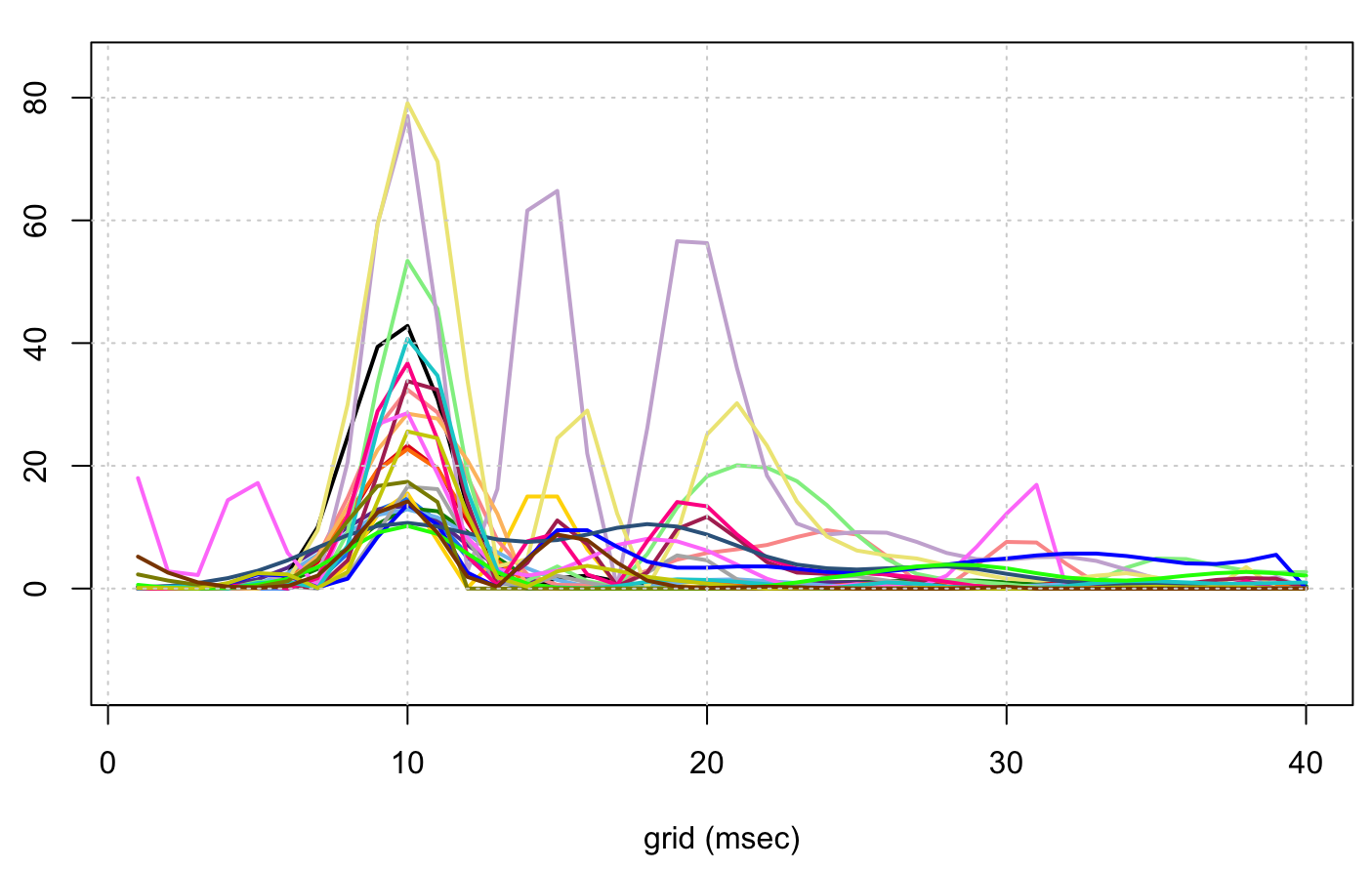}&\includegraphics[height=4.5cm, width=5cm]{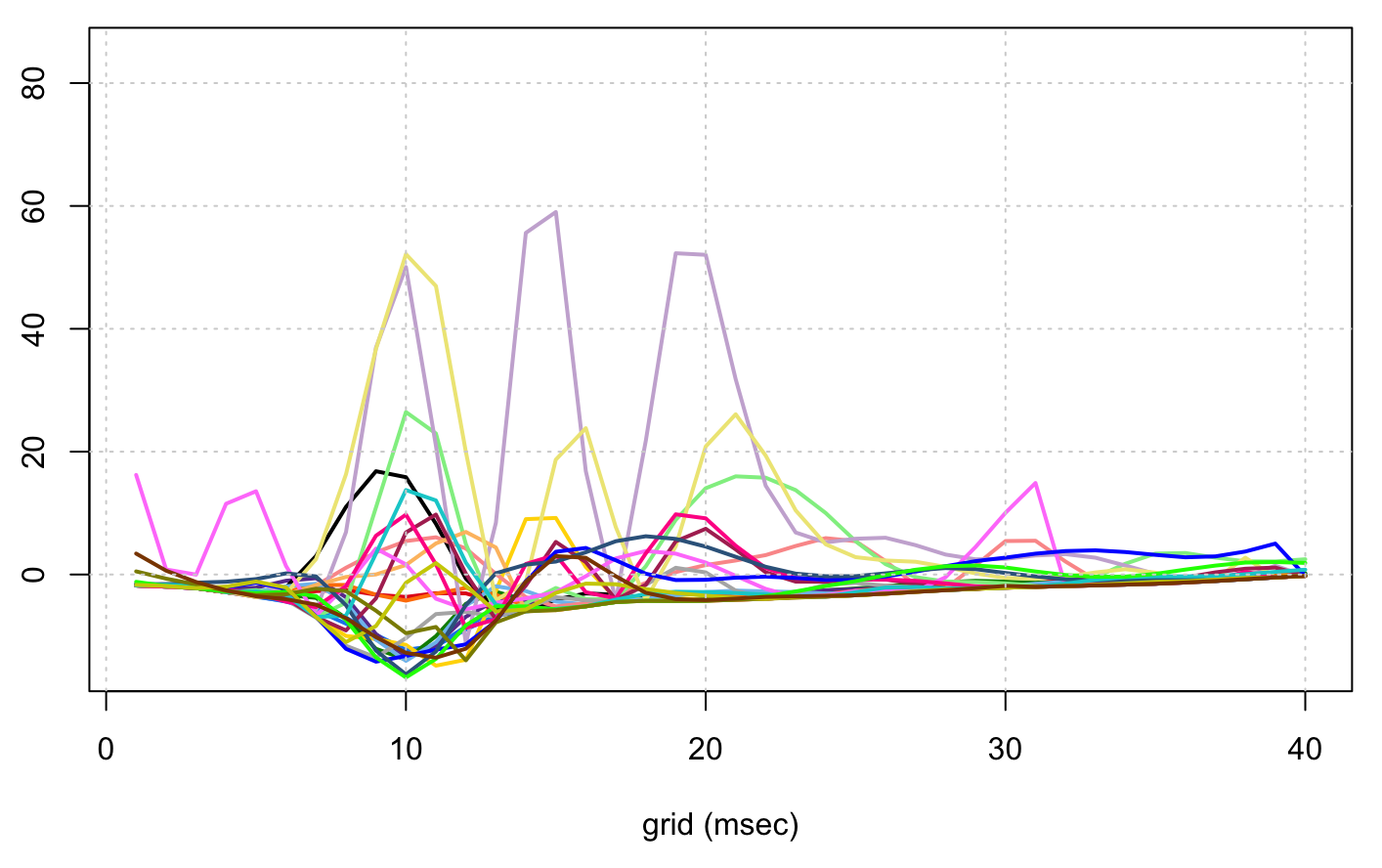}\\
\includegraphics[height=4.5cm, width=5cm]{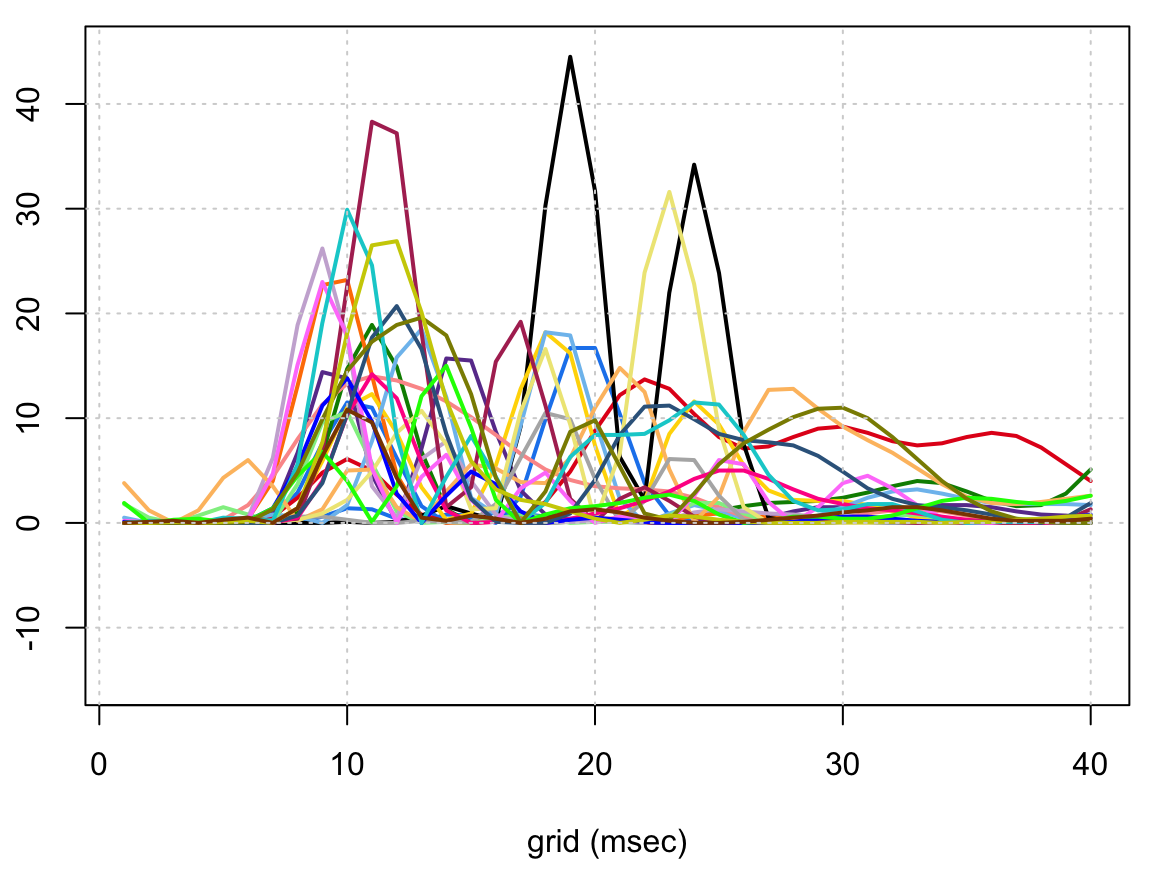}&\includegraphics[height=4.5cm, width=5cm]{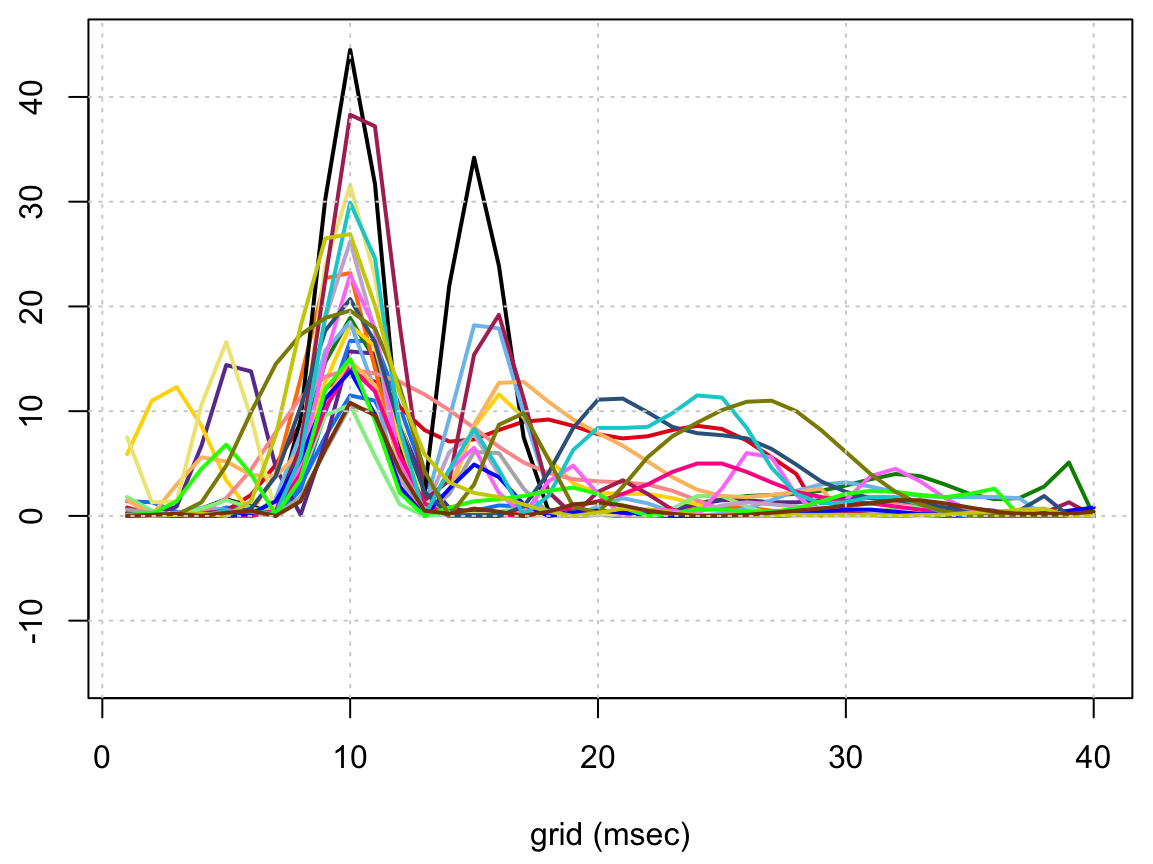}&\includegraphics[height=4.5cm, width=5cm]{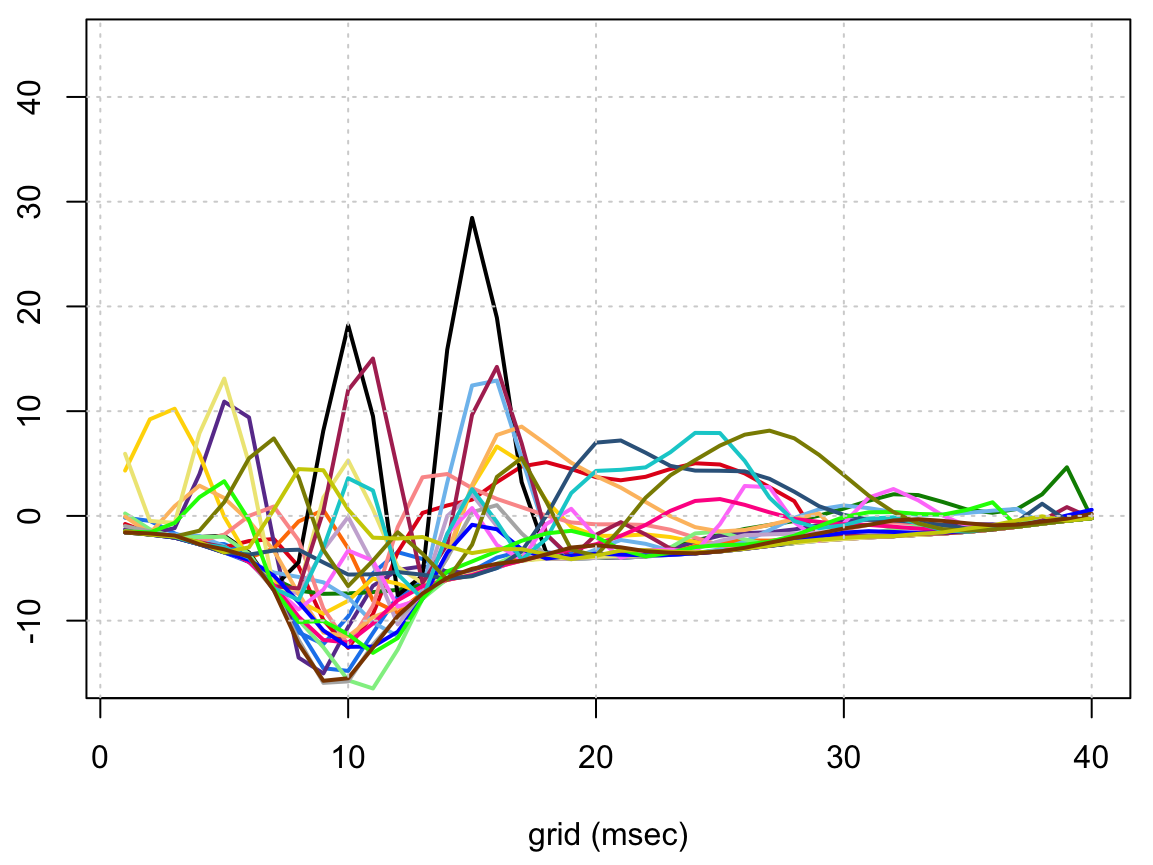}\\
\end{tabular}
\caption{\textbf{Functional observations.} Random player, first 25 impact curves, for each of the three subsets, "Games", "In-season practice", "Preseason practice" (from top to bottom row). From left to right column: (1) Raw data with 8 msec of pre-impact data, (2) Shifted, maximum peak at 10 msec, (3) Shifted and mean-centered.}\label{fig:obs_example}
\end{figure}
\end{center}

To apply the method to real-world data, we use different subsets of the NCAA-DoD Grand Alliance Concussion Assessment,
Research, and Education (CARE) Consortium (2014-2017). Methods are described elsewhere \citep[see][]{broglio2017national}. The study protocol was approved by the Institutional Review Board (IRB) at the Medical College of Wisconsin and local sites used a reliance agreement with the above-named IRB. The participants consented and enrolled in the study are varsity college football players from six NCAA Division I FBS programs. The programs included the following institutions: United States Air Force Academy (AirForce), the United States Military Academy at West Point (Army), the University of California Los Angeles (UCLA), the University of North Carolina at Chapel Hill (UNC), the University of Wisconsin (UW), and Virginia Tech (VT). In 2015, data was collected from three teams (UCLA, UNC, VT) and in 2016 and 2017, from all six teams. In the study, data is gathered using six accelerometers installed inside the football helmets (helmet-base Head Impact Telemetry (HIT) System, Riddell, SRS, Riddell, Rosemont, IL, USA) which measure the head acceleration which is then used to compute the peak component, and resultant linear and rotational accelerations \citep[see][]{duma2005analysis}. Data acquisition was prompted if a head impact was measured to be 9.6 g or greater by any single accelerometer. However, observations were only included in the dataset if the peak resultant linear acceleration exceeded 10 g. The system records 40 msec of high-frequency (1000 Hz) data at each trigger event which includes 8 msec of pre-trigger data. More details about the study can be found for example in \citet{stemper2019comparison}. 
A sequence of 40 msec of head linear accelerations constitutes a single functional observation or impact curve in our framework. As an example, we display 25 of these 40 msec curves for one player for each of the subsets in Figure \ref{fig:obs_example}. The first column shows the raw data where the peak acceleration occurs later for some impact curves. To mitigate inaccuracies in the principal component analysis due to this additional measurement-induced variability, we shifted the impact curves for all players to have their peak component at 10 msec (second column in Figure \ref{fig:obs_example}). The last step in our preprocessing is to mean-center the curves across the sample (third column in Figure (\ref{fig:obs_example})) which is required for PCA as otherwise, FPCA would not decompose the variation of the curves but a combination of their variation and the mean structure. As a result, the first principal component would mostly show the mean function.

We consider three subsets of the CARE/HIT study. Table \ref{tab:subsets} presents the details for each of these subsets which will be called Subset 1, Subset 2 and Subset 3 in the running text. All subsets use data from fall 2017, from five of the six schools, and from five positions. Subset 1 considers data from games only and Subsets 2 and 3 contain data from in-season and preseason practices, respectively.

The four-level nested functional model for the observed data in the considered subsets can be written as follows:
\begin{eqnarray}\label{eq:modeldata}
    Y_{ijkl}(t) = \mu(t) + \texttt{Team}_i(t) + \texttt{Position}_{ij}(t) + \texttt{Player}_{ijk}(t) + \texttt{ImpactCurve}_{ijkl}(t) + \varepsilon_{ijkl}(t)
\end{eqnarray}
where $Y_{ijkl}(t)$ denotes the observed functional outcome for \texttt{ImpactCurve} $l$ within \texttt{Player} $k$, \texttt{Position} $j$, and \texttt{Team} $i$ at time point (millisecond) $t$. The term $\mu(t)$ represents the overall mean function. The random processes capture variability at the team, position, player, and (impact) curve levels, respectively, and $\varepsilon_{ijkl}(t)$ denotes observation noise.
    
Based on this model specification, the corresponding design-specific $n\times n$ matrices are constructed for each subset as the first step of the analysis. As Subset 2 is extremely large (see Table \ref{tab:subsets}, $n = 54,744$ curves), the memory needed for each of the associated $G-$matrices exceeds the limit of most conventional hardware. Thus, we apply the following memory-efficient approach to be able to analyze this subset. Instead of assembling them explicitly, $G_1, \dots, G_4$ are constructed via an operator-based representation. For each $G$, a block evaluation function returns the sub-matrices $G(I,J)$ on demand. Dense components are implicitly treated as low-rank terms; sparse components are accessed via sub-matrix extraction. This approach allows for the memory-efficient HODLR compression of the $G-$matrices without modifying the structure of the matrices or the hierarchical model.

\begin{figure}[H]
\begin{center}
\begin{tabular}{ccc}
\textbf{Games}&\textbf{In-season Practice}&\textbf{Preseason Practice}\\
\vspace{-0.5cm}\\
\includegraphics[height=5cm, width=5cm]{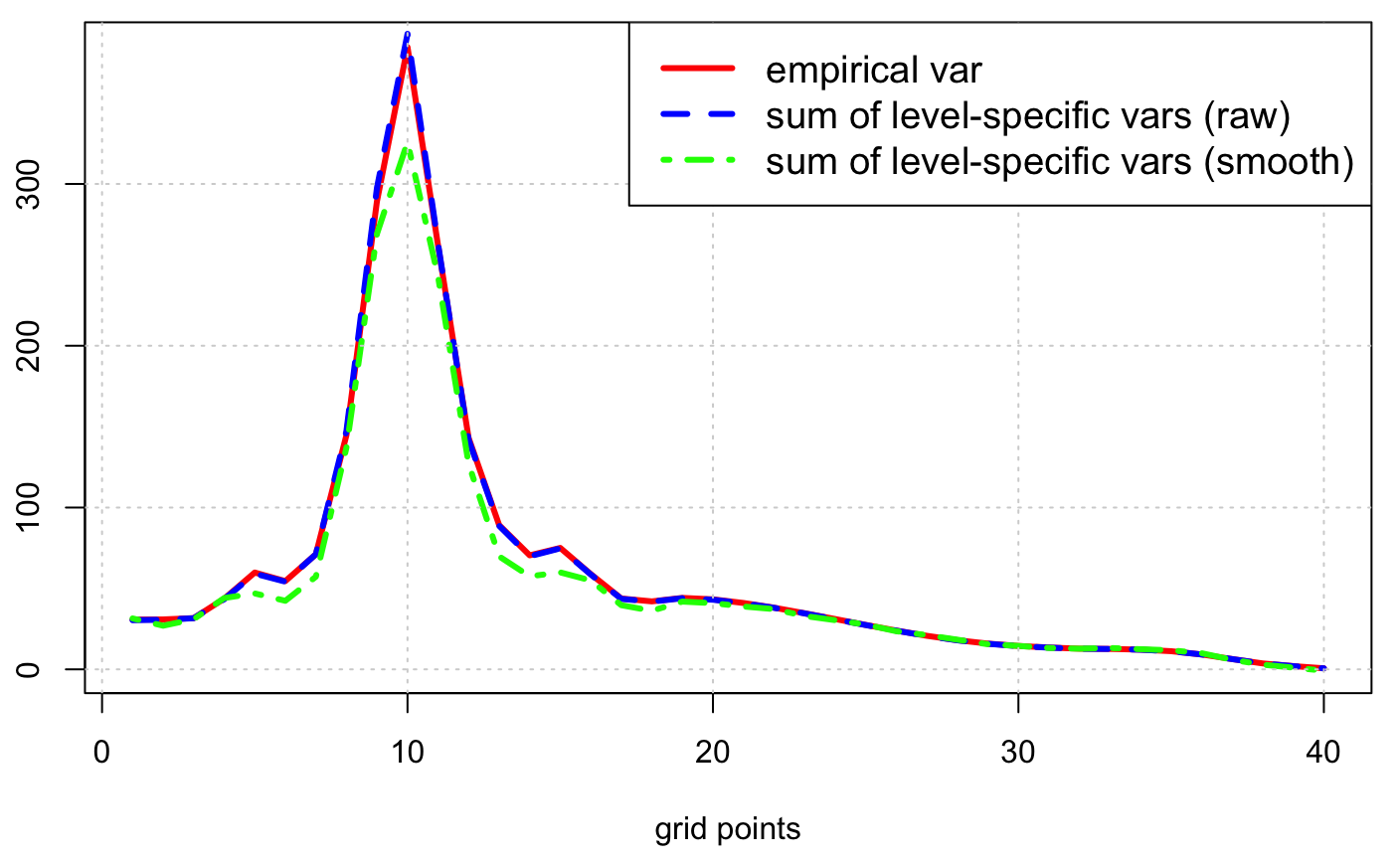}&
\includegraphics[height=5cm, width=5cm]{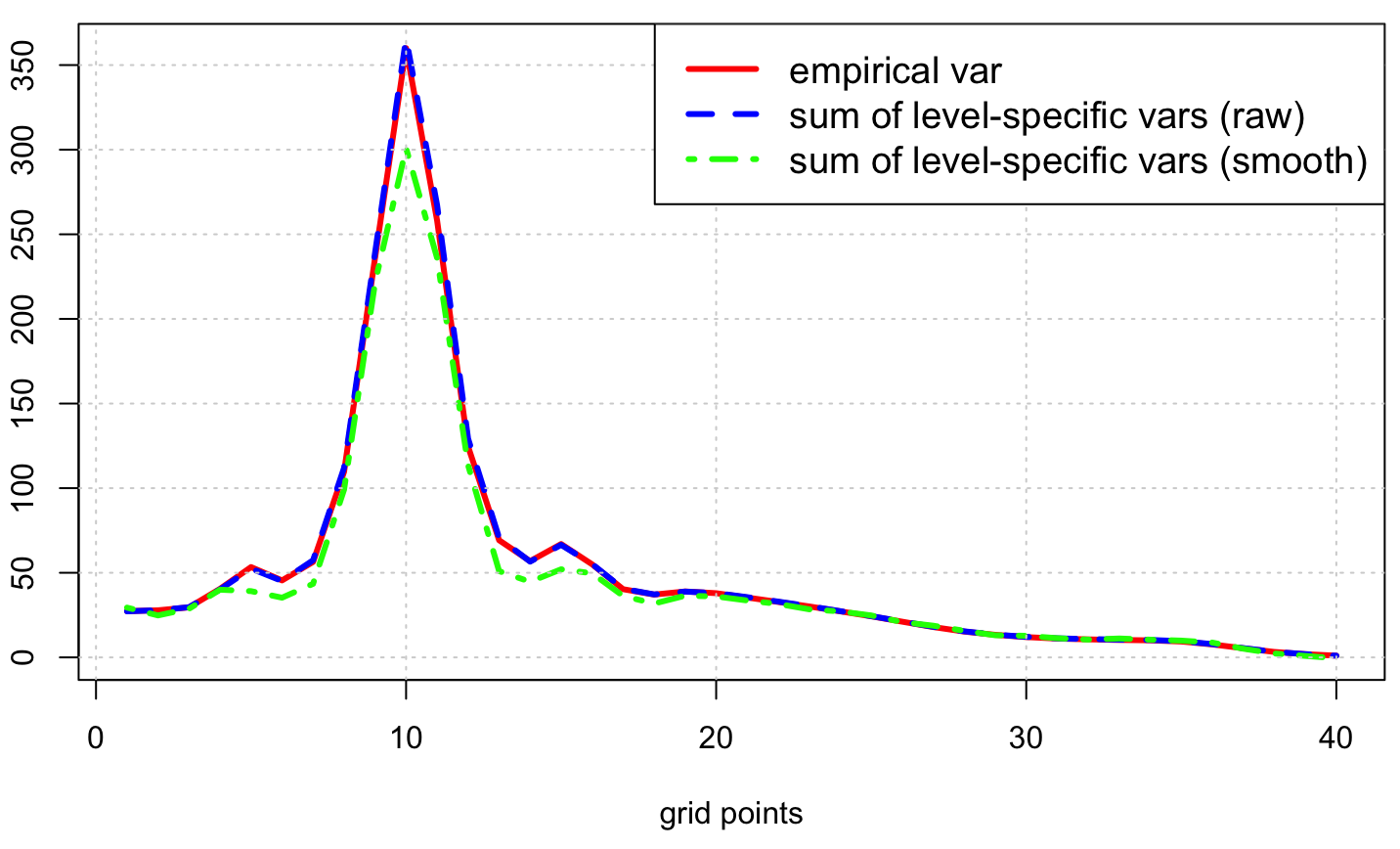}&
\includegraphics[height=5cm, width=5cm]{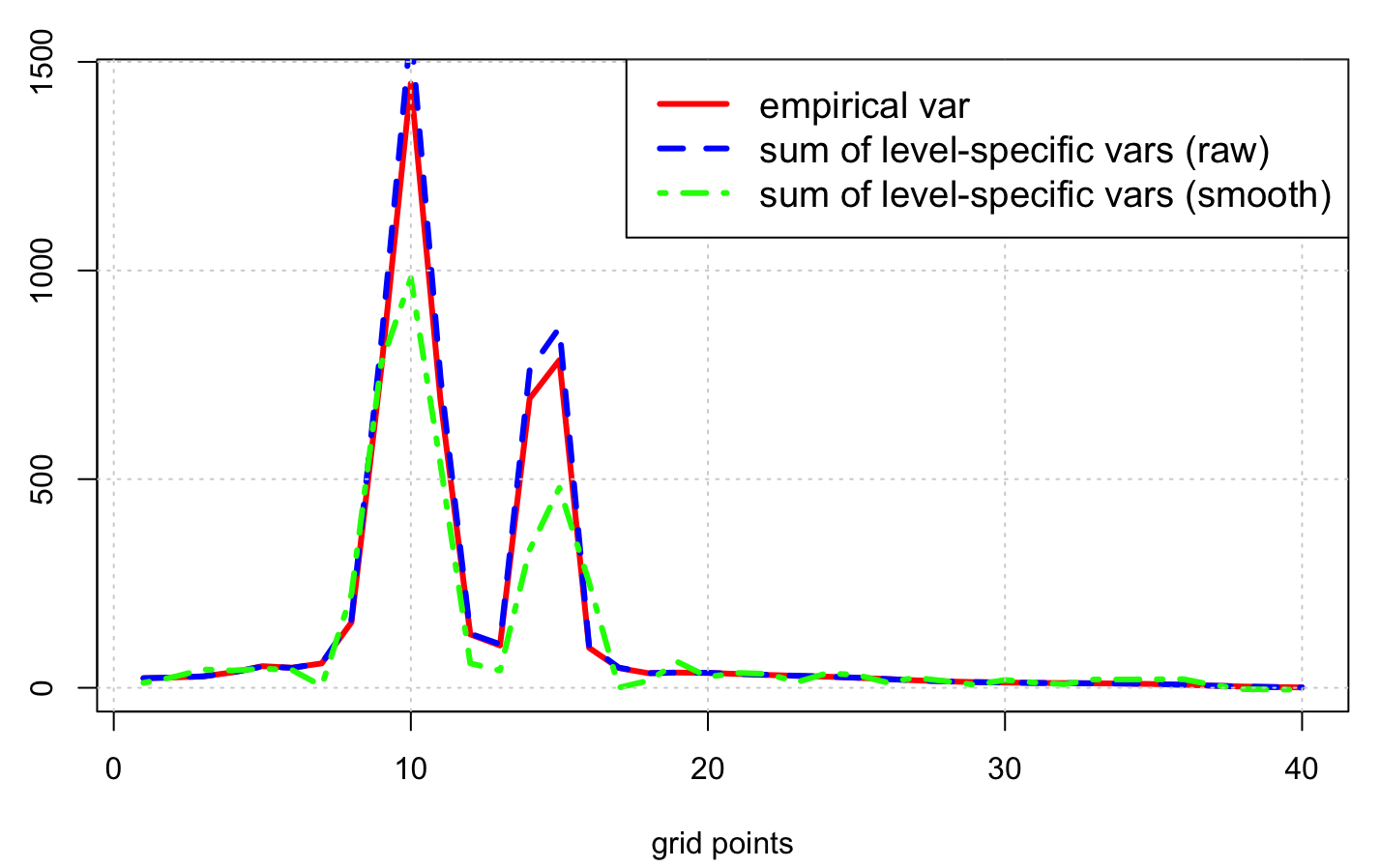}
\end{tabular}
\caption{\textbf{Empirical versus sum of raw level-specific variances versus sum of smoothed level-specific variances.} From left to right: Subset 1 "Games"; Subset 2 "In-Season Practice"; Subset 3 "Preseason Practice". Scales are different for each sample. }\label{fig:var_decomp_graphical_data}
\end{center}	
\end{figure}

\begin{figure}[H]
\begin{center}
\setlength{\tabcolsep}{1.5pt}
\begin{tabular}{cccc}
\textbf{Team}&\textbf{Position}&\textbf{Player}&\textbf{Curve}\\
\includegraphics[height=4cm, width=4cm]{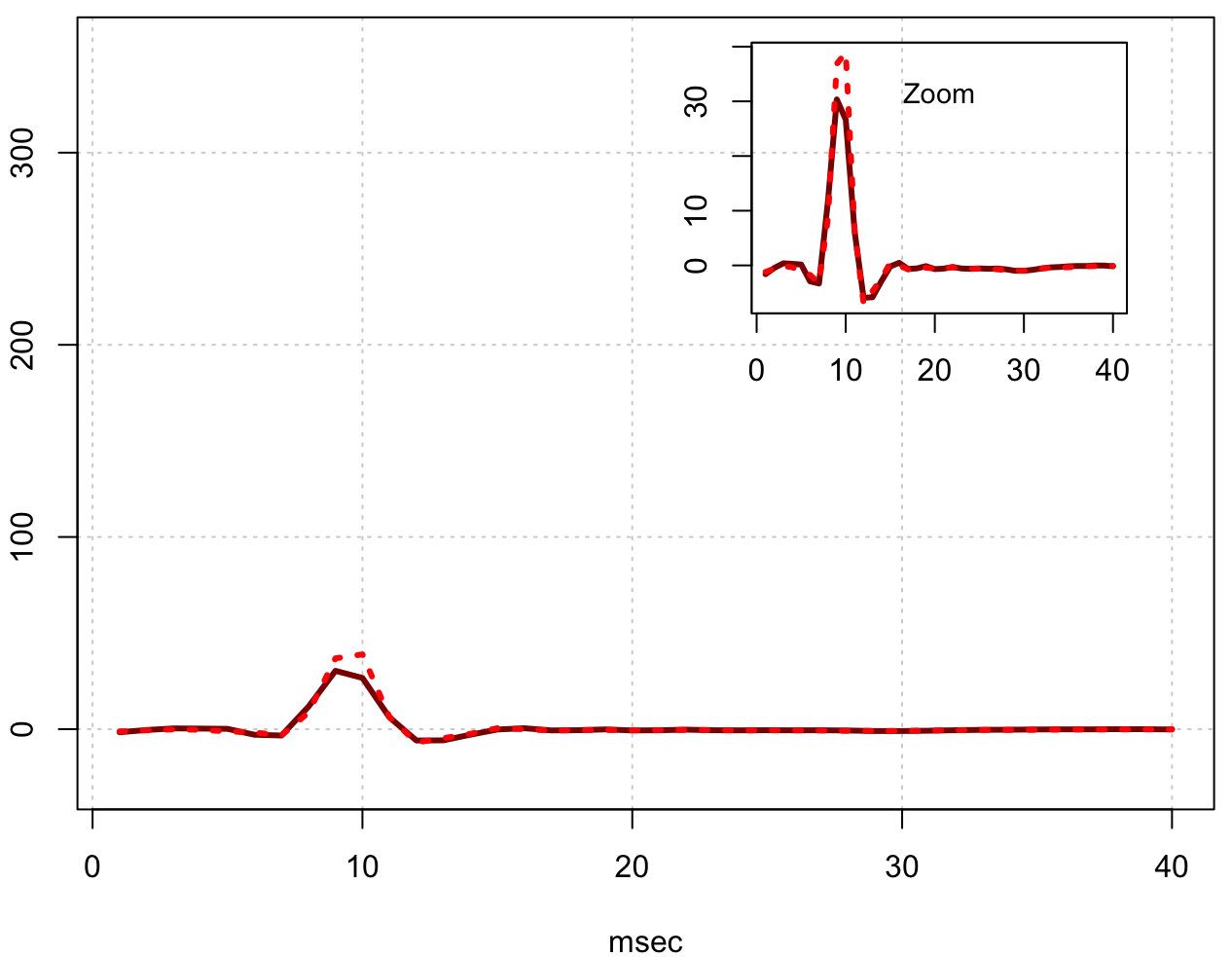}&
\includegraphics[height=4cm, width=4cm]{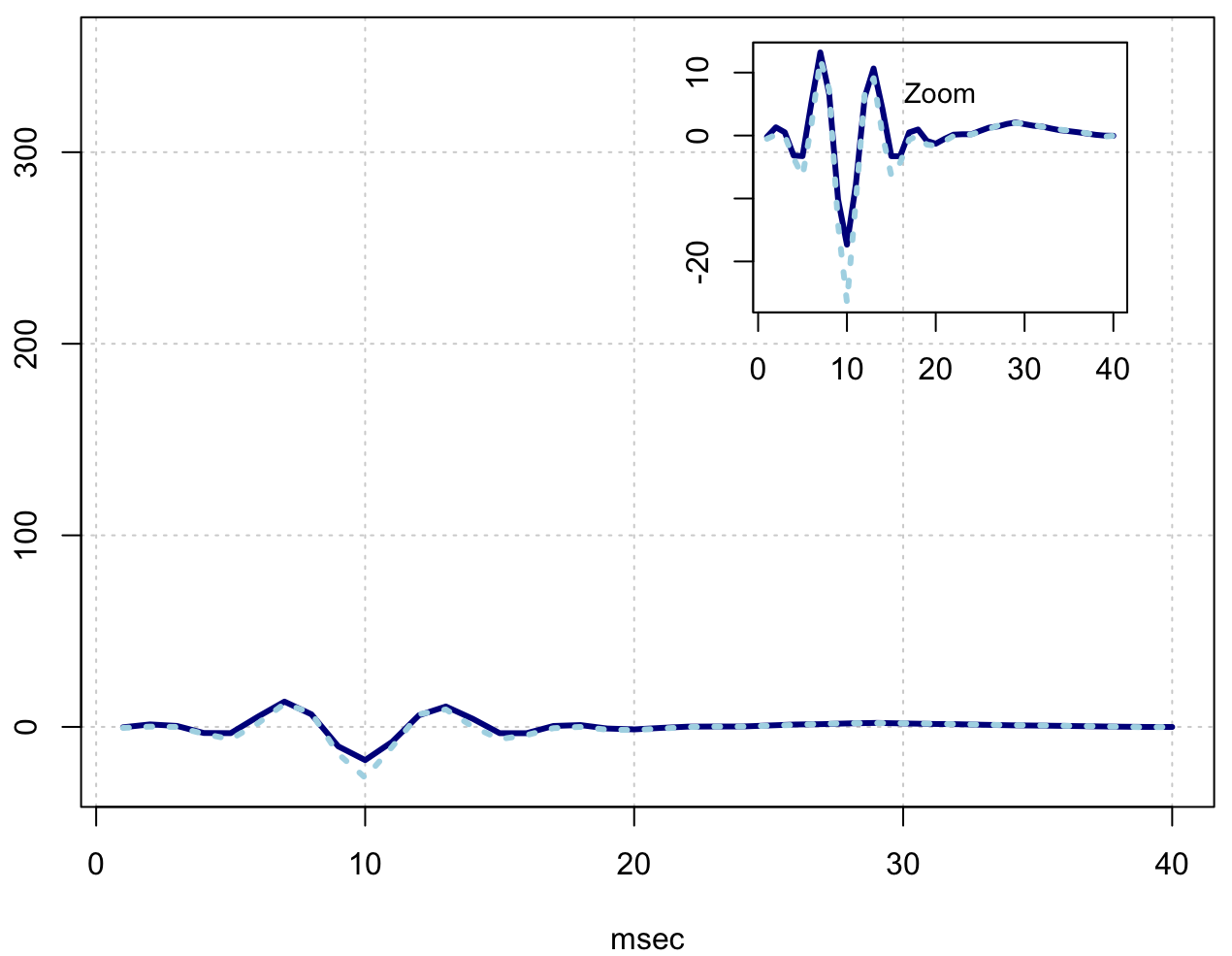}&
\includegraphics[height=4cm, width=4cm]{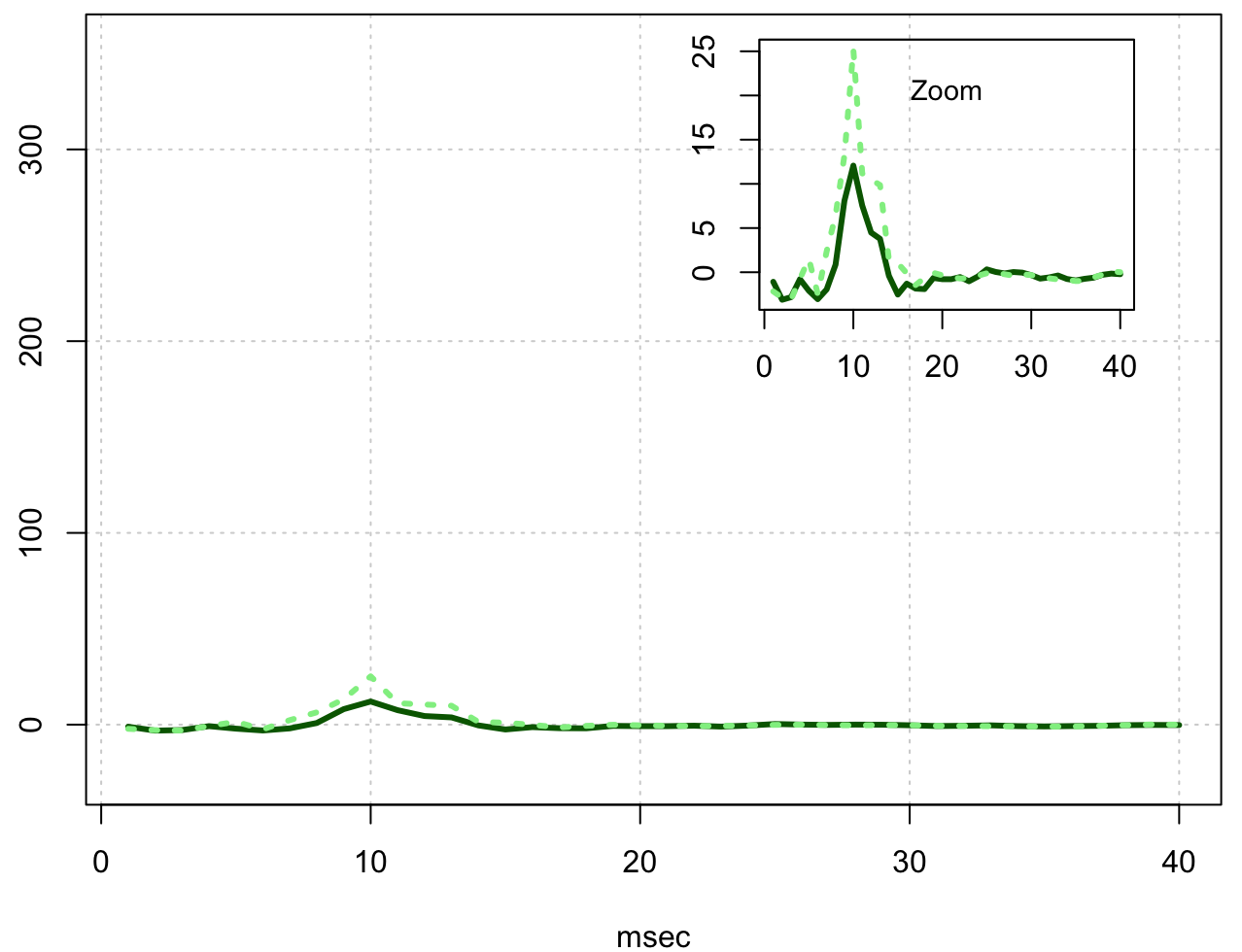}&
\includegraphics[height=4cm, width=4cm]{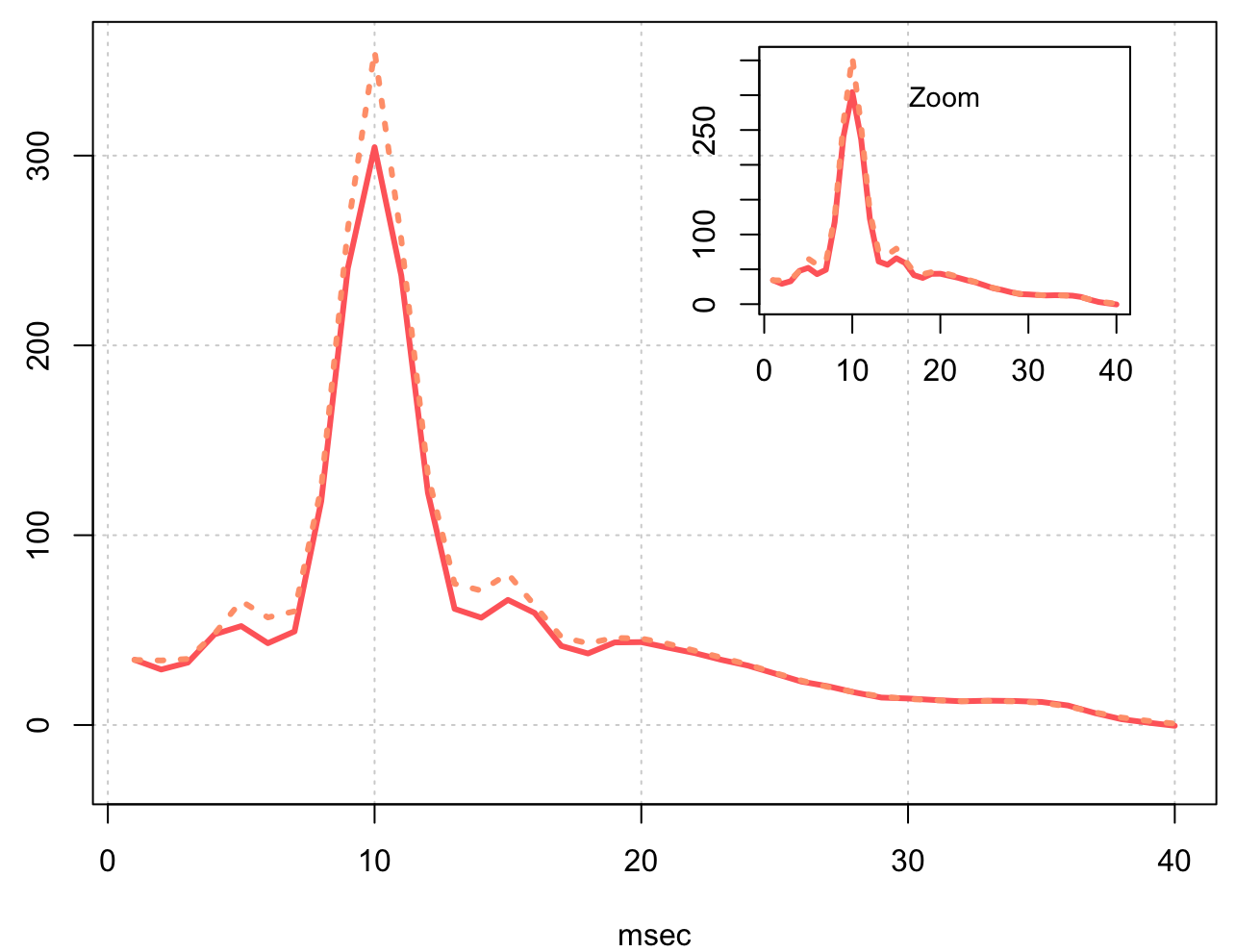}\\
\includegraphics[height=4cm, width=4cm]{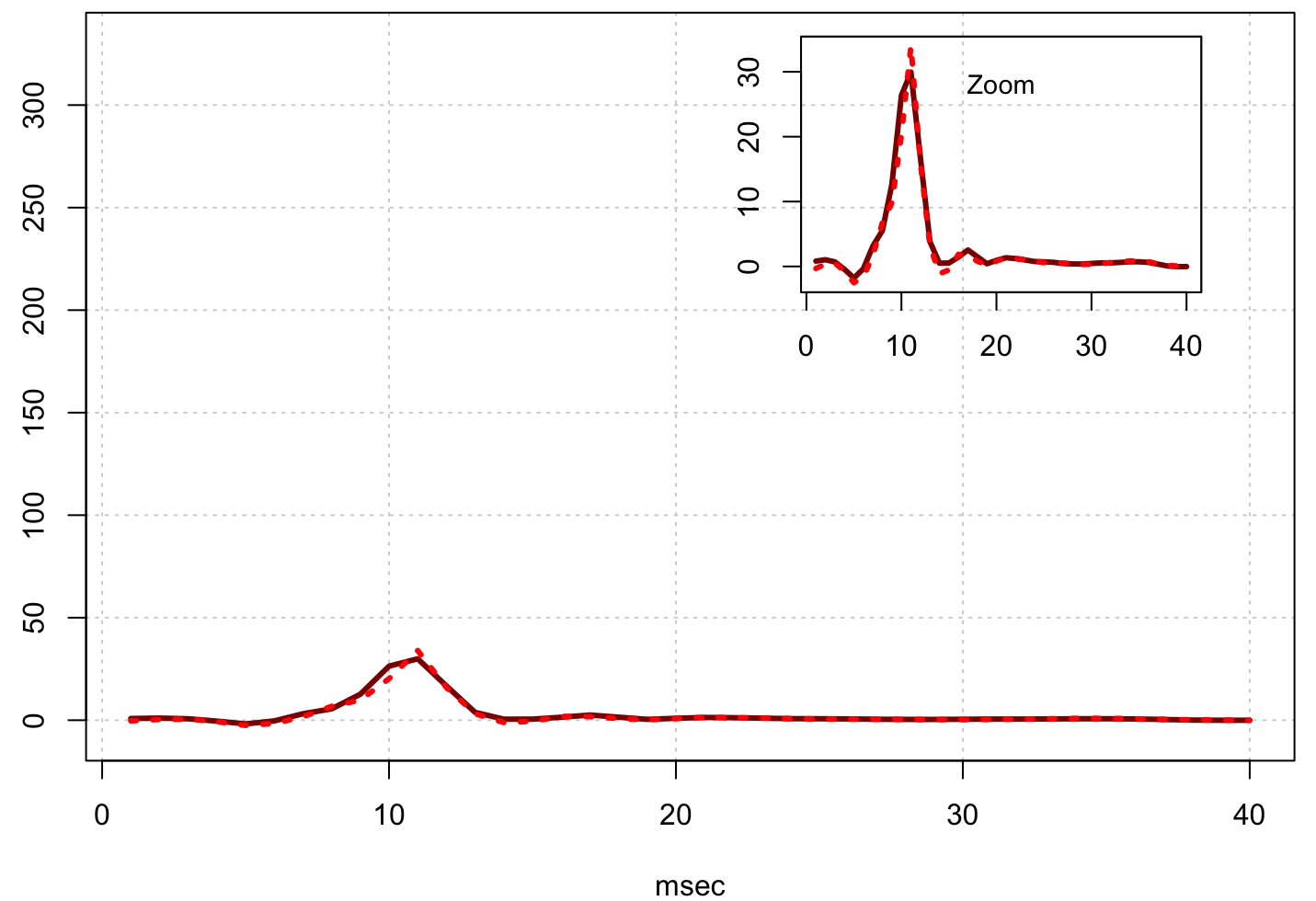}&
\includegraphics[height=4cm, width=4cm]{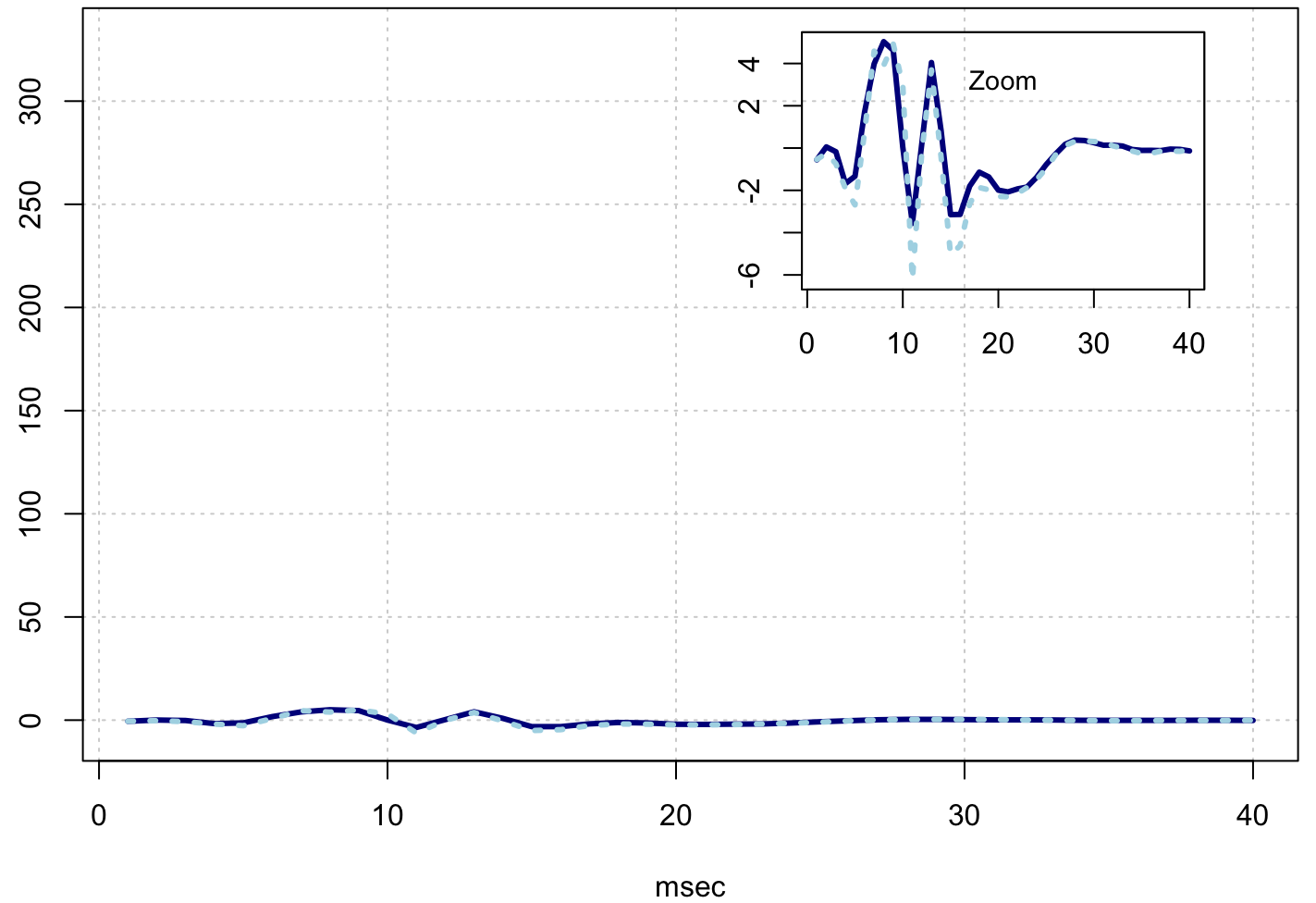}&
\includegraphics[height=4cm, width=4cm]{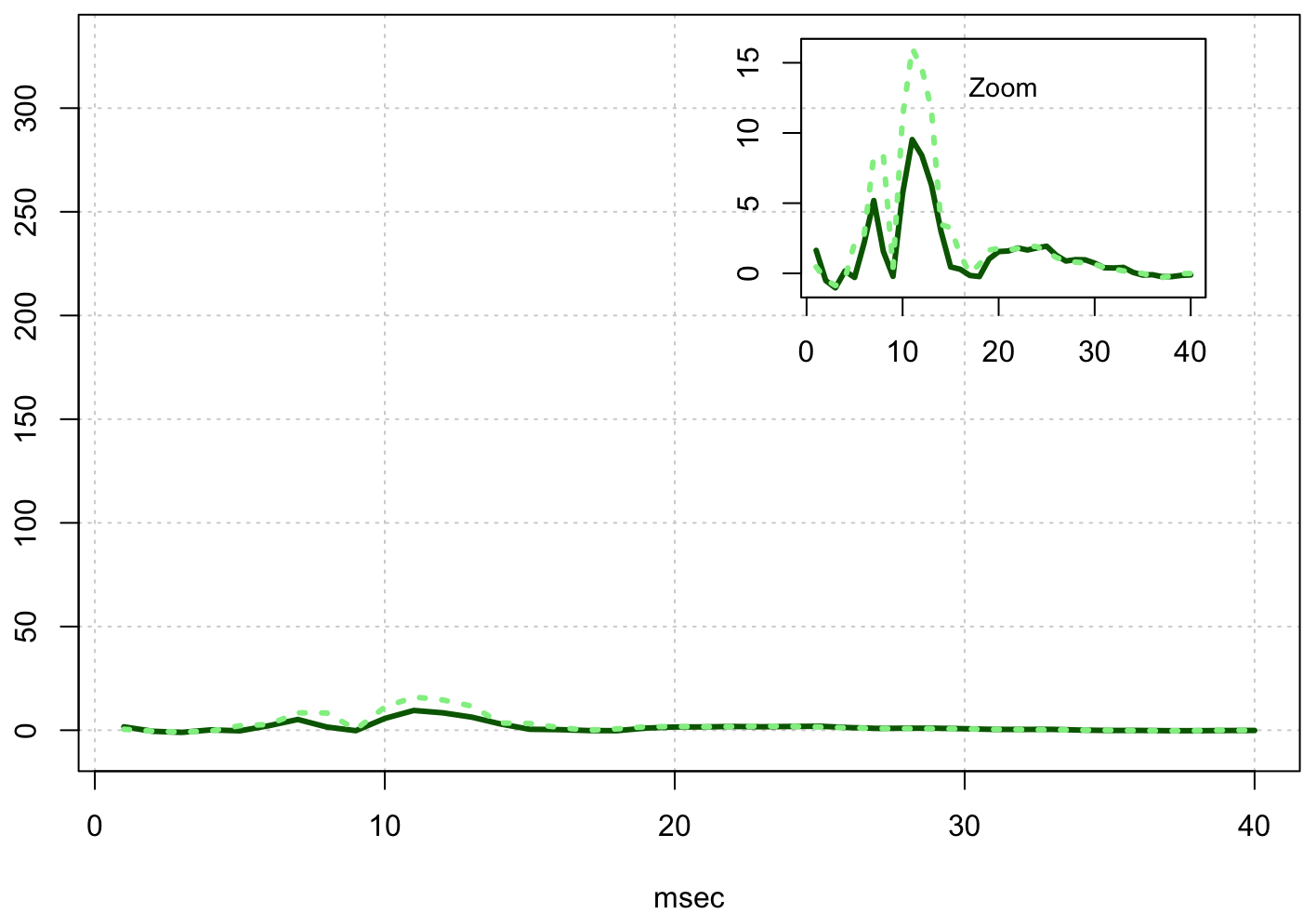}&
\includegraphics[height=4cm, width=4cm]{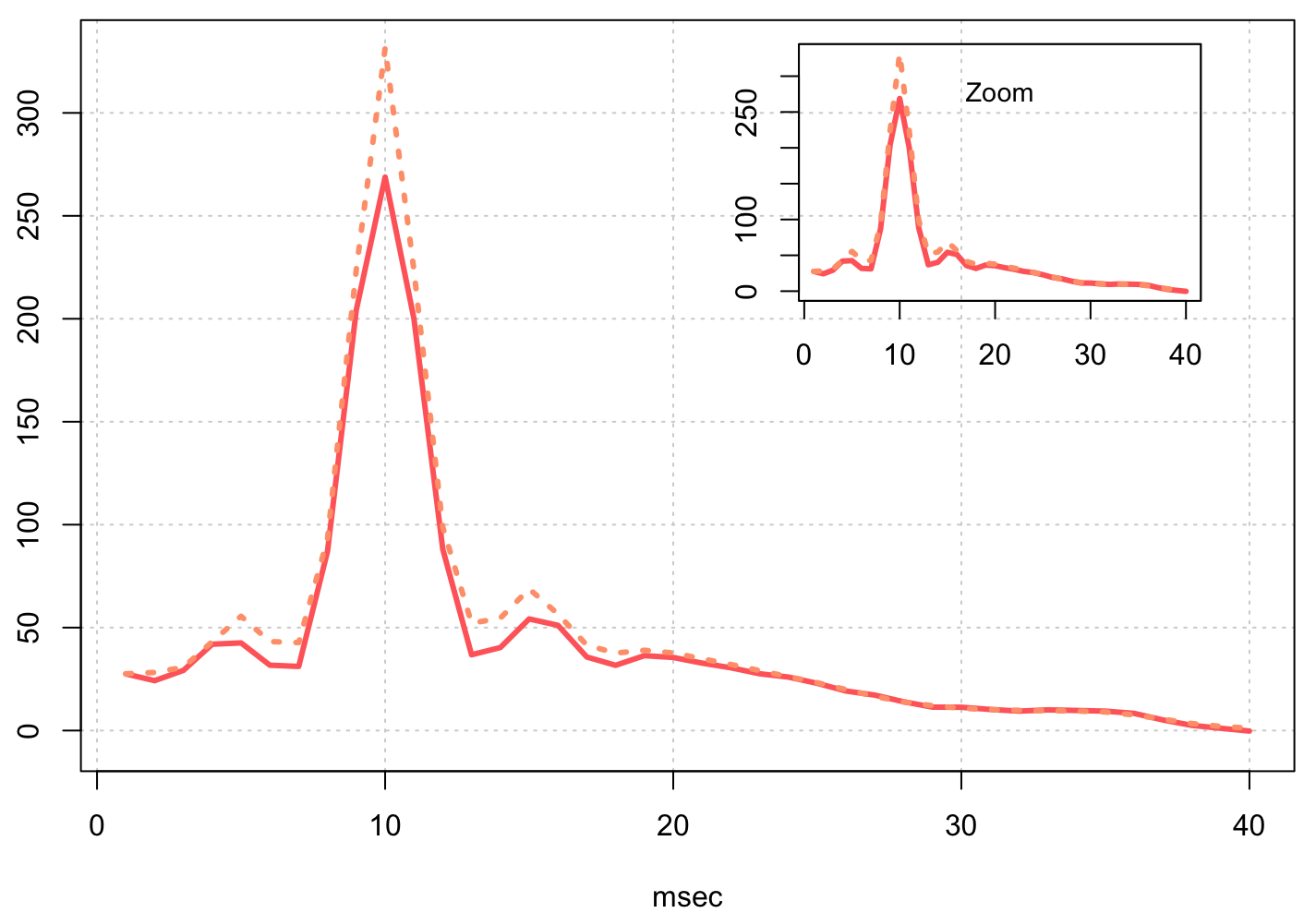}\\
\includegraphics[height=4cm, width=4cm]{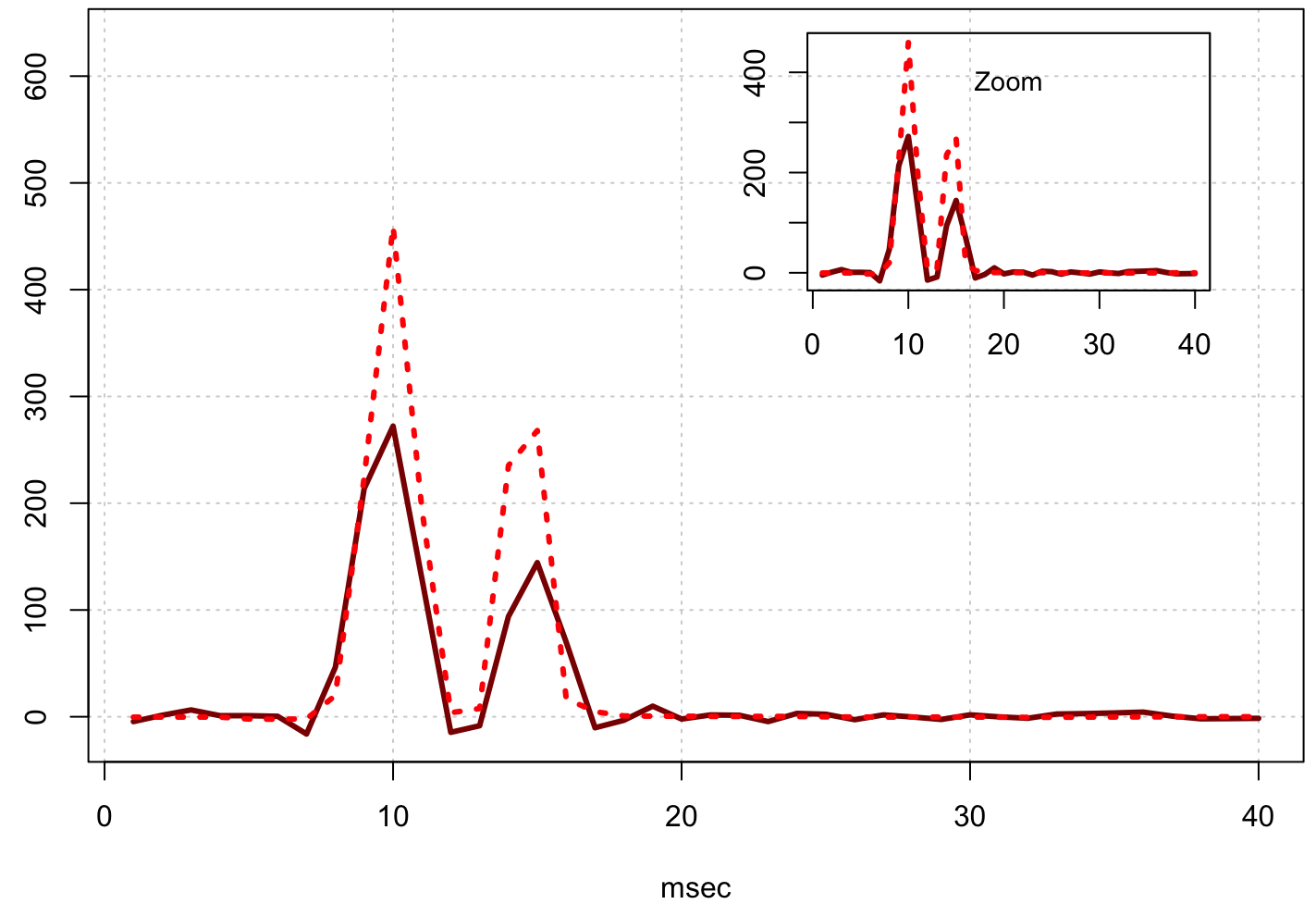}&
\includegraphics[height=4cm, width=4cm]{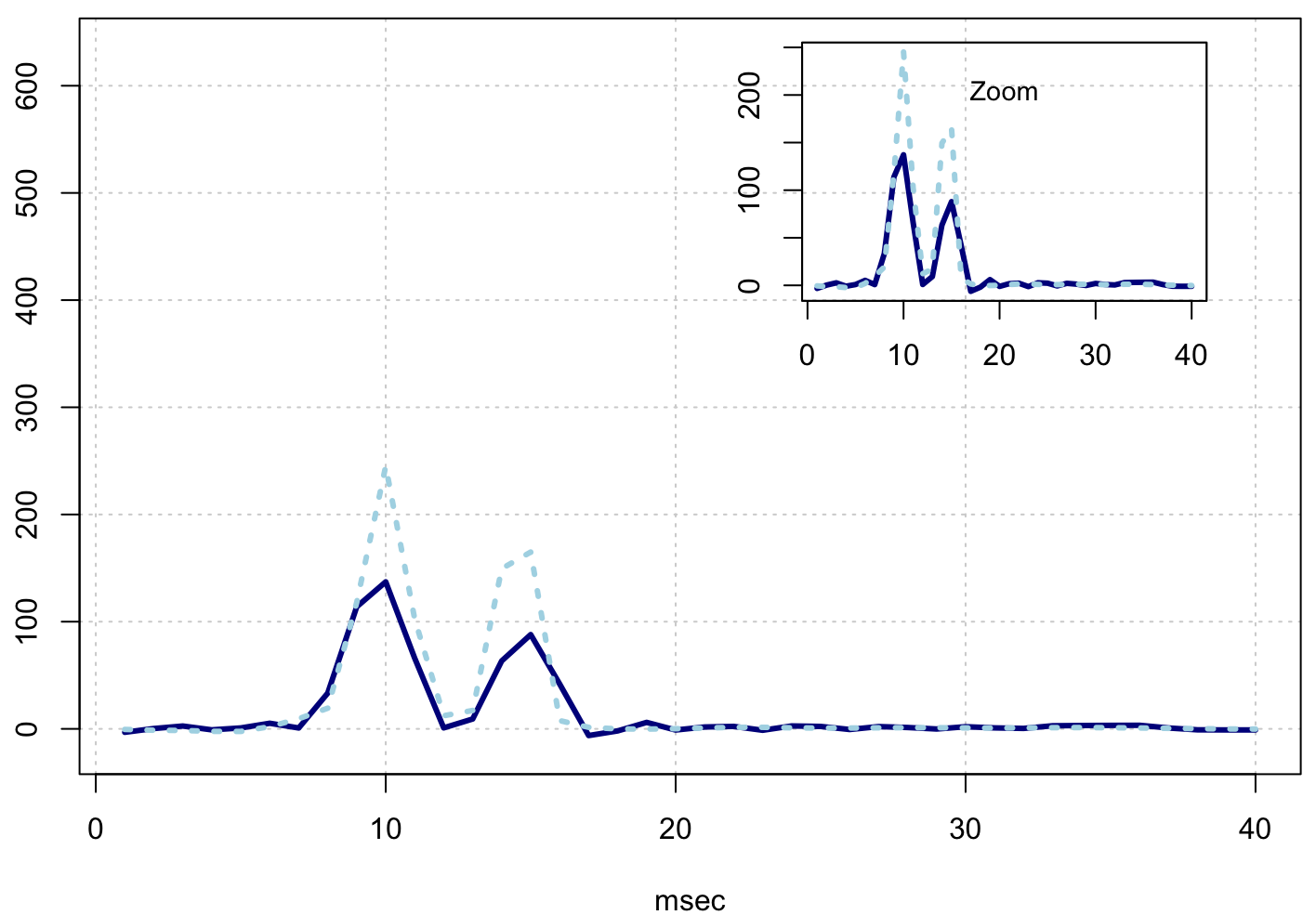}&
\includegraphics[height=4cm, width=4cm]{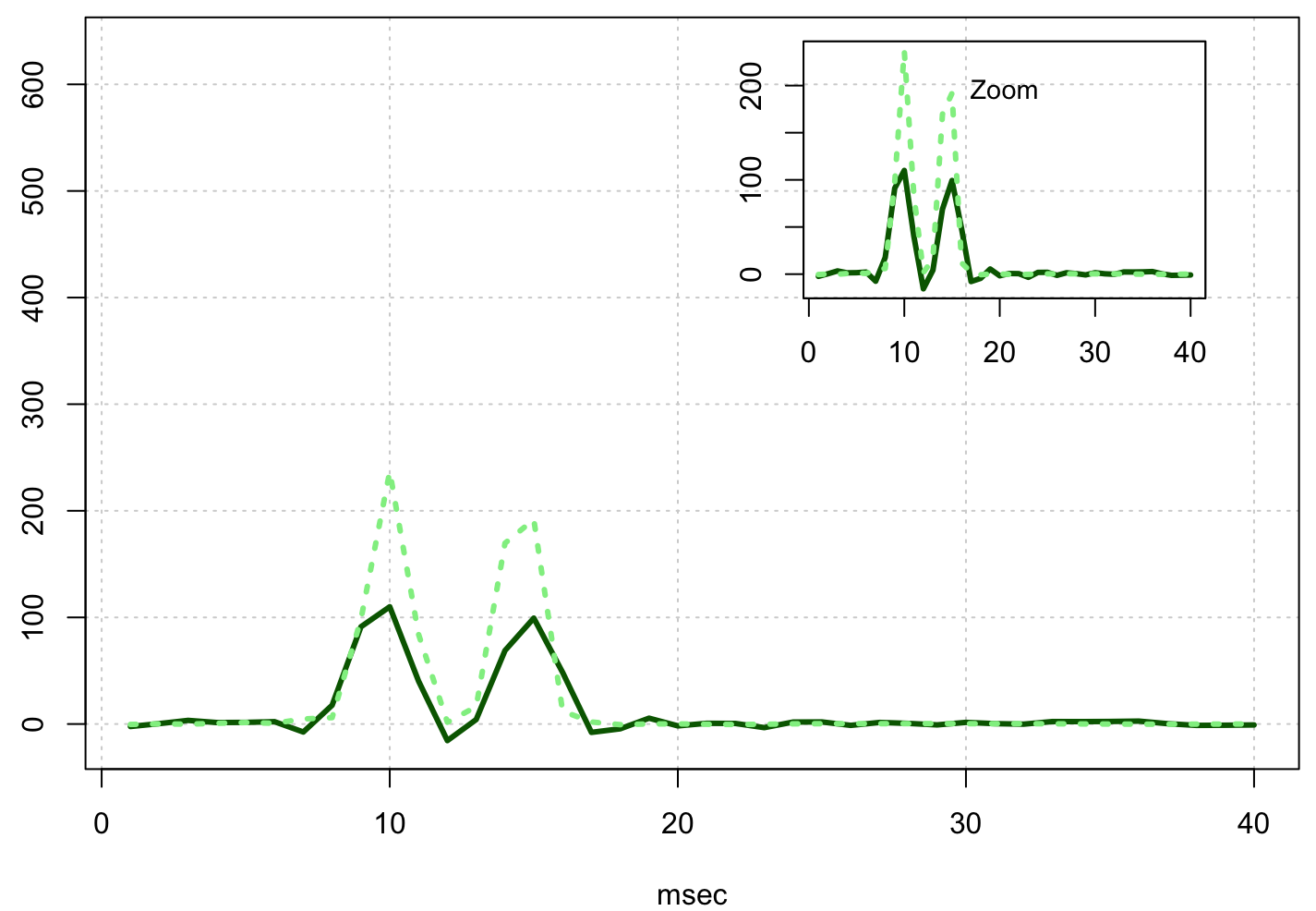}&
\includegraphics[height=4cm, width=4cm]{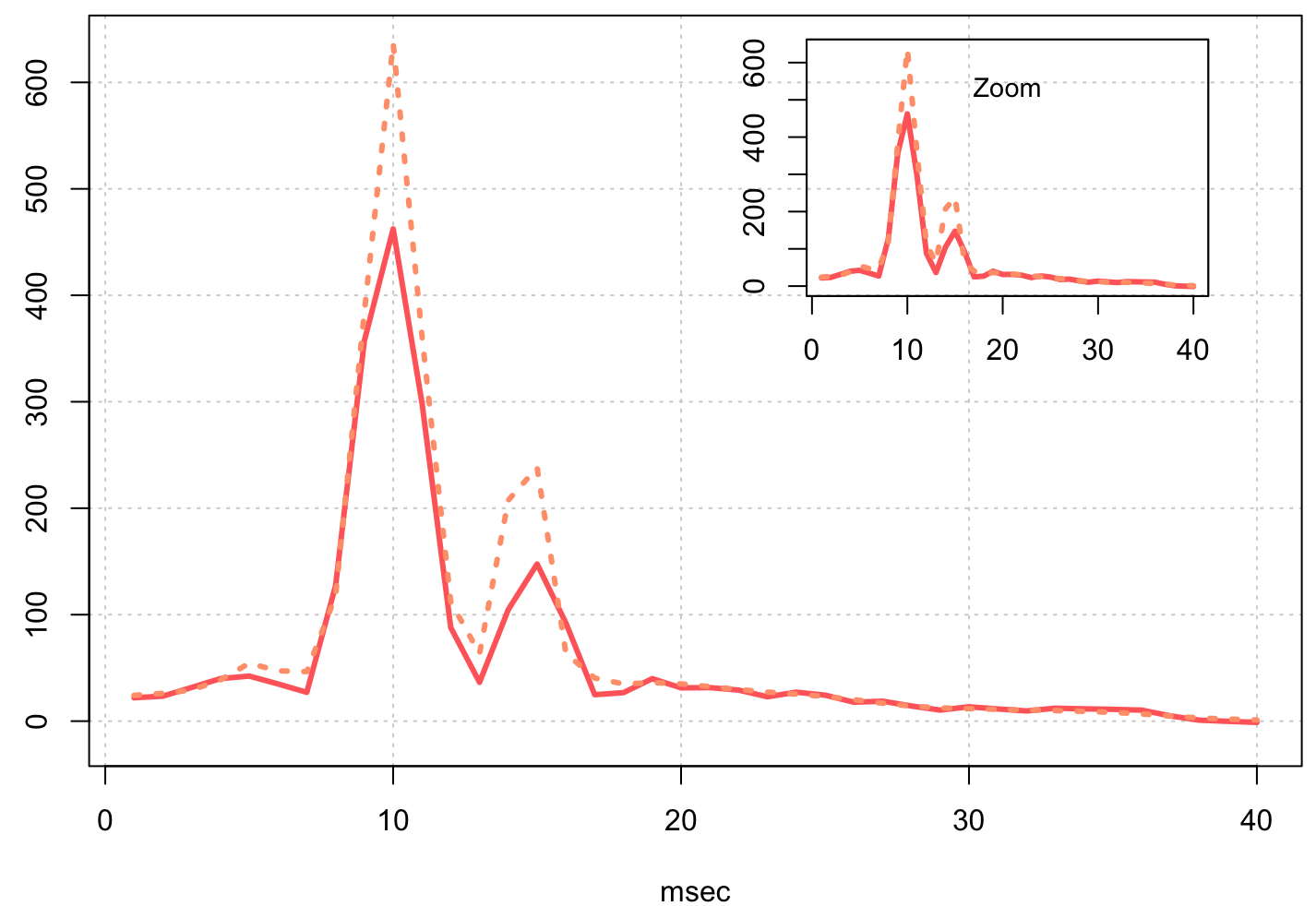}		
\end{tabular}
\caption{\textbf{Raw versus smooth level-specific variance.} First row: Subset 1 "Games". Second row: Subset 2 "In-Season Practice". Third row: Subset 3 "Preseason Practice". Dashed lines depict raw variances, solid lines depict smoothed variances. Inset graphs show zoomed-in versions for shape inspection. Smoothing is performed using penalized regression splines (thin plate splines). Scales are different for each sample.}\label{fig:raw_vs_smooth_levelcov_data}
\end{center}	
\end{figure}

We smooth the covariance surfaces using the same GAM (bivariate thin plate regression spline, smoothing parameters estimated by REML) that we used in the simulation study. Figure \ref{fig:var_decomp_graphical_data} compares the empirical variance of the data matrix $Y$, the diagonal of $YY'/n$, to the sum of the raw level-specific variances and the sum of the smoothed level-specific variances as separated by SFPCA. These comparisons show that the separated variances add up to the empirical variance. Also, the effect of smoothing on the shape of the variance curve it is made visible. Each row of Figure \ref{fig:raw_vs_smooth_levelcov_data} breaks down the raw and smoothed subset variance (blue/red and green curves) in Figure \ref{fig:var_decomp_graphical_data} into the level-specific raw and smoothed variance for the team-, position-, player- and curve-level. The smaller inset graphs display the graphs as zoomed-in versions such that the shape of the curves can be inspected. The majority of the variation of all three subsets is allocated to the finest level, the curve level. The number of principal components needed to explain 90\% of the level-specific variation is given in Table \ref{tab:var_decomp_data} along with the proportion of data variation explained by the levels which is calculated as follows 
\begin{eqnarray*}
	\text {Level-specific \% of Var Explained}=\frac{\sum(\text {Eigenvalues of level-specific PCs })}{\sum(\text {Eigenvalues of all PCs (across levels) })} \times 100.
\end{eqnarray*}
For example, for the team-level of Subset 1 "Games", this equation becomes
\begin{eqnarray*}
	\frac{\sum_{n_1=1}^7 \lambda_{n_1}^{\text{team}}}{\sum_{n_1=1}^7 \lambda_{n_1}^{\text{team}} + \sum_{n_2=1}^8 \lambda_{n_2}^{\text{position}} + \sum_{n_3=1}^{10} \lambda_{n_3}^{\text{player}} + \sum_{n_1=1}^7 \lambda_{n_4}^{\text{curve}}} \times 100.
\end{eqnarray*}

The variance decomposition presented in Table \ref{tab:var_decomp_data} demonstrates marked differences in the hierarchical structure of head-impact variability across competition, in-season practice, and preseason practice settings. In the games and in-season practice subsets, the overwhelming majority of total variability is attributable to the curve level, accounting for approximately 81.5\% and 82.1\% of the total variation, respectively. In contrast, the preseason subset exhibits a substantially different variance structure, with only 41.2\% of the variation assigned to the curve level and greater contributions from the team, position, and player levels. These findings suggest that individual impacts occurring during games and regular-season practices are characterized primarily by impact-specific heterogeneity, whereas preseason activities appear to introduce greater between-team, between-position, and between-player differences. The larger estimated residual variance in the preseason subset further supports the interpretation that preseason activities encompass a broader range of drills, contact situations, and movement patterns, resulting in a more heterogeneous data structure than that observed during competition or in-season practice. 

Examination of the estimated principal components indicates that the first principal component consistently dominates the variability within each hierarchical level and subset. At the team, position, player, and curve levels, the first component explains between 28\% and 68\% of the level-specific variation in the games and in-season practice datasets and exceeds 59\% of the variance at all levels in the preseason dataset. The corresponding eigenfunctions suggest that the primary mode of variation reflects differences in the overall magnitude and shape of the acceleration profiles surrounding the impact peak rather than differences in impact timing, since all curves were aligned to a common peak location during preprocessing. Consequently, the first principal component may be interpreted as representing systematic differences in impact intensity. At the team and position levels, this mode reflects broad differences in average impact characteristics across collegial and positional groups, whereas at the player and curve levels it captures variability in the magnitude and recovery pattern of individual head impacts. The predominance of this component across all levels indicates that variation in impact severity is the principal source of heterogeneity in the observed acceleration trajectories.  

The higher-order principal components explain substantially smaller proportions of variance and therefore represent more localized aspects of the impact curves. In the games and in-season practice subsets, the second and third components typically capture variation in the shape of the impact trajectory before and after the acceleration peak, including differences in the steepness of the initial rise, the rate of post-impact recovery, and secondary oscillatory behavior. These components likely characterize finer distinctions in impact dynamics that are superimposed upon the dominant intensity-related pattern represented by the first principal component. The relatively modest proportions of variance explained by these higher-order components indicate that such shape-related features are secondary contributors to overall variability. Nevertheless, the comparable structure of the leading player- and curve-level components across games and in-season practices suggests that the biomechanical characteristics governing head-impact responses remain broadly similar across these two activity settings despite differences in context and exposure.  

Interpretation of the higher-order components is considerably more challenging in the preseason subset. As shown in the estimated eigenfunctions for the team and position levels, many components beyond the first exhibit oscillatory behavior and account for only a small proportion of the level-specific variance. Such oscillations are often indicative of lower signal-to-noise ratios and reduced interpretability, particularly when the corresponding components explain only a limited amount of variability. Consequently, greater weight should be placed on the variance decomposition itself rather than on the detailed interpretation of individual higher-order preseason components. Collectively, the findings indicate that impact-specific variability dominates the overall variance structure in games and in-season practices, whereas preseason impacts appear to be influenced to a greater extent by differences among teams, positions, and individual players. These results illustrate the ability of the four-level SFPCA framework to separate and quantify distinct sources of functional variation within large and highly unbalanced head-impact datasets, thereby providing insight into how the relative importance of hierarchical factors changes across football activity settings.


\section{Summary}\label{sec:summary}

This paper extends the Structured Functional Principal Component Analysis (SFPCA) framework of \cite{shou2015structfpca} to accommodate four-level nested functional data with arbitrary unbalancedness at every level of the hierarchy. Although the underlying SFPCA methodology is theoretically applicable to hierarchies of arbitrary depth, practical implementation becomes increasingly challenging when the number of functional observations is large and group sizes vary substantially across levels. We developed a comprehensive implementation for estimating level-specific covariance operators, eigenfunctions, eigenvalues, and principal scores in such settings. In addition, we proposed a set of memory-efficient computational techniques that avoid explicit construction of large design matrices and enable the analysis of datasets containing tens of thousands of curves using conventional computing hardware. Simulation studies demonstrated that the proposed approach accurately recovers the underlying covariance structure and principal components across a range of signal-to-noise scenarios while maintaining computational feasibility in settings that would otherwise exceed memory limitations.  

From a statistical perspective, the study provides new evidence regarding the behavior of multilevel functional decompositions under highly unbalanced designs. The simulation results showed that eigenvalue estimation remains relatively robust to observation noise and temporal discretization, whereas the estimation of eigenfunctions and principal scores is more sensitive to level separation, grid resolution, and unequal group sizes. The findings further suggest that moderate observation noise, coupled with covariance smoothing, can improve numerical identifiability by reducing collinearity among level-specific functional subspaces, thereby facilitating more stable estimation of hierarchical components. More broadly, the proposed framework demonstrates that meaningful decomposition of variability across multiple nested levels remains possible even when the number of functional observations exceeds 50,000 and considerable imbalance exists throughout the hierarchy. These characteristics are common in modern biomedical, environmental, digital health, and wearable-sensor applications, highlighting the potential utility of the proposed methodology beyond the motivating application considered here.  

Application of the method to head-impact acceleration data from the NCAA-DoD CARE Consortium illustrated the value of multilevel functional decomposition for understanding the hierarchical sources of biomechanical variability in football-related head impacts. The four-level model separated variation attributable to teams, positions, players, and individual impact events. For both games and in-season practices, approximately 82\% of the total variability was attributed to the curve level, indicating that impact-to-impact differences within individual athletes dominate the overall variation structure. In contrast, preseason practices exhibited substantially greater contributions from team-, position-, and player-level effects and a corresponding reduction in curve-level variability. These findings suggest that preseason activities involve greater heterogeneity in exposure patterns and training environments than either competitions or in-season practices. Furthermore, the leading principal components across all levels predominantly reflected differences in the magnitude and overall shape of the acceleration profiles, whereas higher-order components captured more subtle waveform characteristics and generally explained smaller proportions of variation. Together, these results provide new insight into the relative importance of organizational, positional, player-specific, and impact-specific factors contributing to head-impact biomechanics.  

Despite these contributions, several limitations should be acknowledged. First, estimation at the coarsest hierarchical levels remains challenging when only a small number of independent units are available, as evidenced by the larger variability observed in the team-level components during the simulation study. Second, although the proposed computational strategy accommodates very large numbers of curves, practical feasibility still depends on the temporal resolution of the functional observations. As the number of curves grows, the grid dimension must remain relatively modest to maintain computational tractability, limiting direct application to datasets characterized simultaneously by very large sample sizes and highly dense functional observations. Third, the contrast-based covariance decomposition employed by SFPCA is sensitive to pronounced imbalance in lower-level group sizes, which may induce unequal contributions of group-level information and affect the estimation of higher-level covariance structures. Finally, interpretation of higher-order principal components remains inherently difficult, particularly when they explain only small proportions of variation or exhibit oscillatory behavior suggestive of low signal-to-noise ratios.  

Several directions for future research emerge from this work. Methodologically, future studies could investigate weighting schemes or alternative covariance estimators that explicitly account for severe unbalancedness and reduce the influence of dominant groups on higher-level covariance estimation. Extensions to sparse functional data, irregular observation grids, and deeper hierarchical structures are also of considerable interest. Additional work is needed to develop scalable inferential procedures, uncertainty quantification, and formal hypothesis-testing frameworks for multilevel functional components in large datasets. From an application perspective, future analyses of CARE and related datasets could incorporate additional sources of hierarchy, such as season, practice drill type, athlete characteristics, concussion status, or longitudinal player trajectories, thereby providing a more comprehensive understanding of the factors influencing head-impact exposure. More generally, the proposed framework establishes a foundation for studying complex hierarchical functional data arising in sports biomechanics, digital health monitoring, environmental sensing, neuroimaging, and other domains in which large collections of nested functional observations are increasingly common.


\section*{Acknowledgments}\label{sec:acknowledgments}
\noindent
The authors confirm that there are no known conflicts of interest associated with this publication and there has been no significant financial support for this work that could have influenced its outcome. The work of all authors was partially supported by the NINDS grant R01NS126449.

\clearpage
\pagebreak


\newpage
\selectfont
\setstretch{1.5} 
\bibliographystyle{chicago}
\clearpage
\bibliography{refs}

\clearpage
\pagebreak

\begin{table}[H]
\begin{center}
\footnotesize{
\begin{tabular}{lcccc}
\hline
Matrix & block size & constant, block off-diagonals  & constant, block diagonals & number of blocks\\
\hline
$DEE_4$ & $n \times n $ & $-$1 & $n_{i\cdot\cdot}-$1 & 1\\
\hline
$DEE_3$ & $n_{i\cdot\cdot} \times n_{i\cdot\cdot}$ & $-$1 & $n_{i\cdot\cdot}-$1 & $I$ \\
\hline
$DEE_2$& $n_{ij\cdot} \times n_{ij\cdot}$ & $-$1 & $n_{ij\cdot}-$1& $\sum_{i=1}^I J_{i}$\\
\hline
$DEE_1$& $n_{ijk} \times n_{ijk}$ & $-$1 & $n_{ijk}-$1& $\sum_{i=1}^I \sum_{j=1}^{J_i} K_{ij}$\\
\hline
\end{tabular}
}
	\caption{Structure of second coarsest ($U$) to finest level ($V$) $G-$matrices for four-level nested model.}\label{tab:structGmats}
	\end{center}
\end{table}

\begin{table}[H]
\begin{center}
\footnotesize{
\begin{tabular}{ccl}
\hline
\textbf{matrix} & \textbf{row dimension} & \textbf{column dimension} \\
\hline
$\operatorfont{B}_{X}$ & $p \cdot L_2L_3L_4(i)$ &  $N_1$\\
\hline
$\operatorfont{B}_{U}$ & $p \cdot L_2L_3L_4(i)$ &  $N_2\cdot L_2(i)$\\
\hline
$\operatorfont{B}_{W}$ & $p \cdot L_2L_3L_4(i)$ &  $N_3\cdot L_2L_3(i)$\\
\hline
$\operatorfont{B}_{V}$ & $p \cdot L_2L_3L_4(i)$ &  $N_4\cdot L_2L_3L_4(i)$\\
\hline
\end{tabular}
}
\caption{Dimensions of the level-specific $\operatorfont{B}$ matrices. $L_2 = J = \left[J_1, \dots, J_I \right]$, $L_2L_3 = \left[\sum_{j=1}^{J_1} K_{1j}, \dots \sum_{j=1}^{J_I} K_{Ij}\right]$, $L_2L_3L_4 = \left[ \sum_{j=1}^{J_1}\sum_{k=1}^{K_{1j}} L_{1,j,k}, \dots, \sum_{j=1}^{J_I}\sum_{k=1}^{K_{Ij}}  L_{I,j, k} \right]$.}\label{tab:Bdimensions}
\end{center}
\end{table}

\begin{table}[H]
\begin{center}
\footnotesize{
\begin{tabular}{lcc}
\hline
\textbf{Matrix size ($n \times n$)} & \textbf{Memory required} & \textbf{Fits in 16 GB?} \\
\hline
40,000 $\times$ 40,000 & 11.9 GB & {yes} \\
\hline
44,000 $\times$ 44,000 & 15.5 GB & {borderline} \\
\hline
45,000 $\times$ 45,000 & 16.2 GB & {no} \\
\hline
46,000 $\times$ 46,000 & 17.0 GB & {no} \\
\hline
\end{tabular}
}
\caption{Approximate memory usage for \underline{dense double-precision} matrices on a 16GB system (MATLAB). Note: 45,000 $\times$ 45,000 is often referred to as ``theoretical" limit. However, this might only be feasible with sparse structures and OS buffering.}
\end{center}
\end{table}

\begin{table}[H]
  \centering
  {\tablesize
    \begin{tabular}{lcccccccccccc}
    \hline
    \hline
    &\multicolumn{12}{c}{Principal components (Median MSE)}\\
    \hline
    $\boldsymbol\sigma$&$\boldsymbol\phi_1^X$&$\boldsymbol\phi_2^X$&$\boldsymbol\phi_3^X$&$\boldsymbol\phi_1^U$&$\boldsymbol\phi_2^U$&$\boldsymbol\phi_3^U$&$\boldsymbol\phi_1^W$&$\boldsymbol\phi_2^W$&$\boldsymbol\phi_3^W$&$\boldsymbol\phi_1^V$&$\boldsymbol\phi_2^V$&$\boldsymbol\phi_3^V$\\
    \hline
    \textbf{0} & 0.0078 & 0.0198 & 0.0263 & 0.0022 & 0.0033 & 0.0026 & 0.0001 & 0.0002 & 0.0001 & 0.0000 & 0.0000 & 0.0000 \\
    \textbf{0.5} & 0.0078 & 0.0197 & 0.0261 & 0.0021 & 0.0032 & 0.0025 & 0.0001 & 0.0003 & 0.0002 & 0.0000 & 0.0000 & 0.0001 \\
    \textbf{1} & 0.0079 & 0.0193 & 0.0261 & 0.0020 & 0.0033 & 0.0027 & 0.0002 & 0.0003 & 0.0002 & 0.0001 & 0.0001 & 0.0004 \\
     \hline
    &\multicolumn{12}{c}{Principal scores (Median MSE)}\\
    \hline
 $\boldsymbol\sigma$&$\boldsymbol\xi_1$&$\boldsymbol\xi_2$&$\boldsymbol\xi_3$&$\boldsymbol\theta_1$&$\boldsymbol\theta_2$&$\boldsymbol\theta_3$&$\boldsymbol\zeta_1$&$\boldsymbol\zeta_2$&$\boldsymbol\zeta_3$&$\boldsymbol\eta_1$&$\boldsymbol\eta_2$&$\boldsymbol\eta_3$\\
 \hline
    \textbf{0} & 0.0111 & 0.0197 & 0.0469 & 0.0447 & 0.0129 & 0.0034 & 0.0011 & 0.0006 & 0.0004 & 0.0178 & 0.0089 & 0.0044 \\
    \textbf{0.5} & 0.0045 & 0.0118 & 0.0126 & 0.0136 & 0.0076 & 0.0025 & 0.0010 & 0.0006 & 0.0004 & 0.0240 & 0.0152 & 0.0107 \\
    \textbf{1} & 0.0045 & 0.0110 & 0.0117 & 0.0119 & 0.0076 & 0.0023 & 0.0012 & 0.0009 & 0.0007 & 0.0429 & 0.0343 & 0.0298 \\
   \hline
   &\multicolumn{12}{c}{Eigenvalues (MSE)}\\
    \hline
 $\boldsymbol\sigma$&$\boldsymbol\lambda_1^X$&$\boldsymbol\lambda_2^X$&$\boldsymbol\lambda_3^X$&$\boldsymbol\lambda_1^U$&$\boldsymbol\lambda_2^U$&$\boldsymbol\lambda_3^U$&$\boldsymbol\lambda_1^W$&$\boldsymbol\lambda_2^W$&$\boldsymbol\lambda_3^W$&$\boldsymbol\lambda_1^V$&$\boldsymbol\lambda_2^V$&$\boldsymbol\lambda_3^V$\\ 
 \hline
    \textbf{0} & 0.5065 & 0.0473 & 0.0123 & 0.1350 & 0.0167 & 0.0057 & 0.0068 & 0.0018 & 0.0004 & 0.0001 & 0.0000 & 0.0000 \\
    \textbf{0.5} & 0.5077 & 0.0472 & 0.0123 & 0.1345 & 0.0168 & 0.0058 & 0.0067 & 0.0018 & 0.0004 & 0.0002 & 0.0001 & 0.0001 \\
    \textbf{1} & 0.5092 & 0.0471 & 0.0123 & 0.1340 & 0.0169 & 0.0058 & 0.0067 & 0.0020 & 0.0004 & 0.0006 & 0.0002 & 0.0002 \\
   \hline    
   \hline
    \end{tabular}
    }
  \caption{\textbf{Comparison true and estimated principal components, principal scores and eigenvalues.} 100 simulation repetitions, different signal-to-noise ratios. MSE for estimating $\sigma$: $1.1124\times10^{-5}$ for $\sigma=0.5$, $5.0396\times10^{-6}$ for $\sigma=1$. Simulation for $\sigma = 0$ performed without smoothing}\label{tab:comparison_true_vs_est}%
\end{table}

\begin{center}
\begin{table}[H]
\centering
\footnotesize{
	\begin{tabular}{cccllcc}
	\hline
	\hline
\textbf{Subset}	&\textbf{Season} & \textbf{Activity} & \textbf{Teams} & \textbf{Positions} & \textbf{\# of Players} & \textbf{\# of Curves}\\
\hline
1 & Fall 2017 & Competition& UW, UNC, VT,& OL, DL, LB, & 142 & 27,899\\
  &           &            & AirForce,   & TE, DB  &&\\
  &           &            & Army        &&&\\
 \hline
 2 & Fall 2017 & In-Season Practice& UW, UNC, VT,& OL, DL, LB, & 160 & 54,744\\
  &           &            & AirForce,   & TE, DB  &&\\
  &           &            & Army        &&&\\
 \hline
3 & Fall 2017 & Preseason Practice& UW, UNC, VT,& OL, DL, LB, & 160 & 37,858\\
  &           &            & AirForce,   & TE, DB  &&\\
  &           &            & Army        &&&\\
 \hline
 \hline
 \end{tabular}
 }
	\caption{\textbf{Data subsets.} Abbreviations for teams as used in text. Abbreviations for positions - OL: Offensive lineman, DL: Defensive lineman, LB: Linebacker, TE: Tight end, DB: Defensive back.}\label{tab:subsets}
	\end{table}
\end{center}

\begin{table}[H]
\begin{center}
\footnotesize{
\begin{tabular}{llccccc}
    \hline
	\hline
\textbf{Subset}&& \textbf{Team} & \textbf{Position} & \textbf{Player} & \textbf{Curve}& $\boldsymbol{\widehat\sigma}$\\
\hline
\textbf{1 Games}&\textbf{Estimated \# of PCs}& 7 & 8 & 10 & 7 & \\
&\textbf{Variation explained} (in \%) & 5.3203&  6.8983&  6.2754& 81.5060& 2.5024 \\
\hline	
\textbf{2 In-season Practice}&\textbf{Estimated \# of PCs}& 7& 9 & 8 & 8&\\
& \textbf{Variation explained} (in \%) & 7.0596&  4.4703&  6.3373& 82.1327& 2.5182\\
\hline
\textbf{3 Preseason Practice}&\textbf{Estimated \# of PCs}& 9& 9& 9& 8&\\
& \textbf{Variation explained} (in \%) & 26.1410& 15.9300& 16.7643& 41.1648&  6.4511 \\
\hline
\hline
\end{tabular}
}
\caption{\textbf{Variance decomposition for smoothed Covariance.}  Estimated number of principal components (PCs) for 90\% of level-specific variation explained.}\label{tab:var_decomp_data}
\end{center}
\end{table}

\begin{figure}[H]
\centering
\setlength{\tabcolsep}{1.5pt}
\begin{tabular}{cccc}
\multicolumn{4}{c}{\textbf{\scriptsize{Team level}}}\\
\includegraphics[height=4cm, width=4cm]{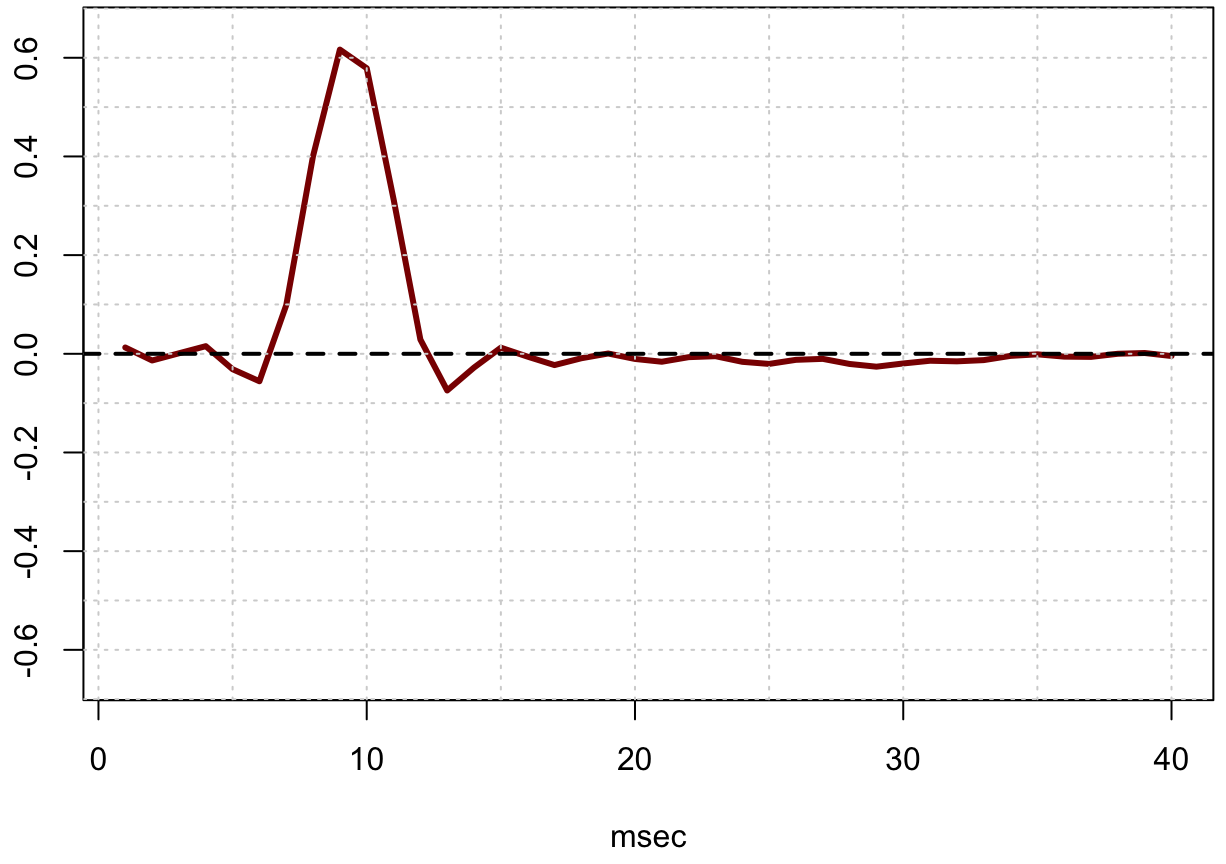}&\includegraphics[height=4cm, width=4cm]{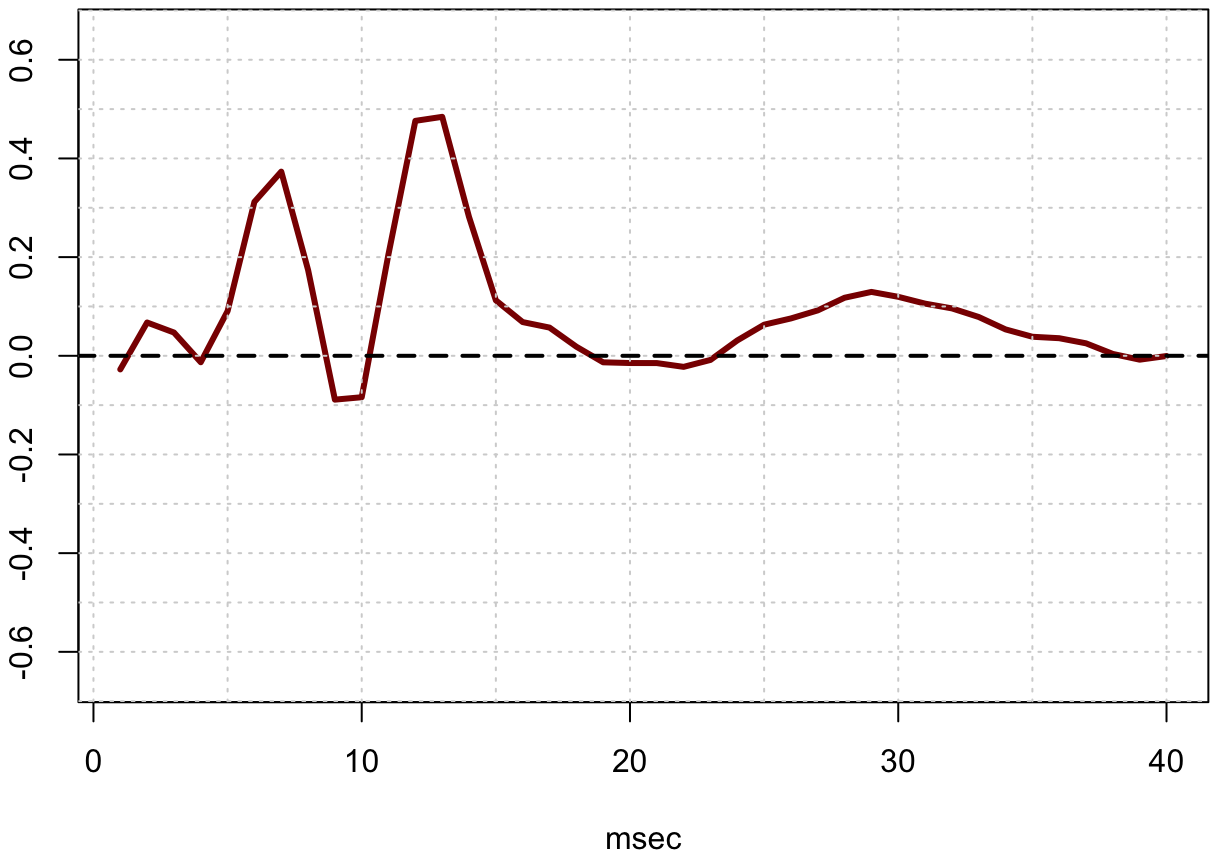}&
\includegraphics[height=4cm, width=4cm]{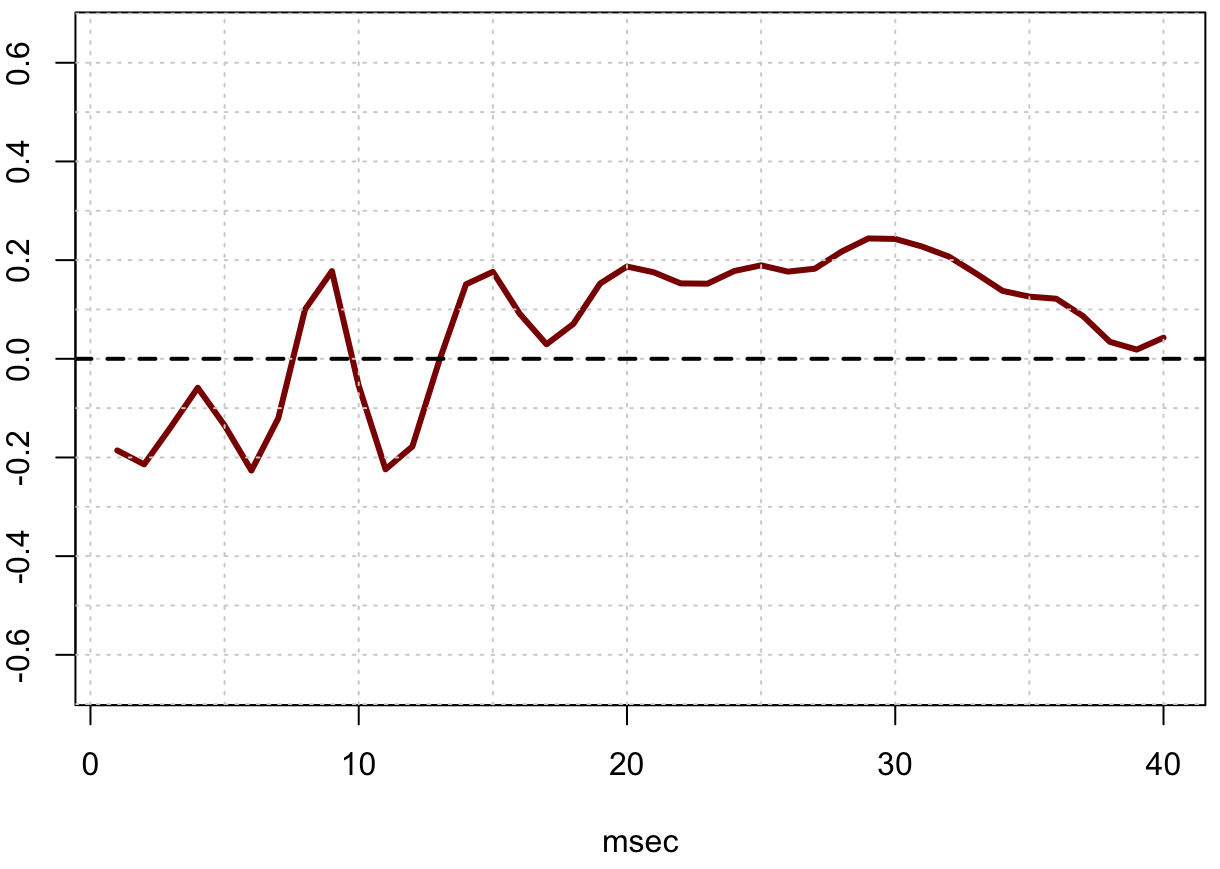}&
\includegraphics[height=4cm, width=4cm]{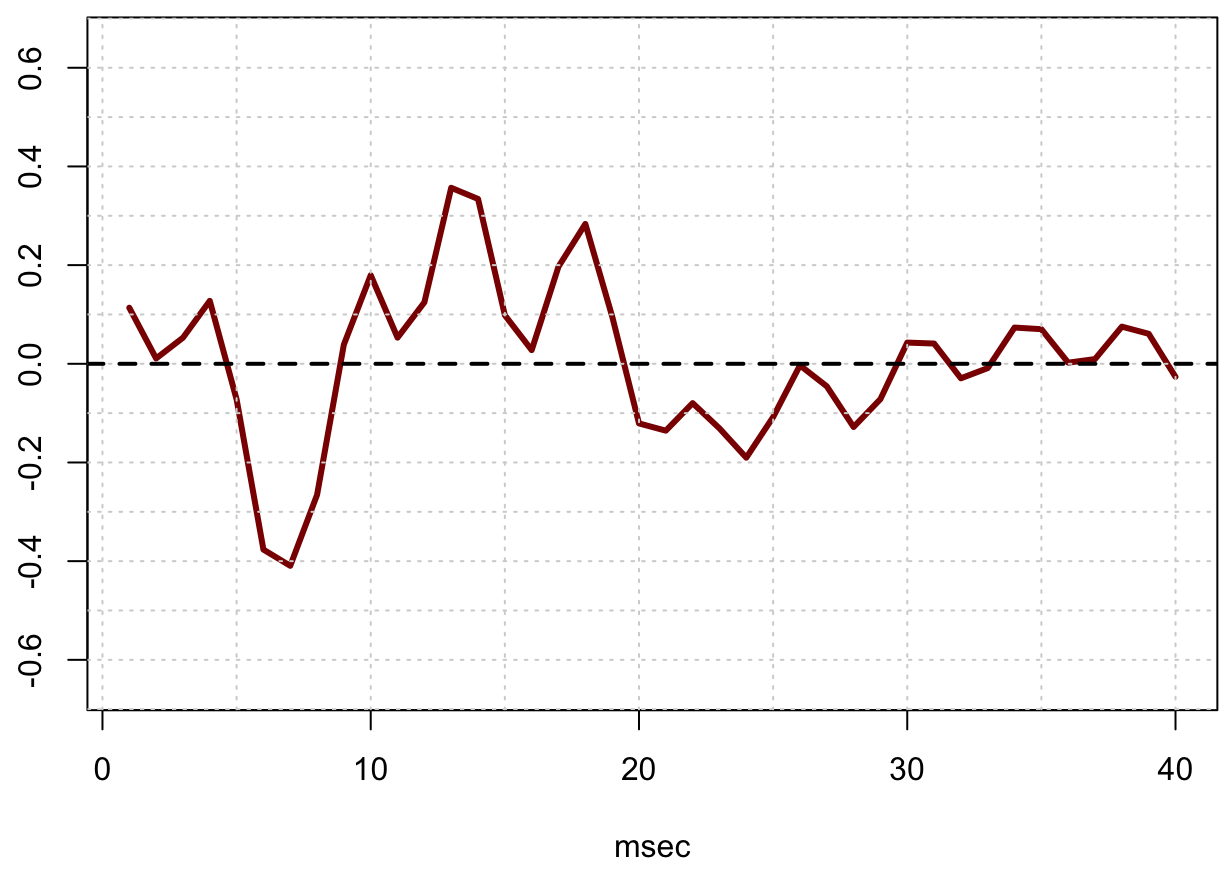}\\
\scriptsize{58.19\%}& \scriptsize{17.49\%}& \scriptsize{4.89\%}& \scriptsize{2.95\%}\\
\multicolumn{4}{c}{\textbf{\scriptsize{Position level}}}\\
\includegraphics[height=4cm, width=4cm]{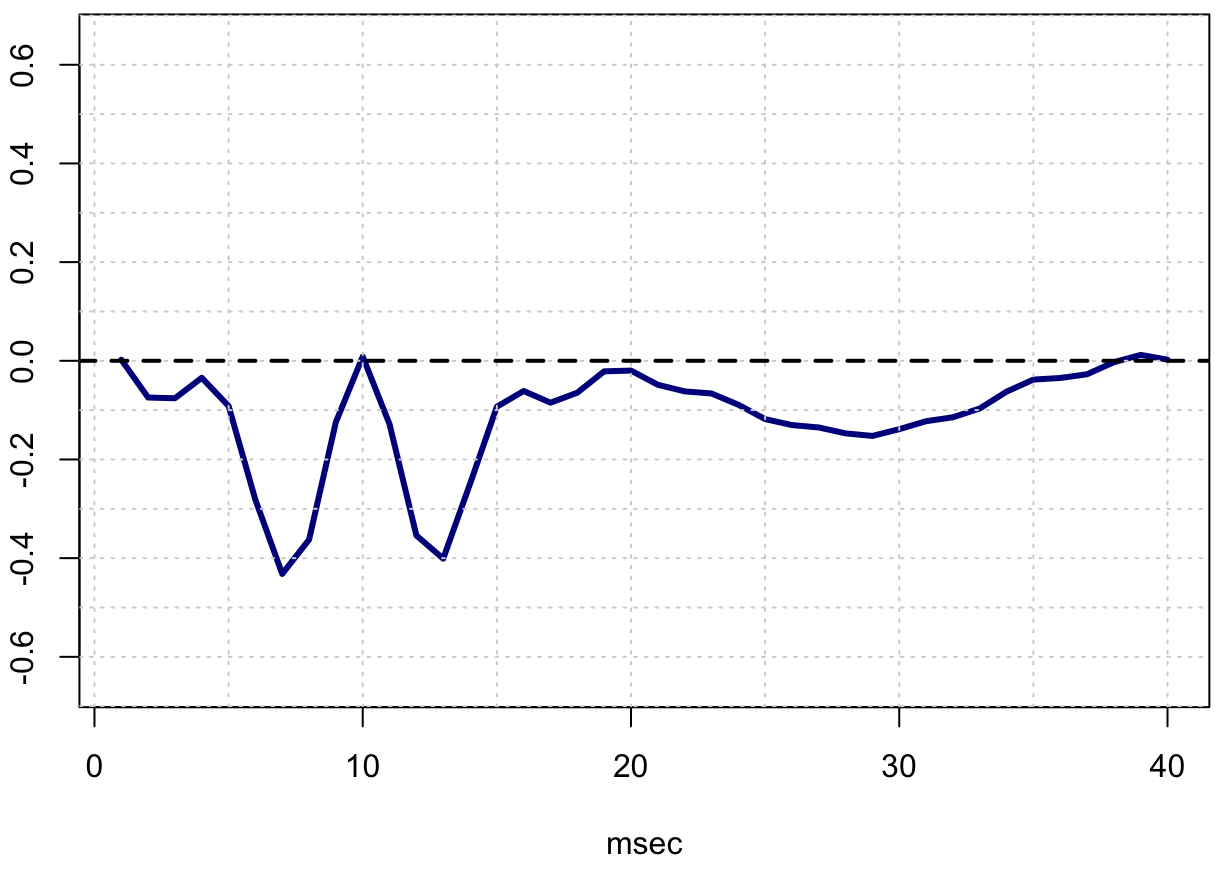}&\includegraphics[height=4cm, width=4cm]{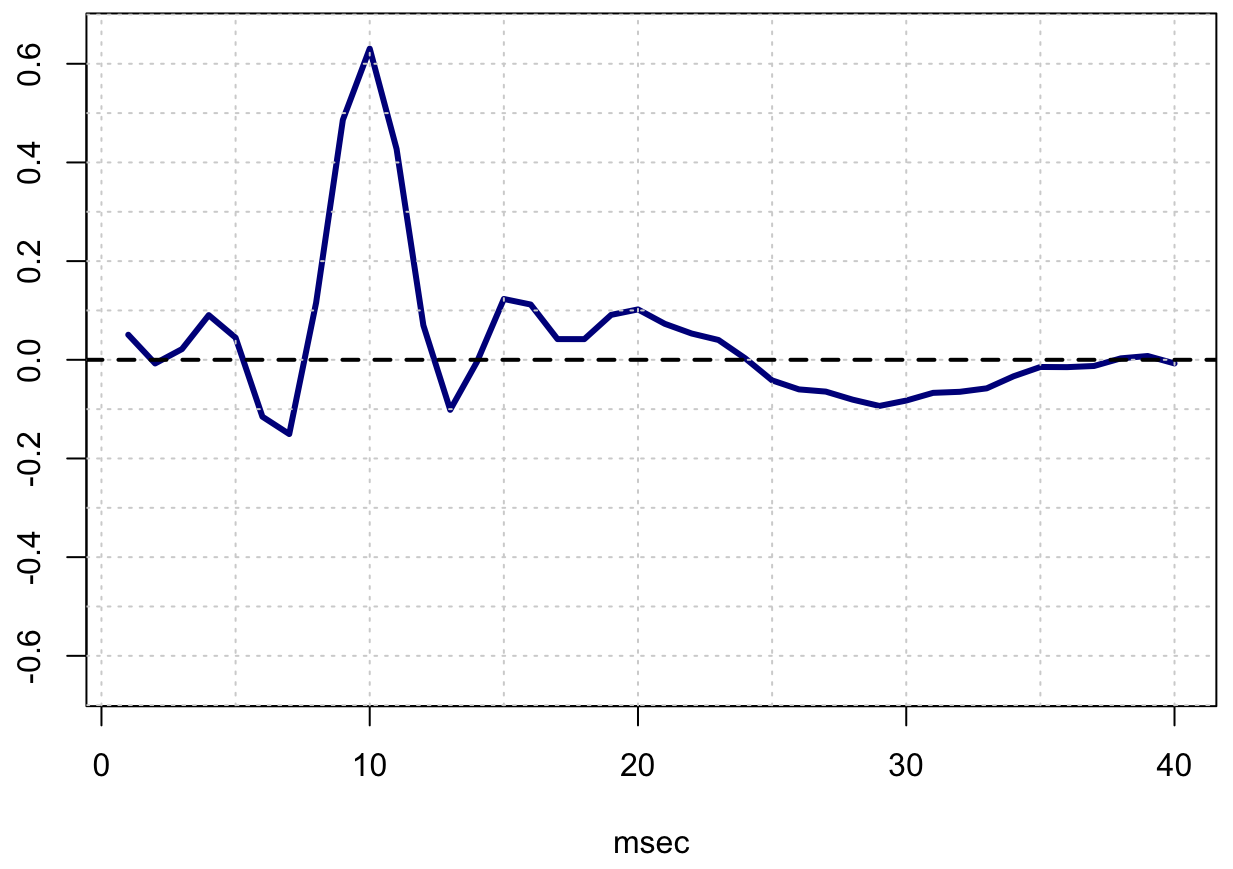}&
\includegraphics[height=4cm, width=4cm]{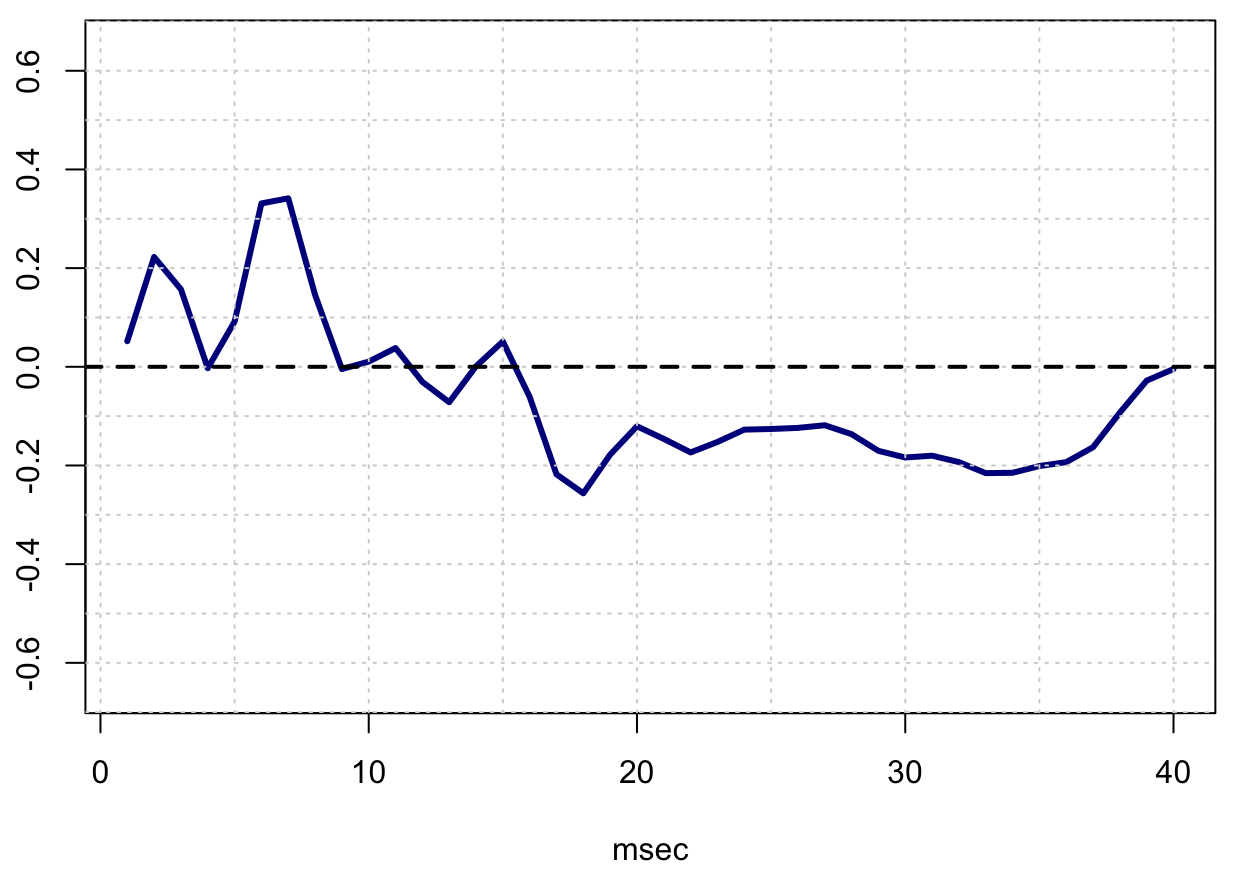}&
\includegraphics[height=4cm, width=4cm]{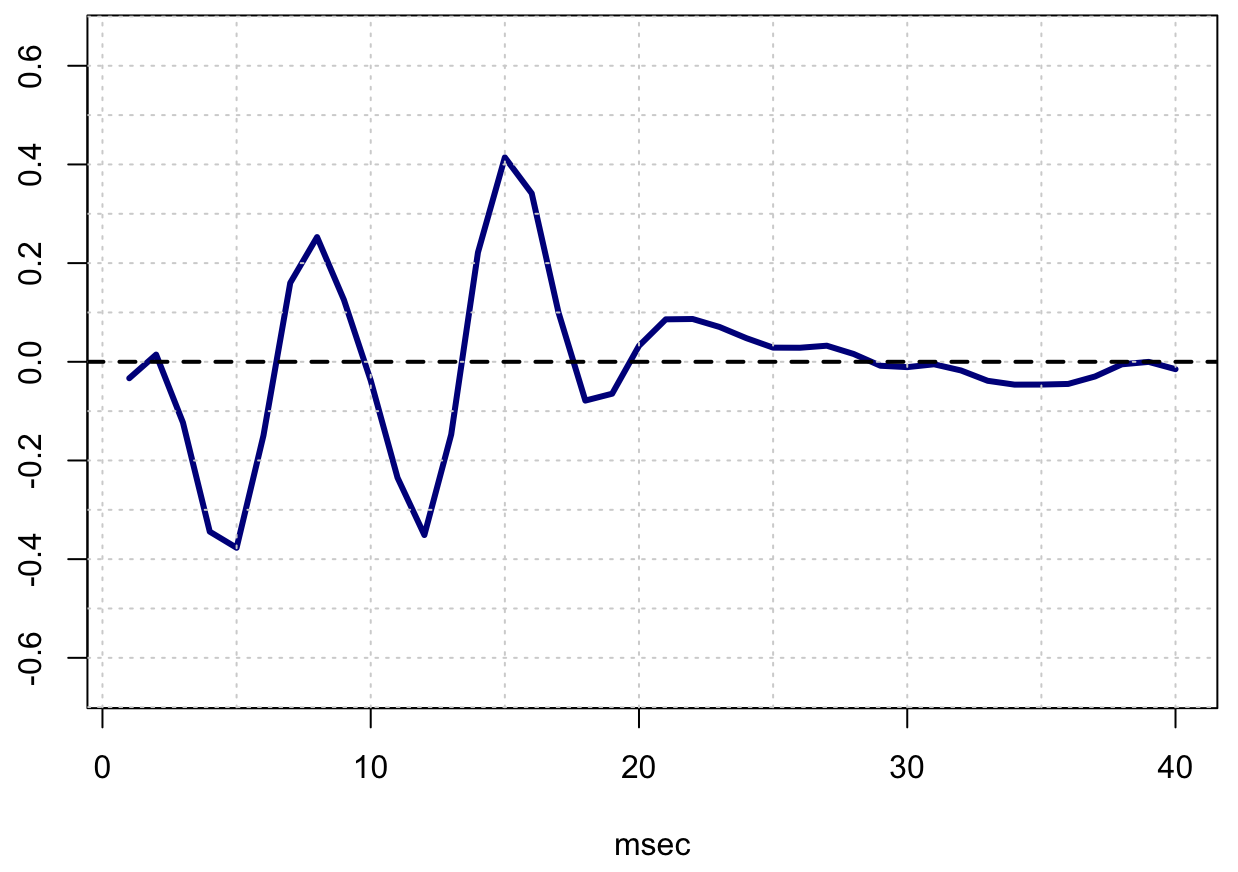}\\
\scriptsize{39.56\%}& \scriptsize{24.98\%}& \scriptsize{6.74\%}& \scriptsize{ 6.58\%}\\
\multicolumn{4}{c}{\textbf{\scriptsize{Player level}}}\\
\includegraphics[height=4cm, width=4cm]{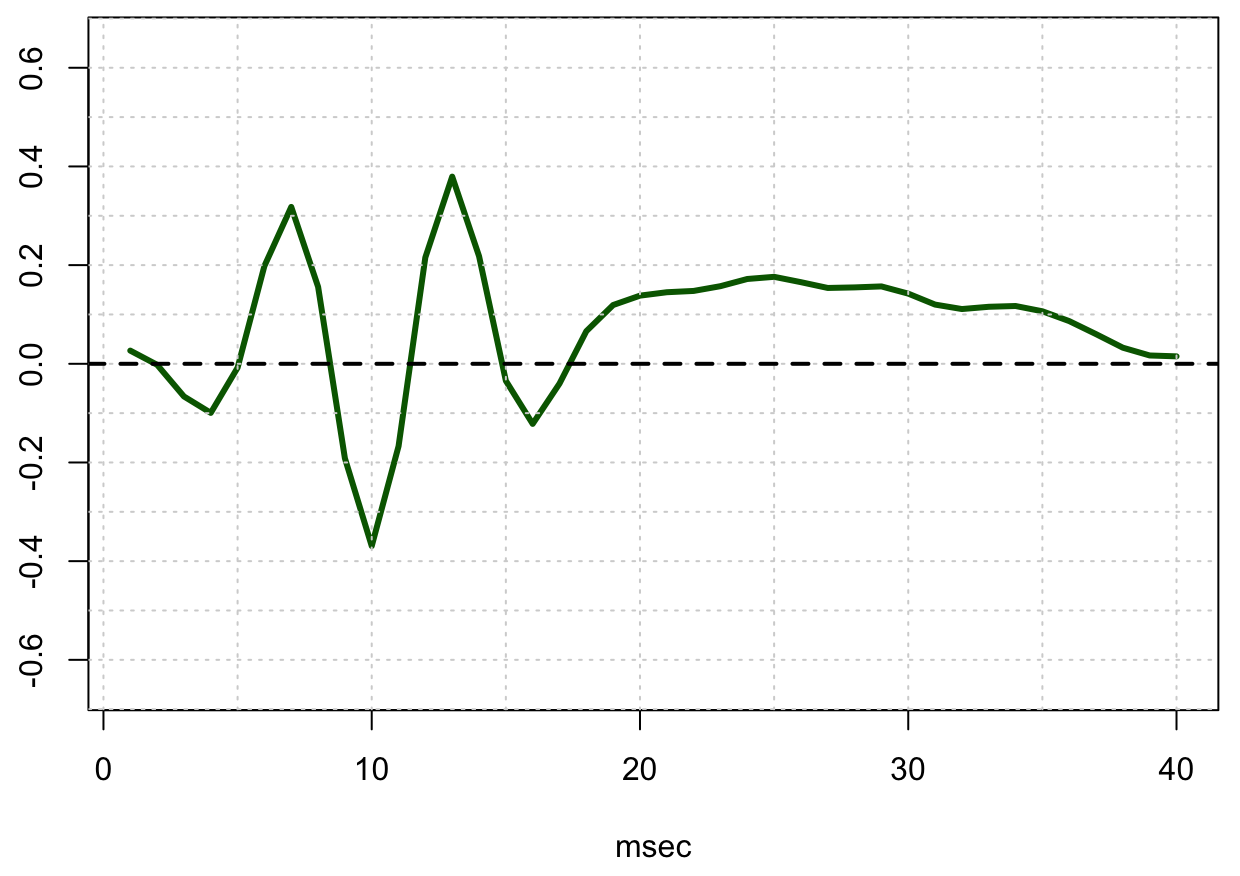}&\includegraphics[height=4cm, width=4cm]{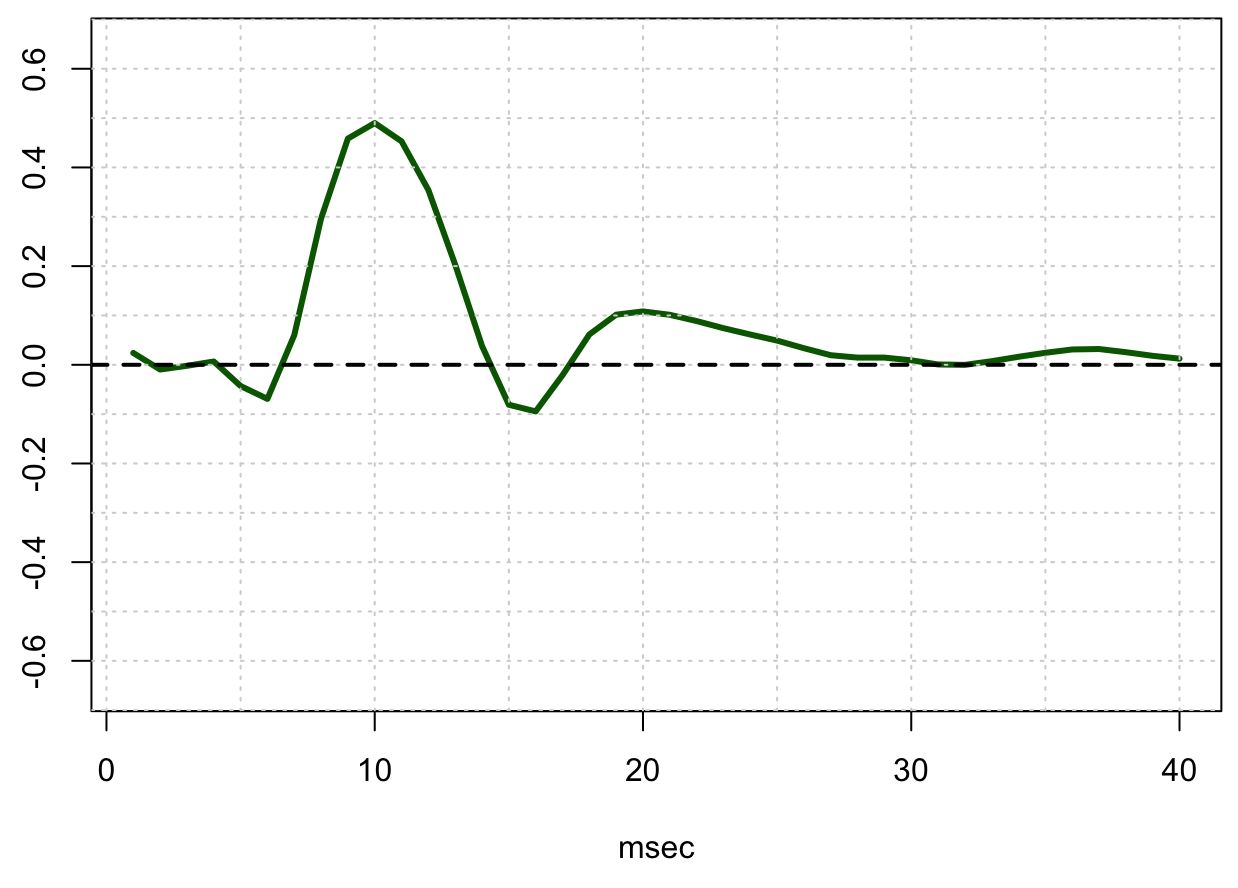}&
\includegraphics[height=4cm, width=4cm]{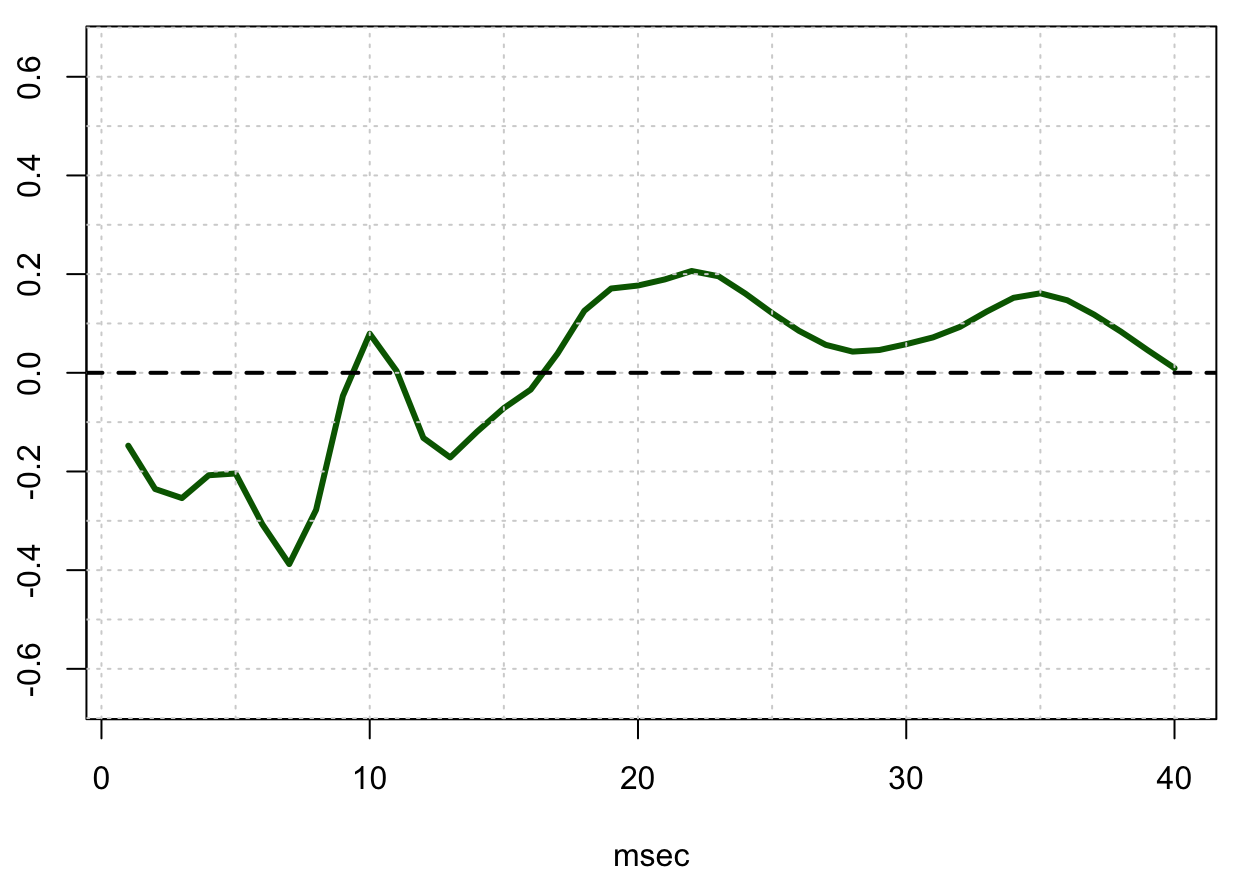}&
\includegraphics[height=4cm, width=4cm]{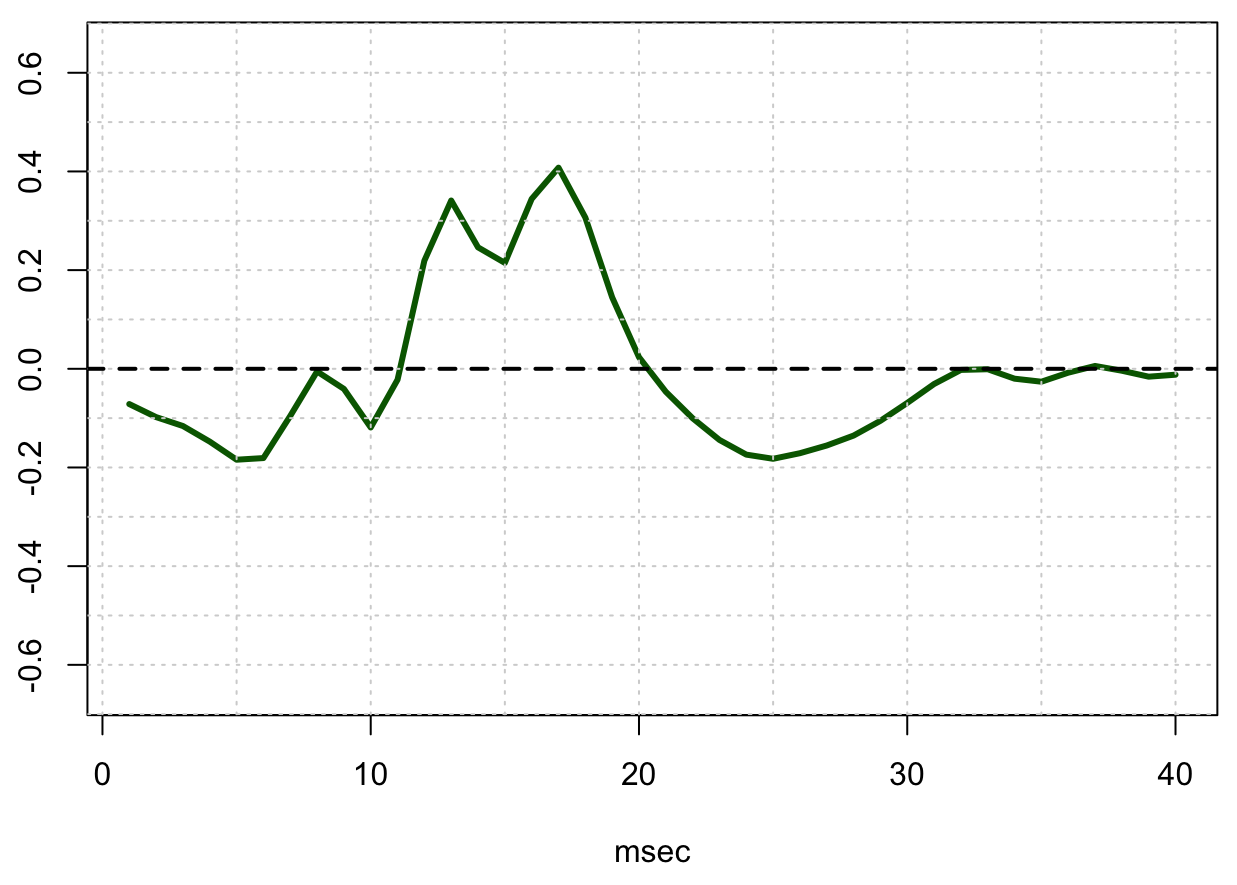}\\
\scriptsize{22.97\%}& \scriptsize{20.51\%}& \scriptsize{19.34 \%}& \scriptsize{7.38\%}\\
\multicolumn{4}{c}{\textbf{\scriptsize{Curve level}}}\\
\includegraphics[height=4cm, width=4cm]{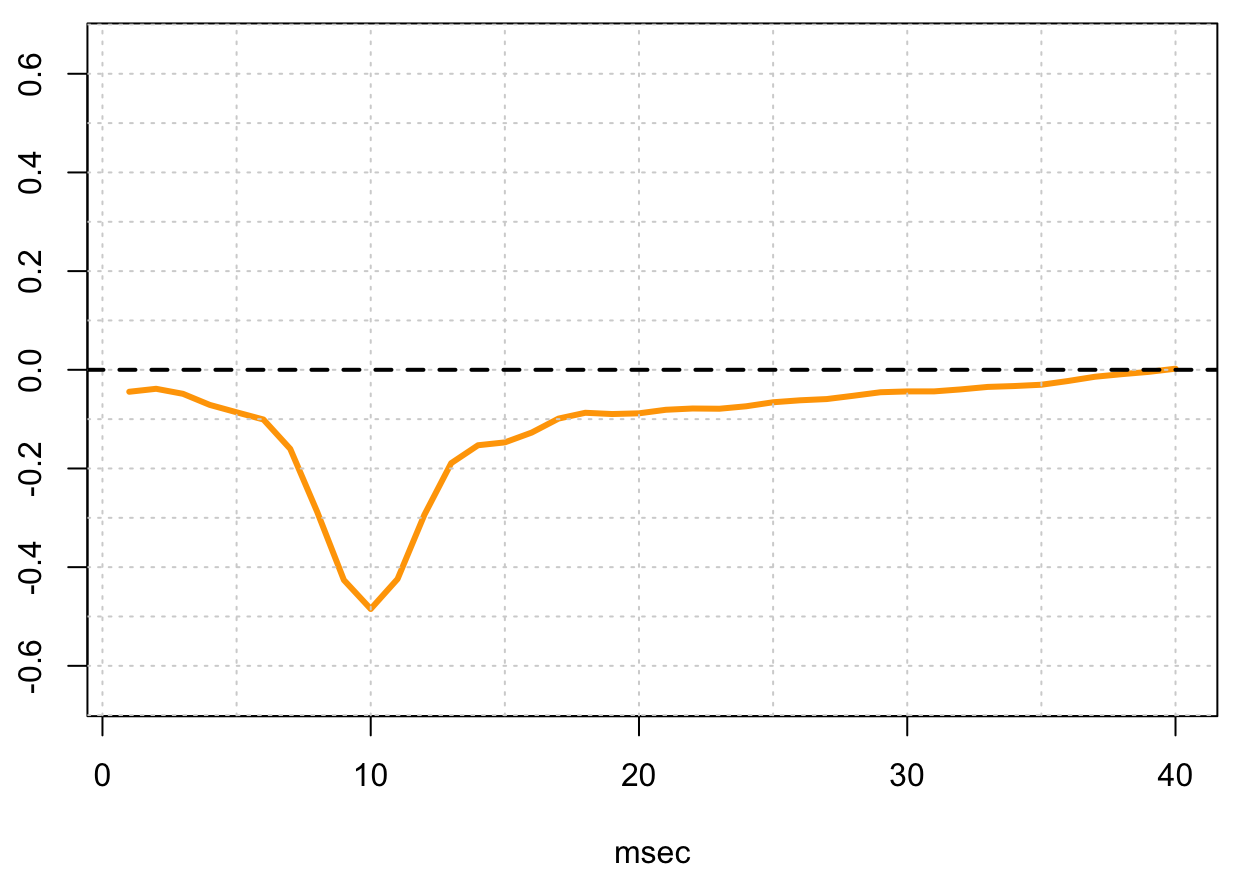}&\includegraphics[height=4cm, width=4cm]{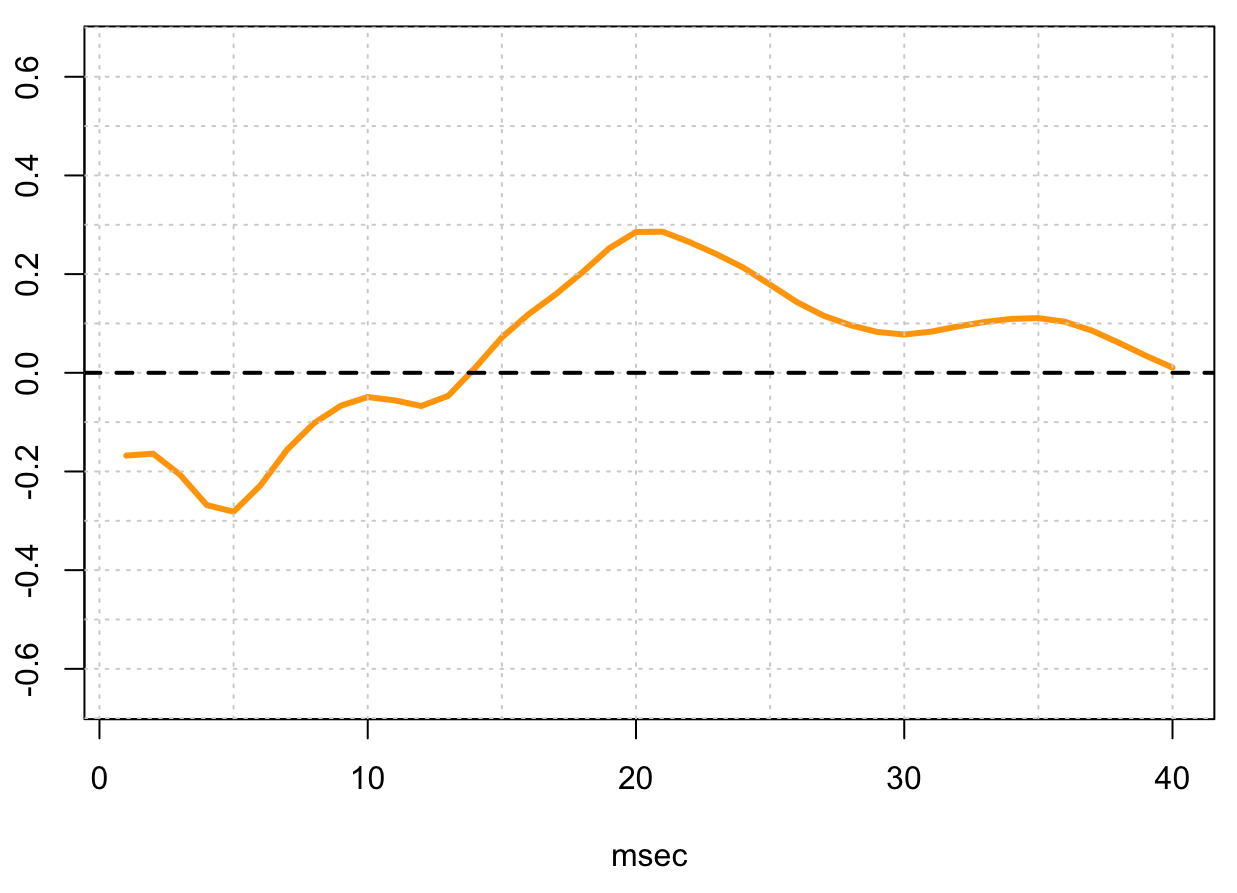}&
\includegraphics[height=4cm, width=4cm]{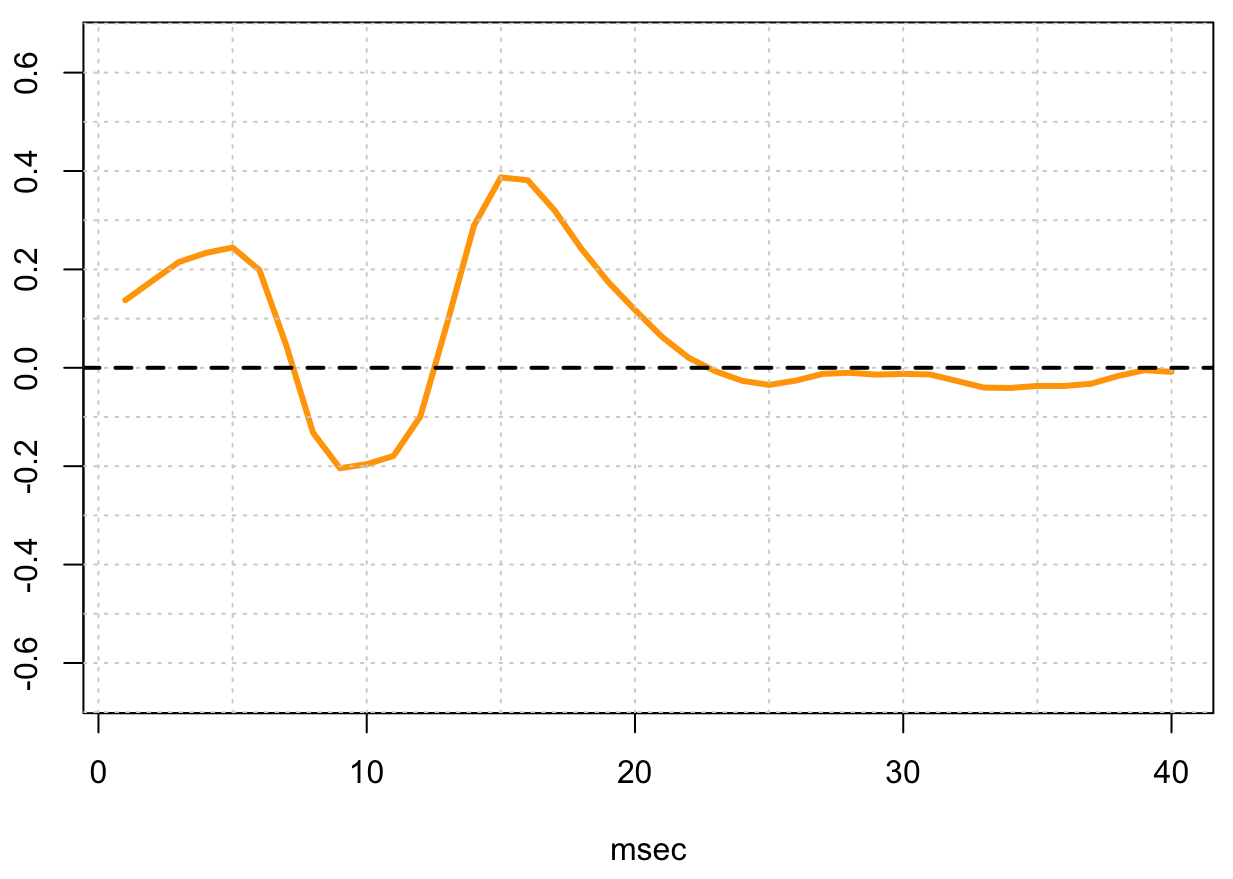}&
\includegraphics[height=4cm, width=4cm]{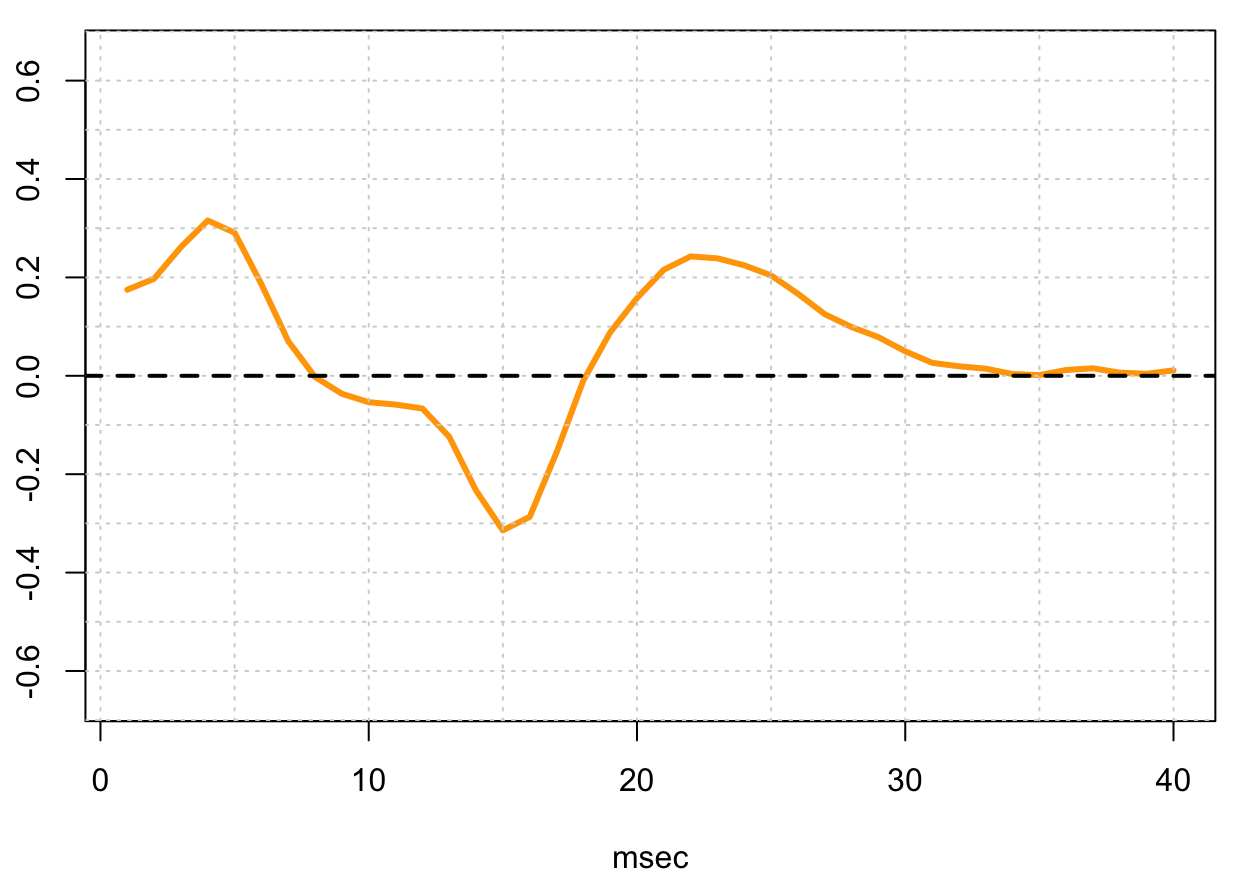}\\
\scriptsize{58.95\%}& \scriptsize{11.06\%}& \scriptsize{7.02\%}& \scriptsize{5.49\%}\\
\end{tabular}
\caption{\textbf{Eigenfunctions/principal components: Subset 1 "Games".} Four strongest components out of (7, 8, 10, 7) = (team level, position level, player level, curve level) estimated components. Percentage of level-specific variation explained below graphs.}\label{fig:eigfcts_games}
\end{figure}

\begin{figure}[H]
\centering
\setlength{\tabcolsep}{1.5pt}
\begin{tabular}{cccc}
\multicolumn{4}{c}{\textbf{\scriptsize{Team level}}}\\
\includegraphics[height=4cm, width=4cm]{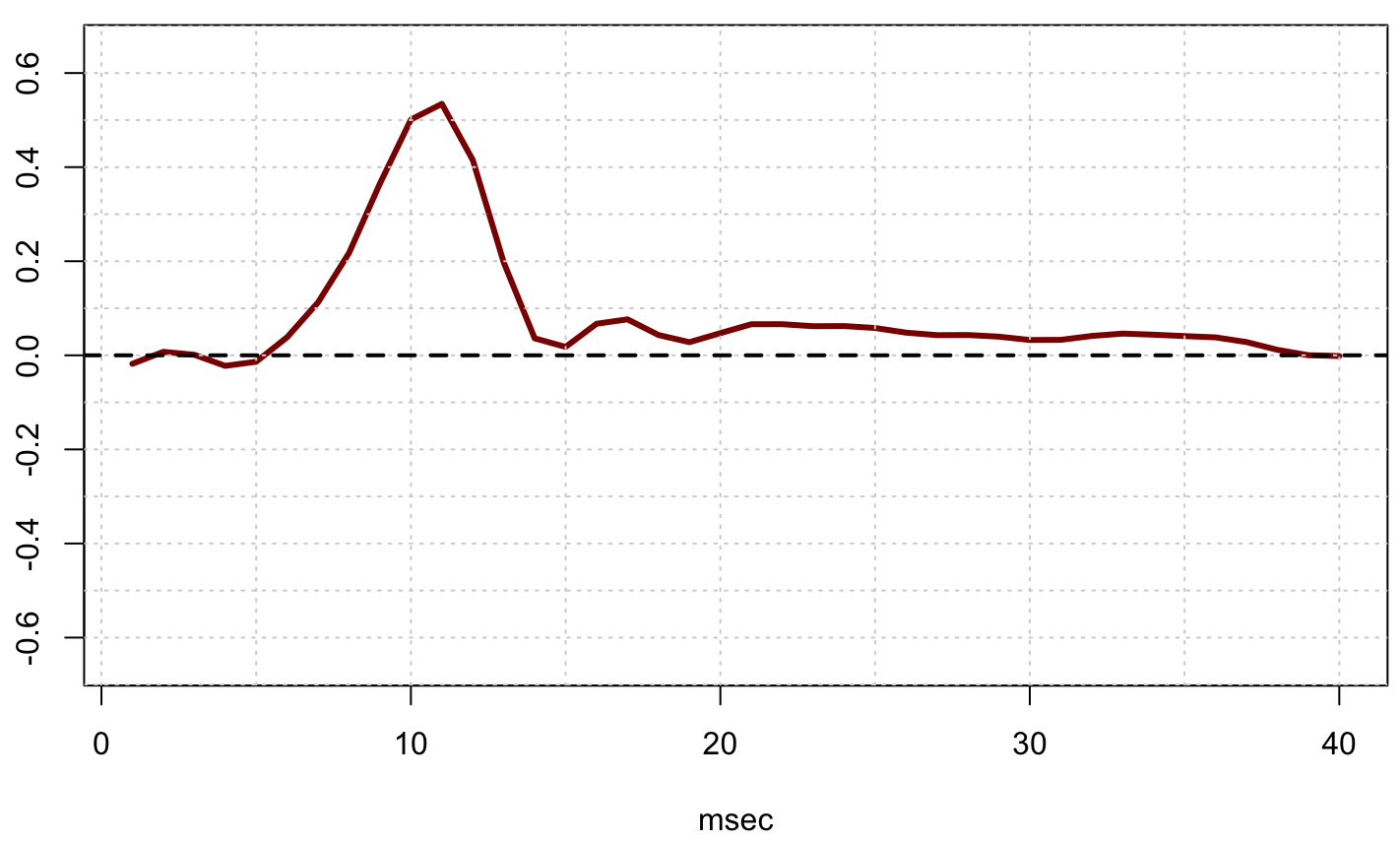}&\includegraphics[height=4cm, width=4cm]{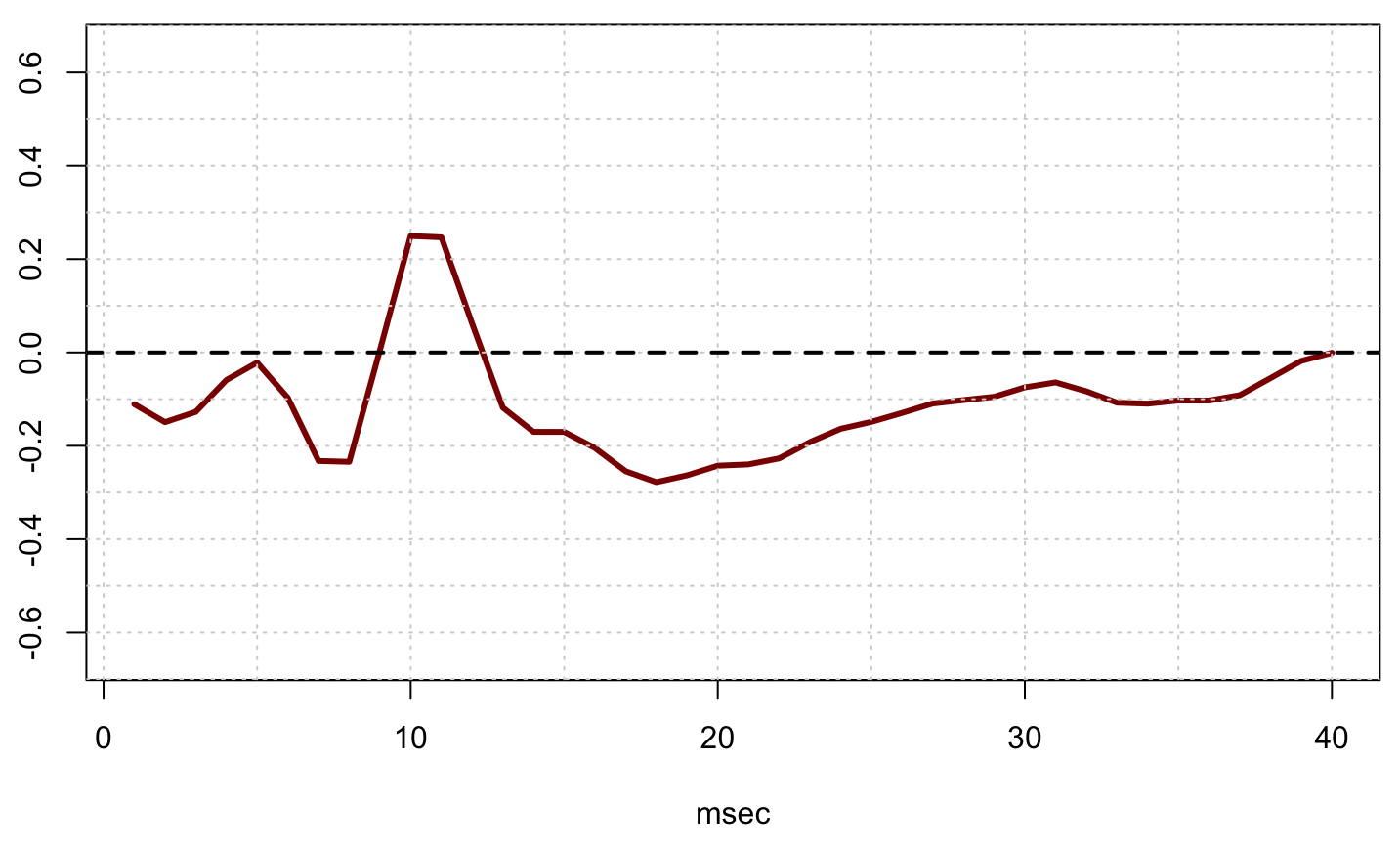}&
\includegraphics[height=4cm, width=4cm]{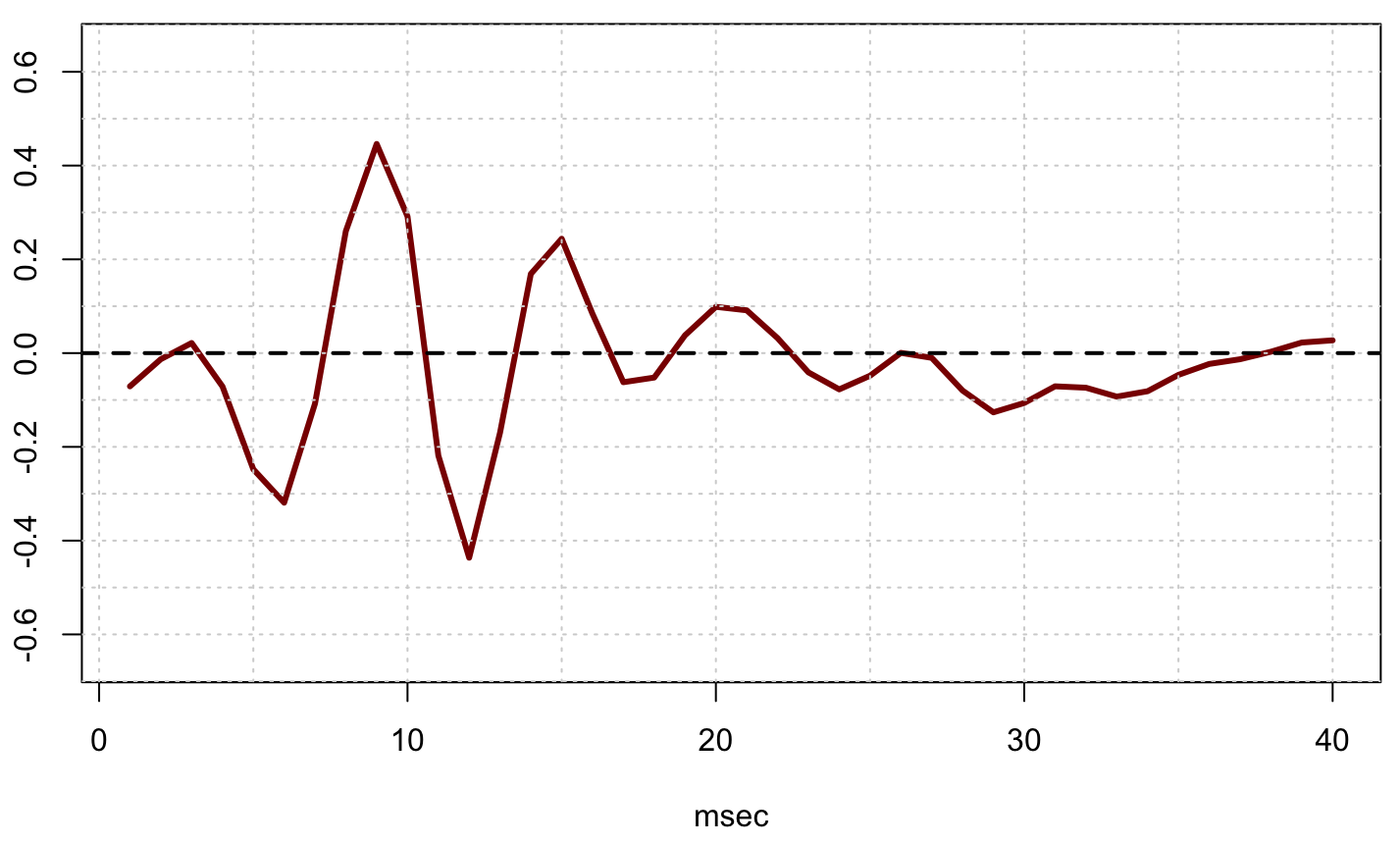}&
\includegraphics[height=4cm, width=4cm]{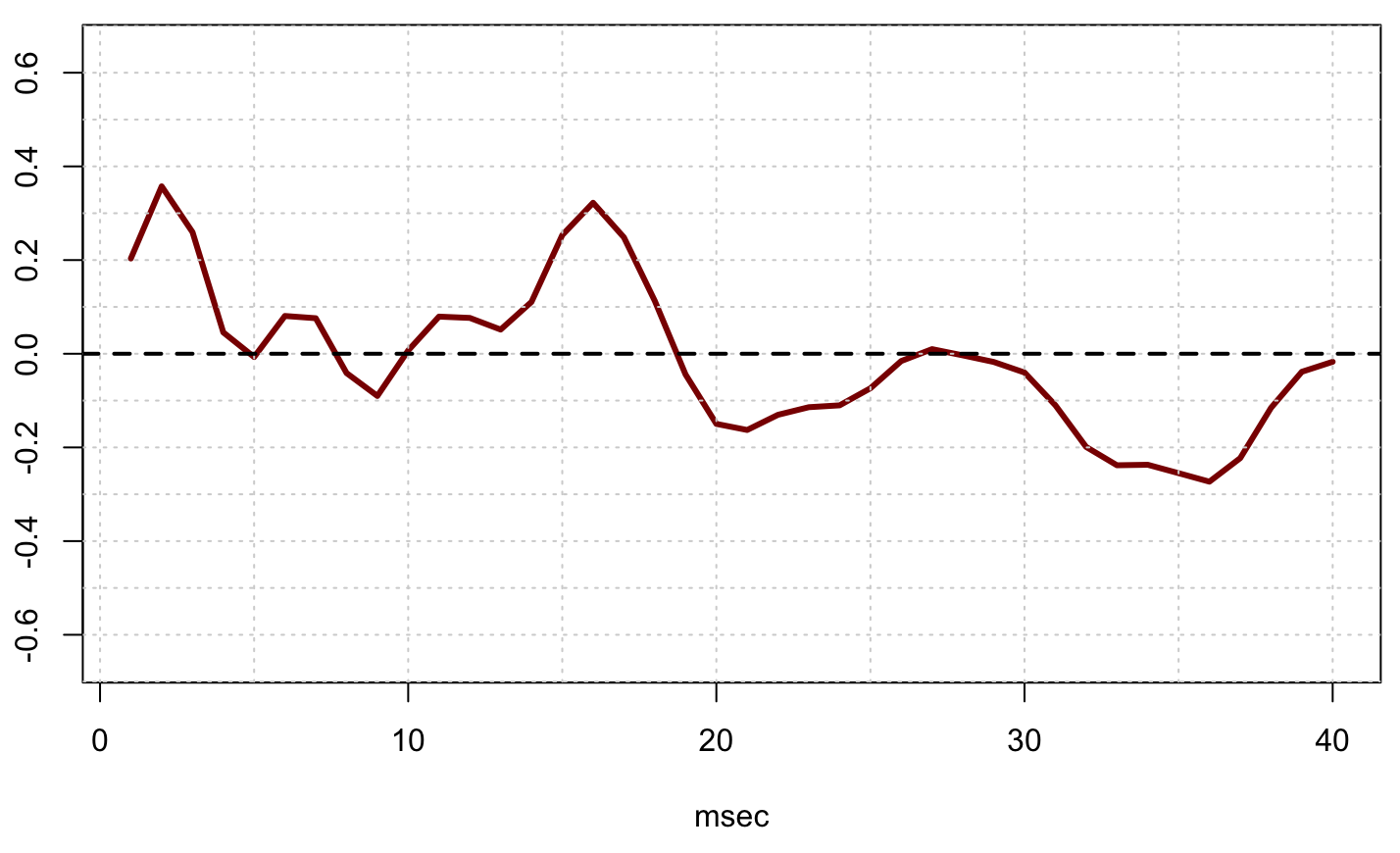}\\
\scriptsize{67.67\%}& \scriptsize{8.90\%}& \scriptsize{3.70\%}& \scriptsize{3.10\%}\\
\multicolumn{4}{c}{\textbf{\scriptsize{Position level}}}\\
\includegraphics[height=4cm, width=4cm]{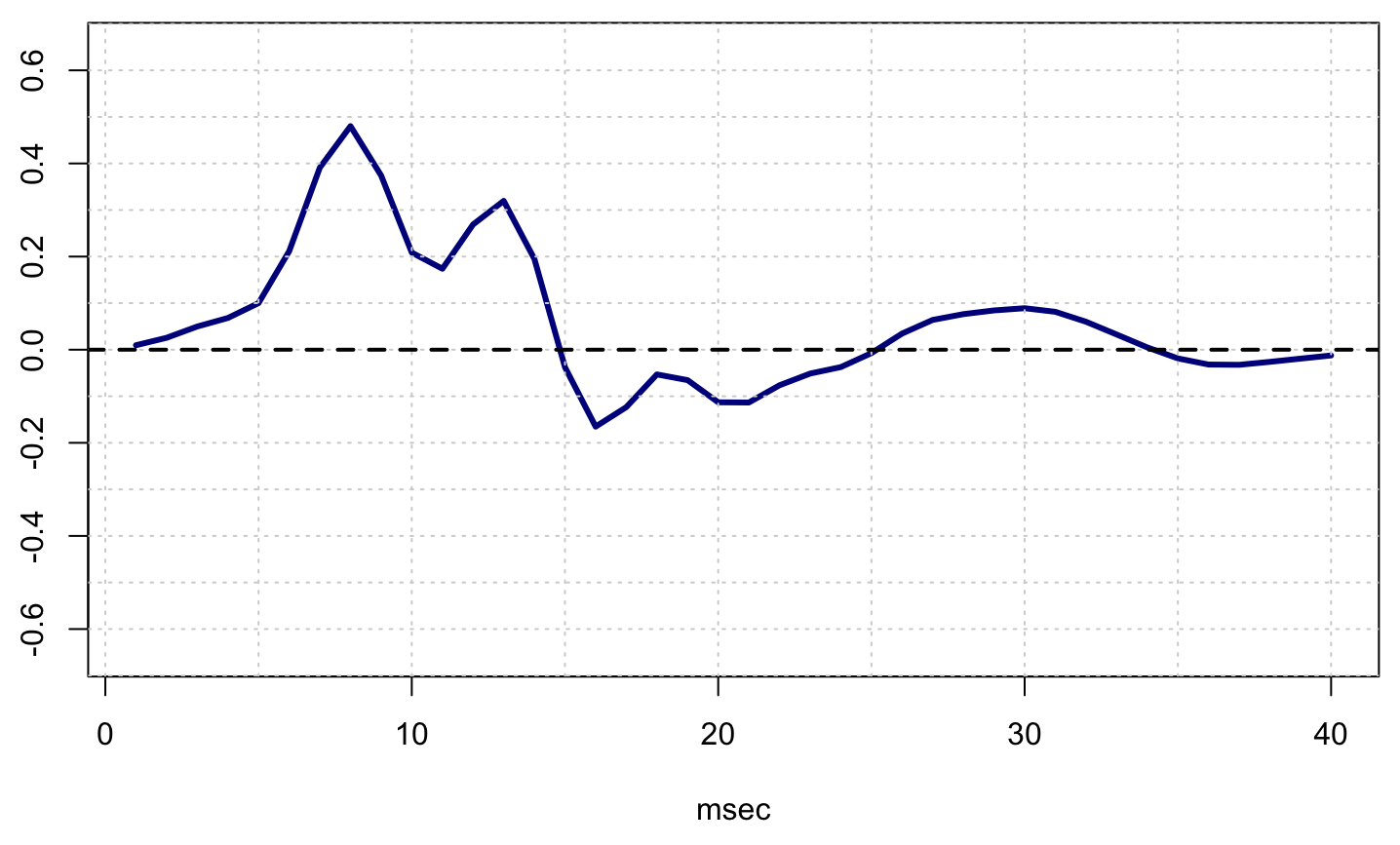}&\includegraphics[height=4cm, width=4cm]{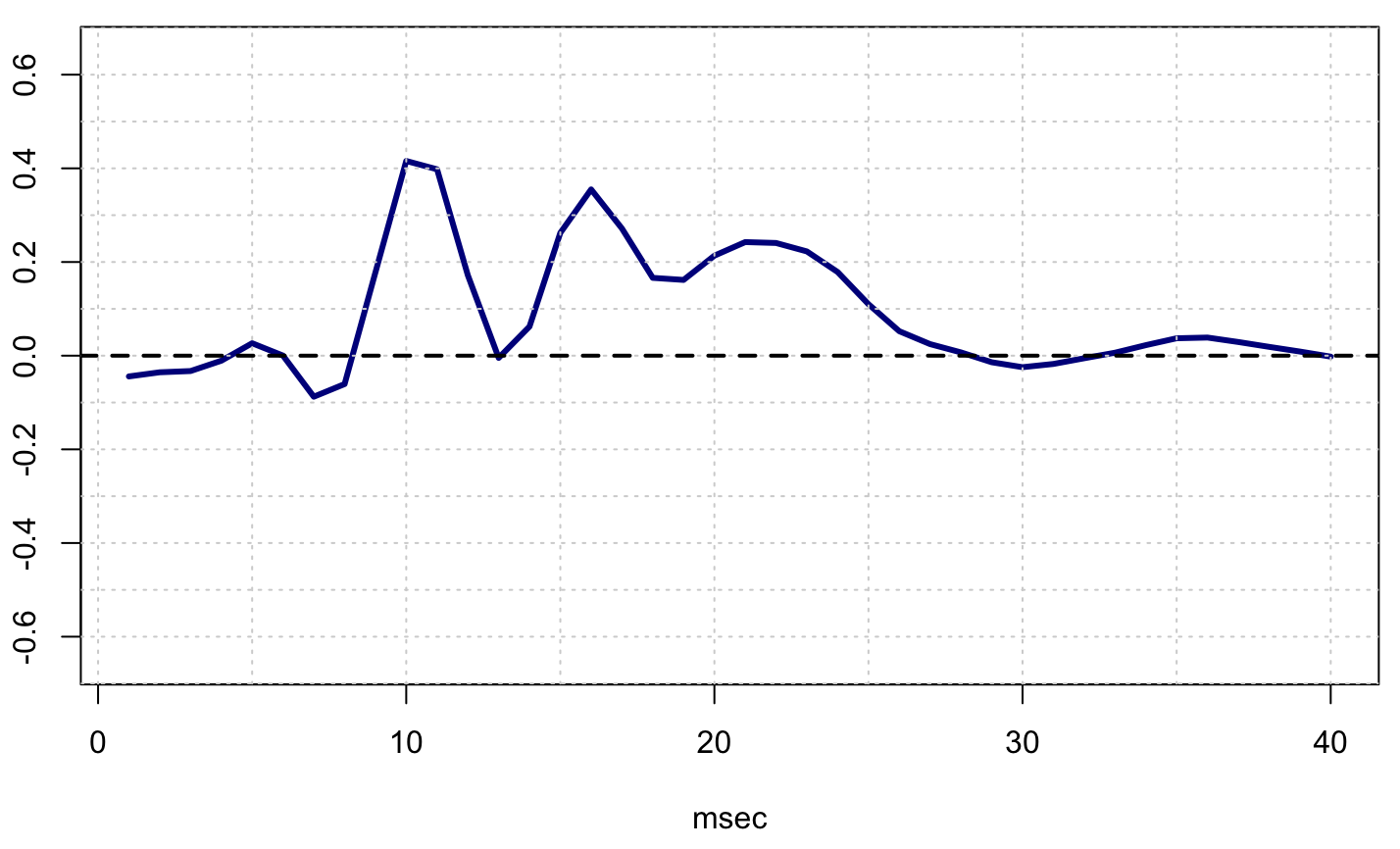}&
\includegraphics[height=4cm, width=4cm]{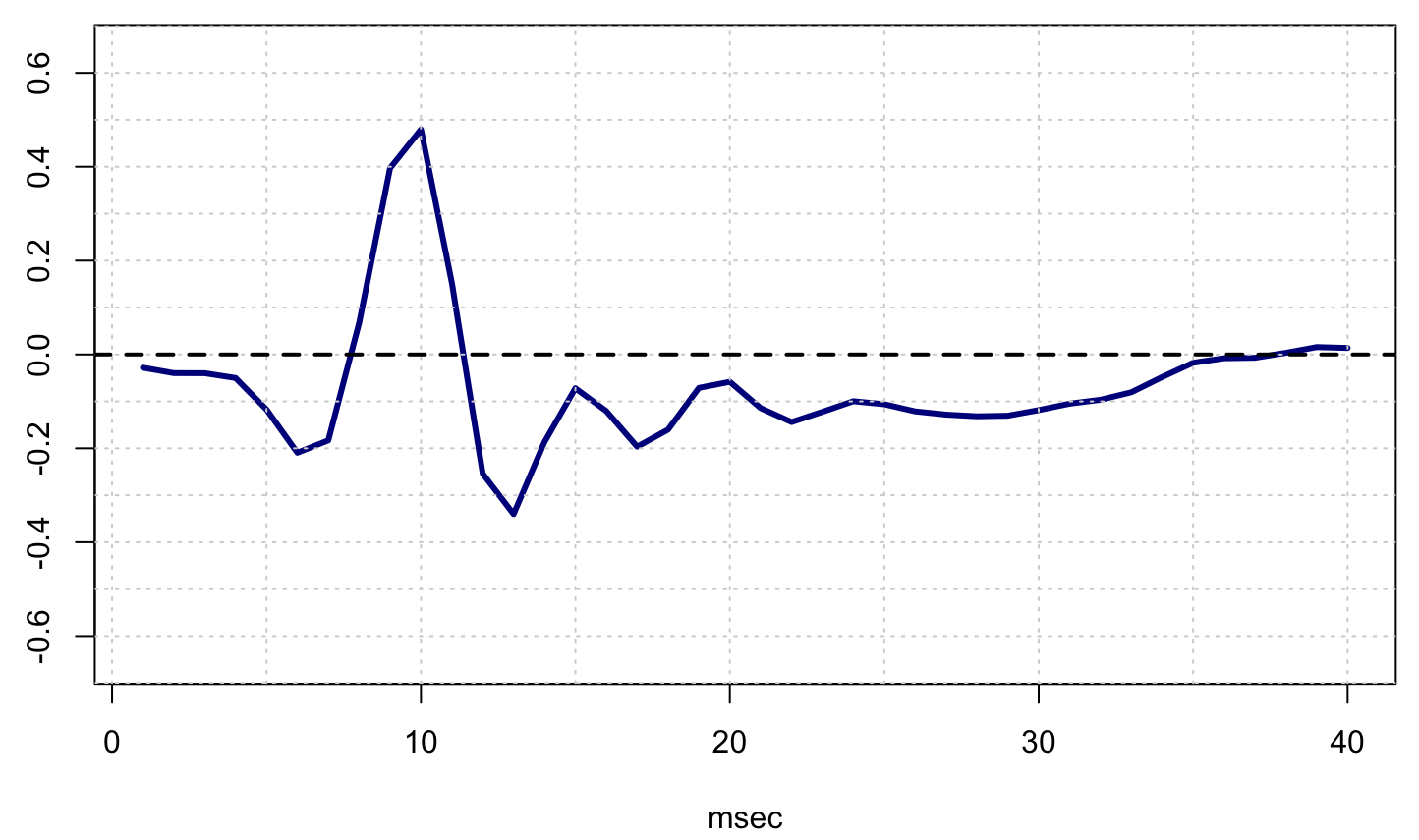}&
\includegraphics[height=4cm, width=4cm]{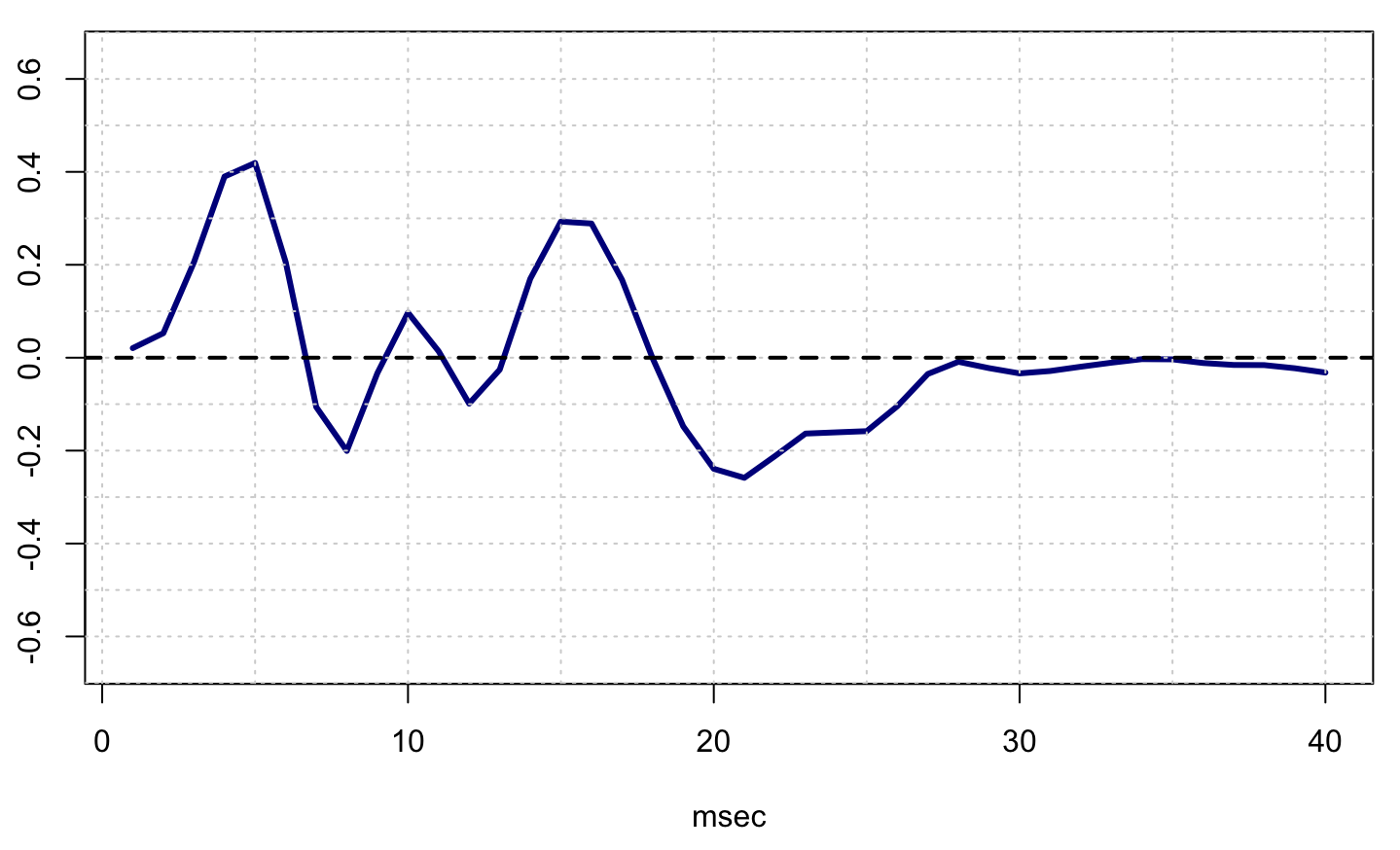}\\
\scriptsize{27.55\%}& \scriptsize{24.07 \%}& \scriptsize{13.62\%}& \scriptsize{8.05\%}\\   
\multicolumn{4}{c}{\textbf{\scriptsize{Player level}}}\\
\includegraphics[height=4cm, width=4cm]{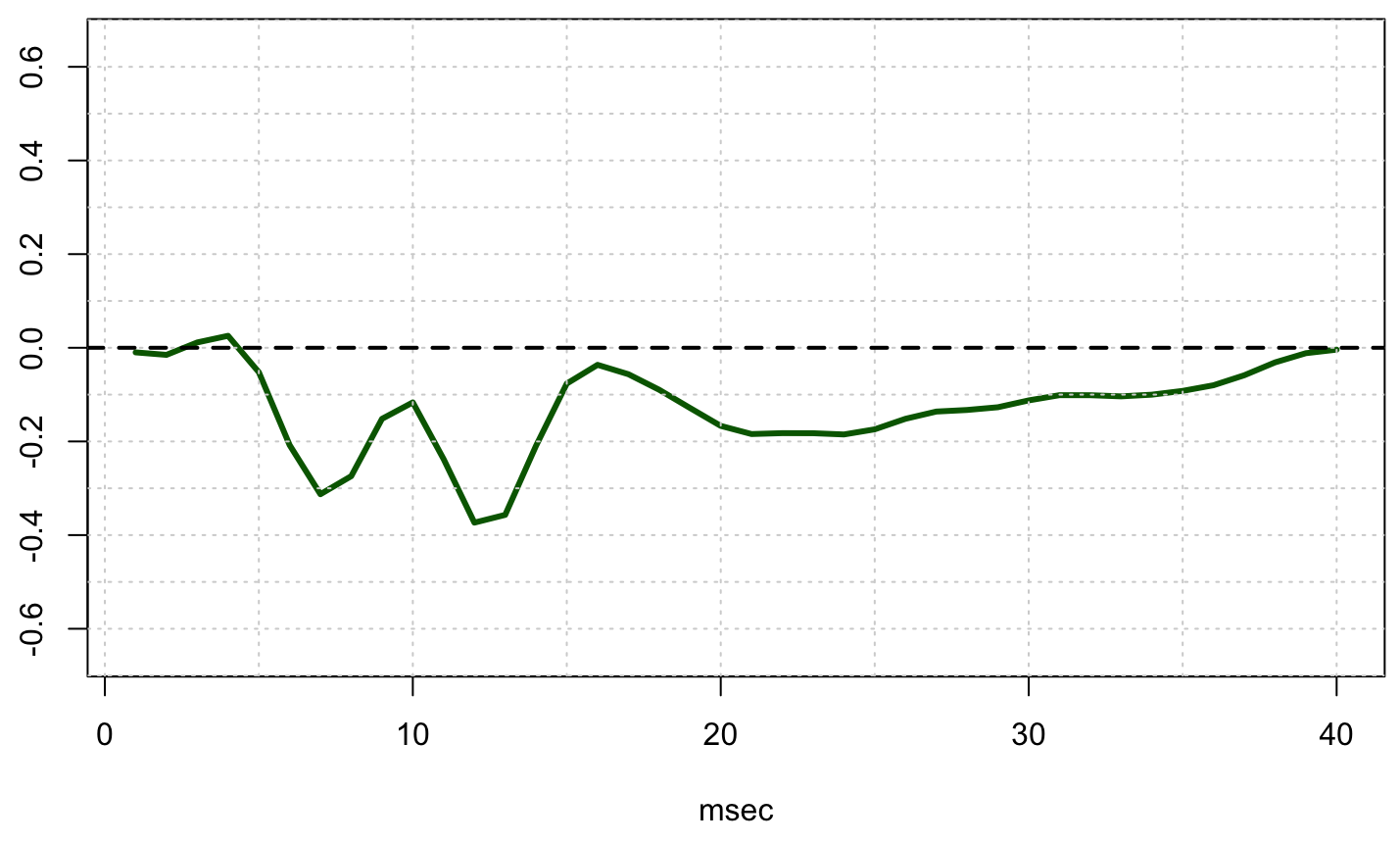}&\includegraphics[height=4cm, width=4cm]{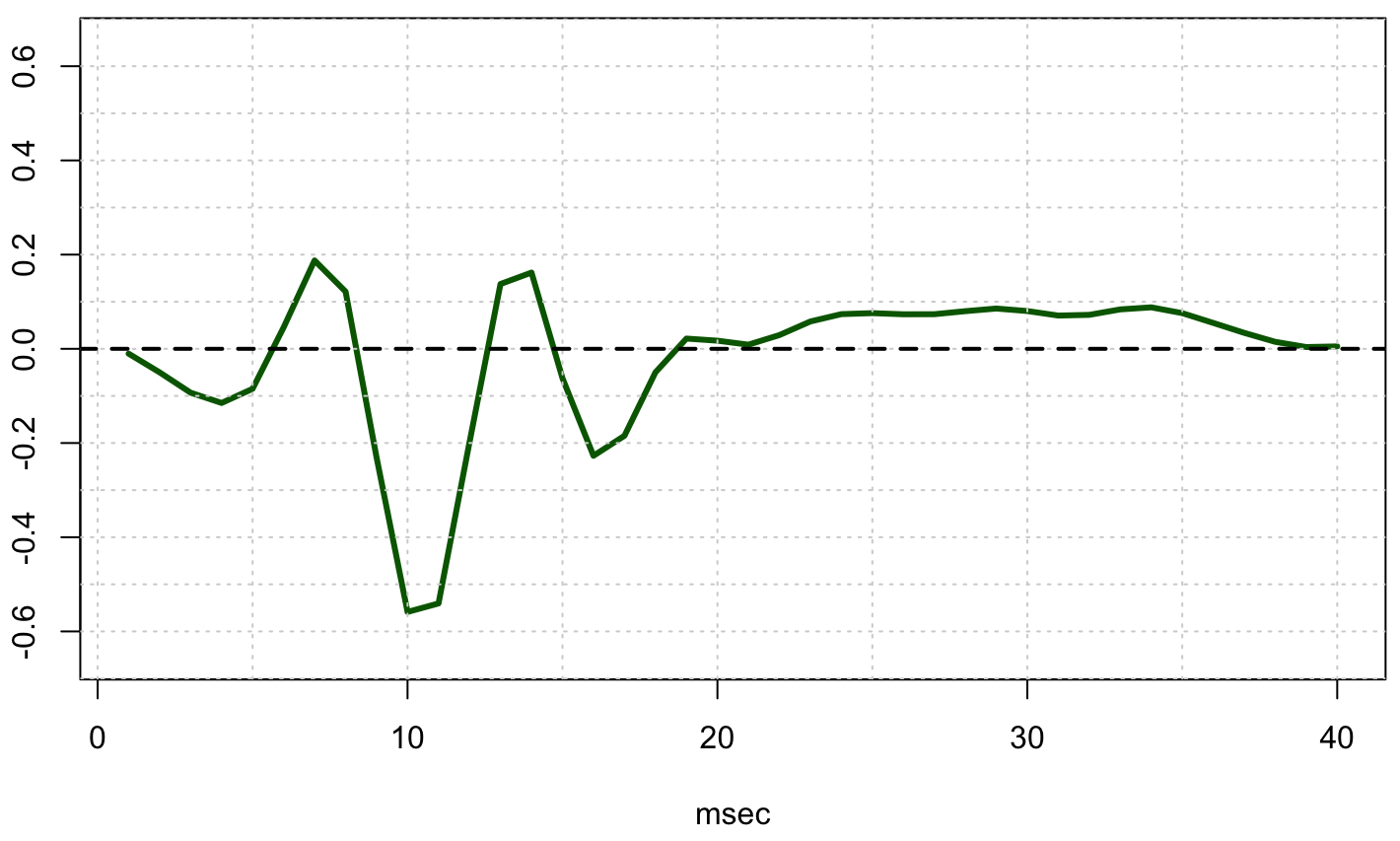}&
\includegraphics[height=4cm, width=4cm]{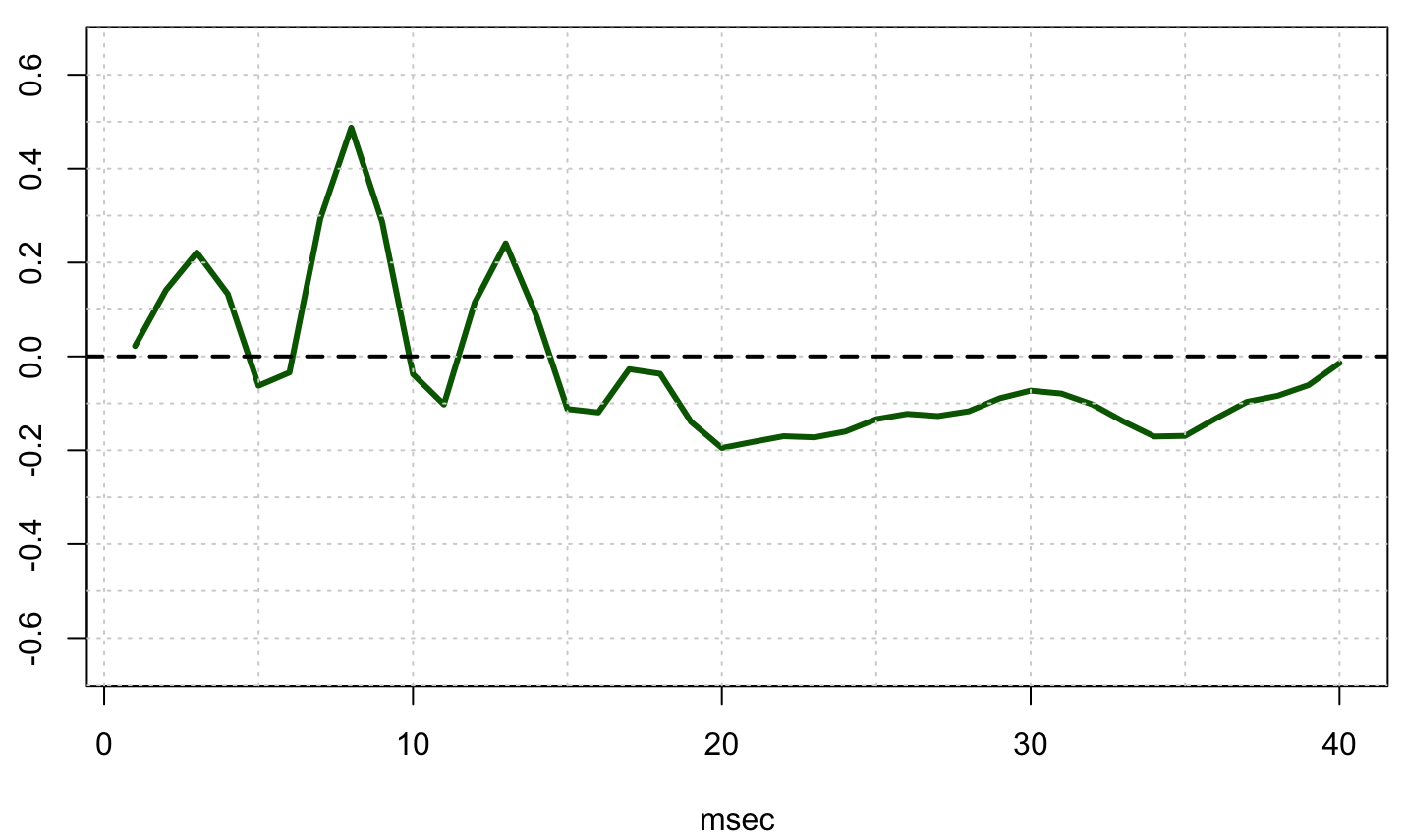}&
\includegraphics[height=4cm, width=4cm]{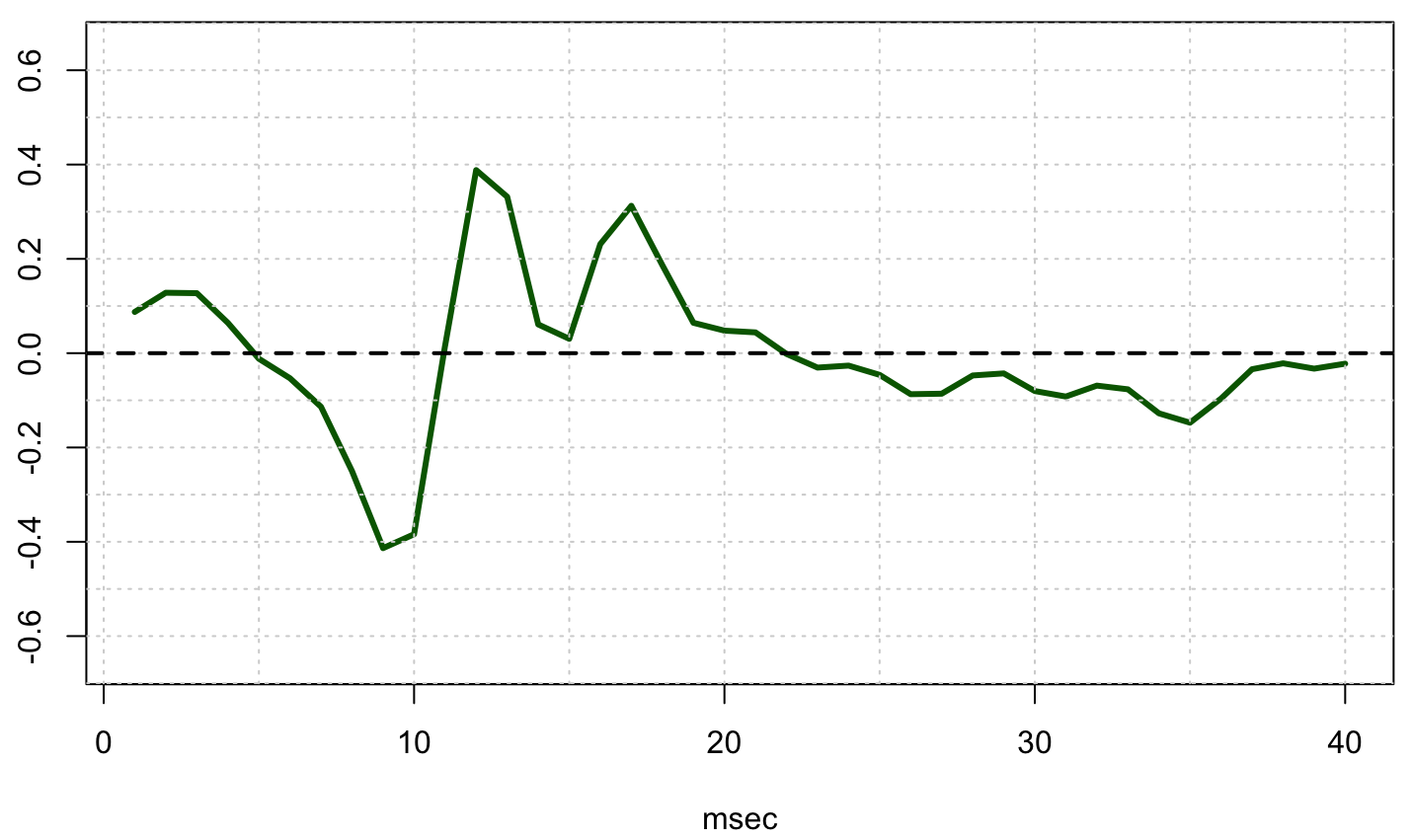}\\
\scriptsize{48.25\%}& \scriptsize{14.64\%}& \scriptsize{ 9.23 \%}& \scriptsize{7.06 \%}\\   
\multicolumn{4}{c}{\textbf{\scriptsize{Curve level}}}\\
\includegraphics[height=4cm, width=4cm]{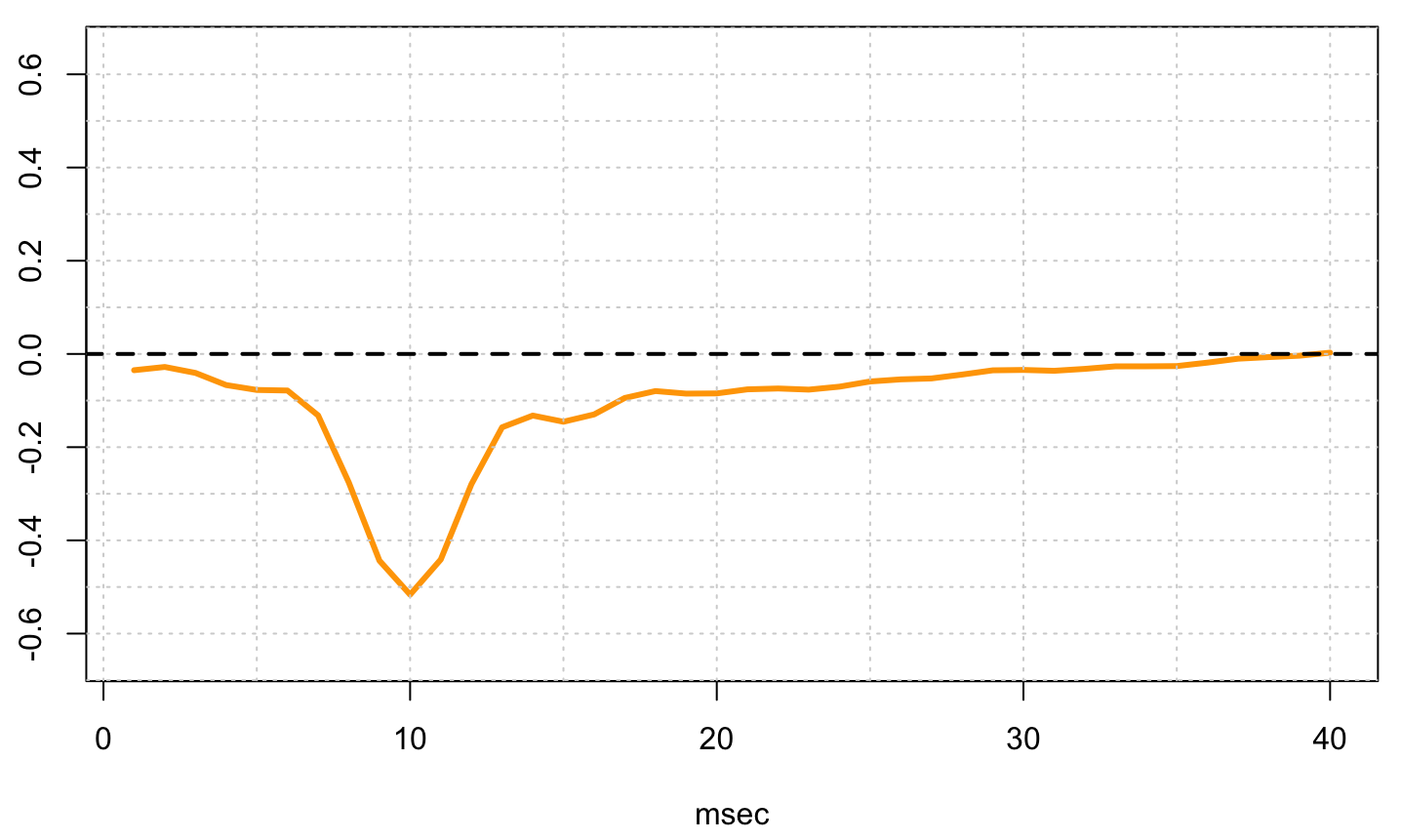}&\includegraphics[height=4cm, width=4cm]{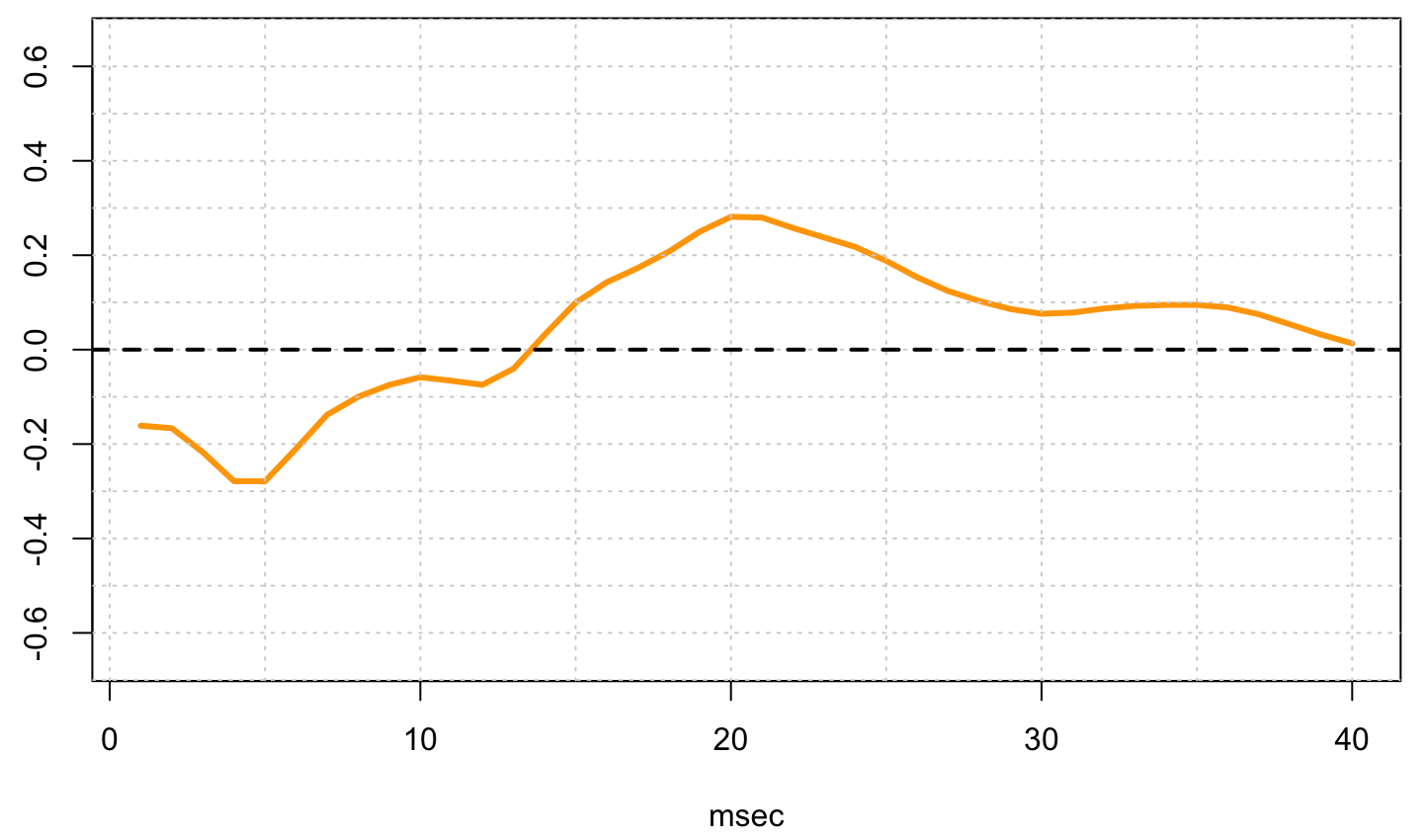}&
\includegraphics[height=4cm, width=4cm]{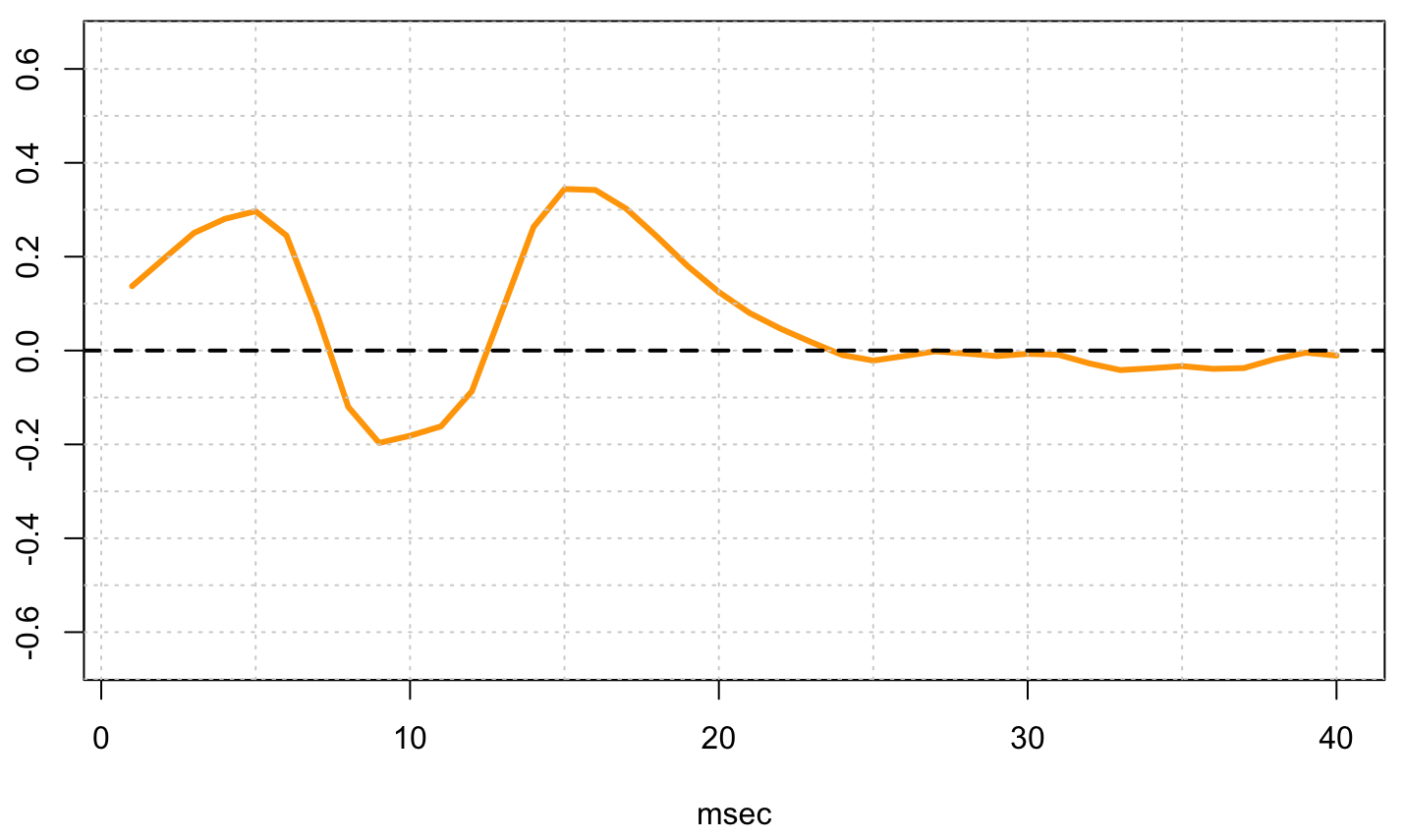}&
\includegraphics[height=4cm, width=4cm]{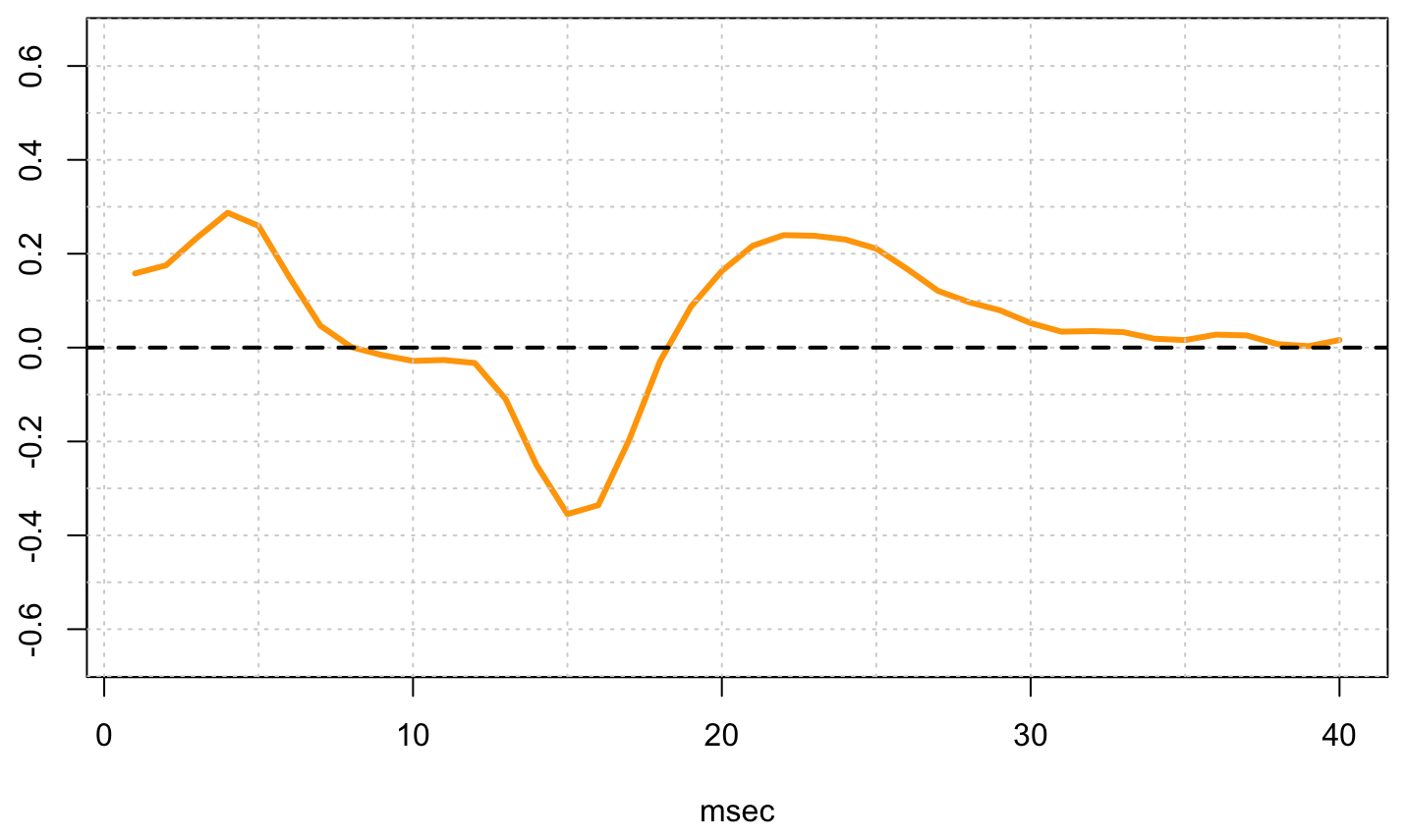}\\
\scriptsize{55.99\%}& \scriptsize{10.72\%}& \scriptsize{7.73 \%}& \scriptsize{5.56\%}\\    
\end{tabular}
\caption{\textbf{Eigenfunctions/principal components: Subset 2 "In-season Practice".} Four strongest components out of (7, 9, 8, 8) = (team level, position level, player level, curve level) estimated components. Percentage of level-specific variation explained below graphs.}\label{fig:eigfcts_inseason}
\end{figure}

\begin{figure}[H]
\centering
\setlength{\tabcolsep}{1.5pt}
\begin{tabular}{cccc}
\multicolumn{4}{c}{\textbf{\scriptsize{Team level}}}\\
\includegraphics[height=4cm, width=4cm]{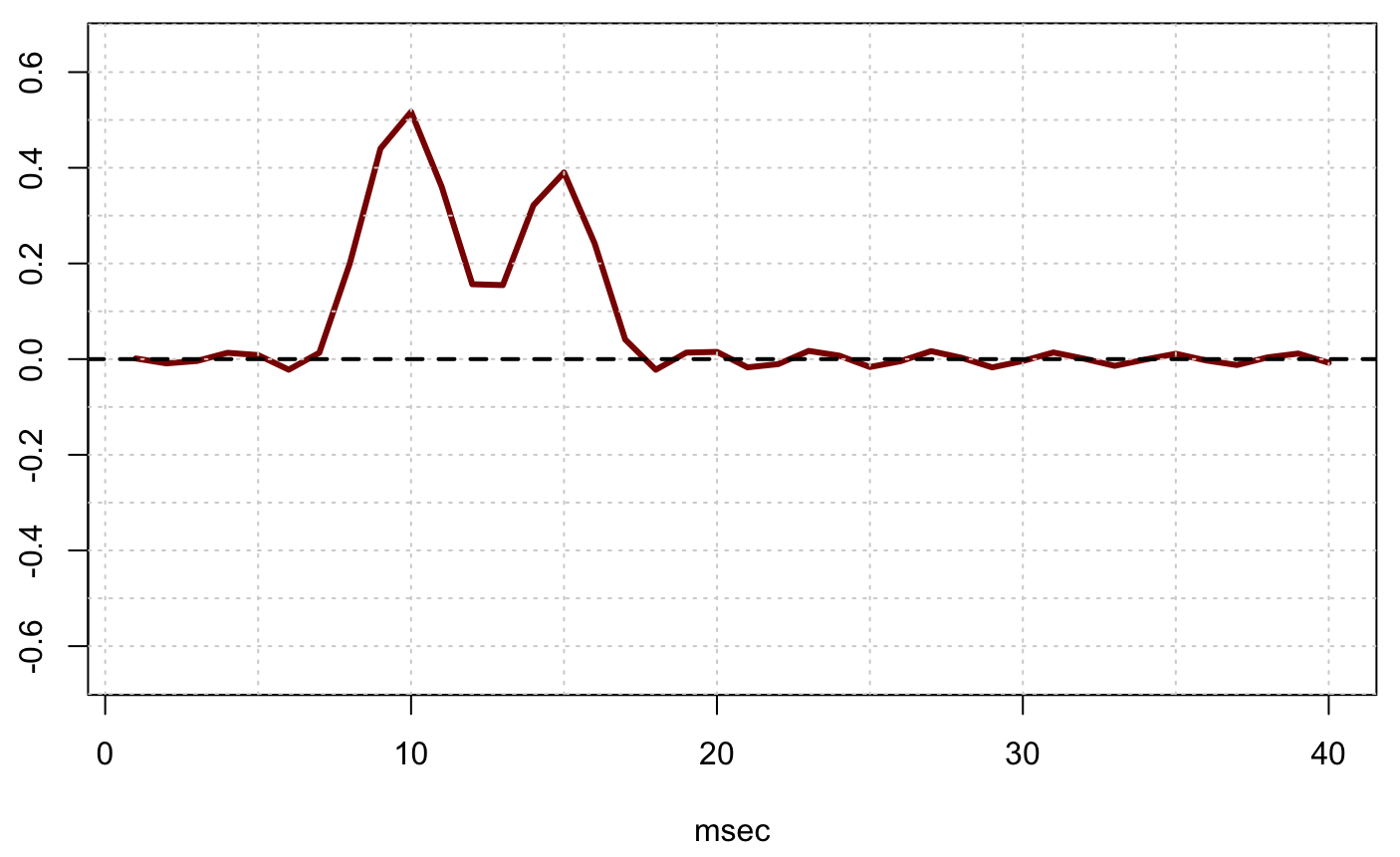}&\includegraphics[height=4cm, width=4cm]{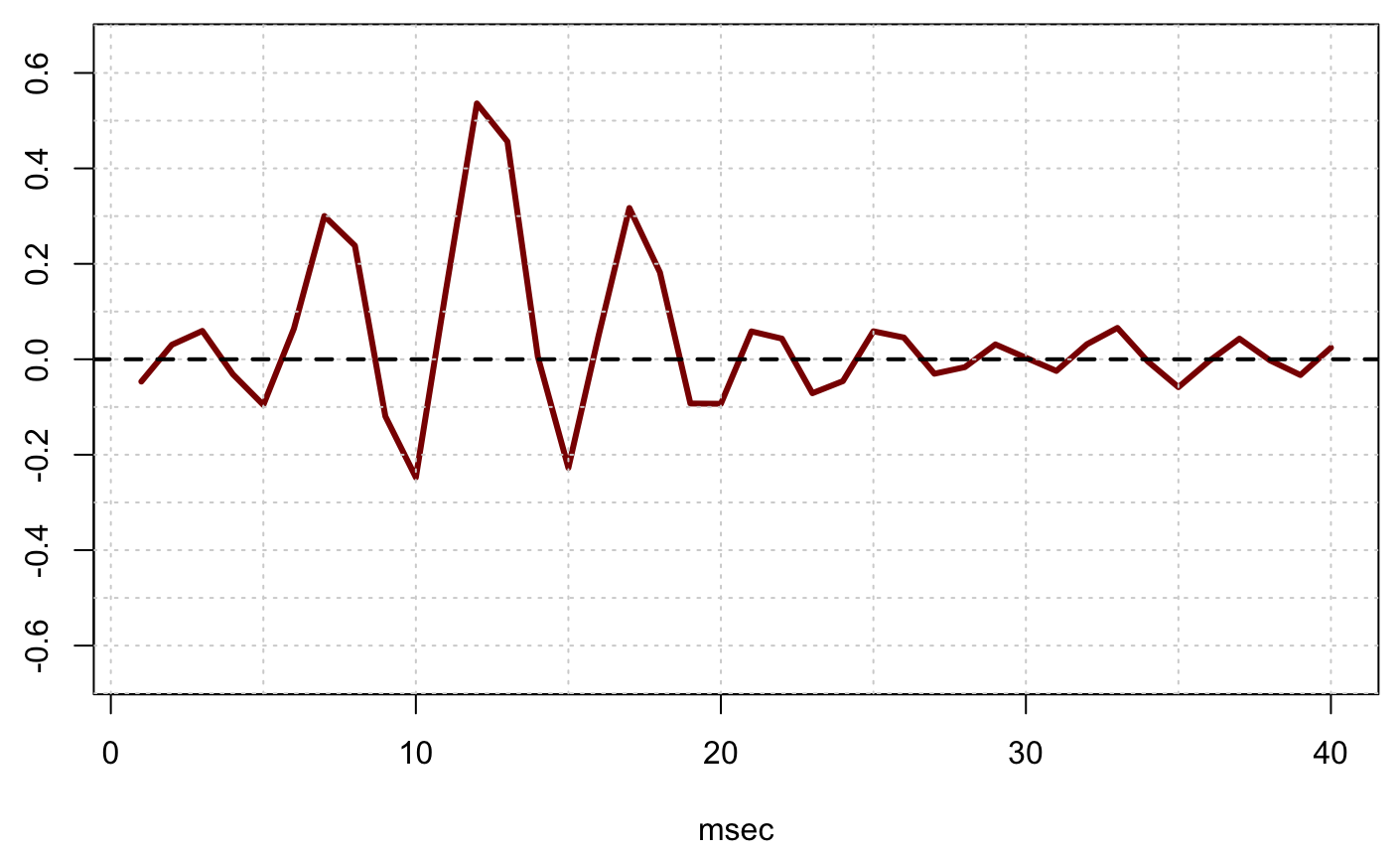}&
\includegraphics[height=4cm, width=4cm]{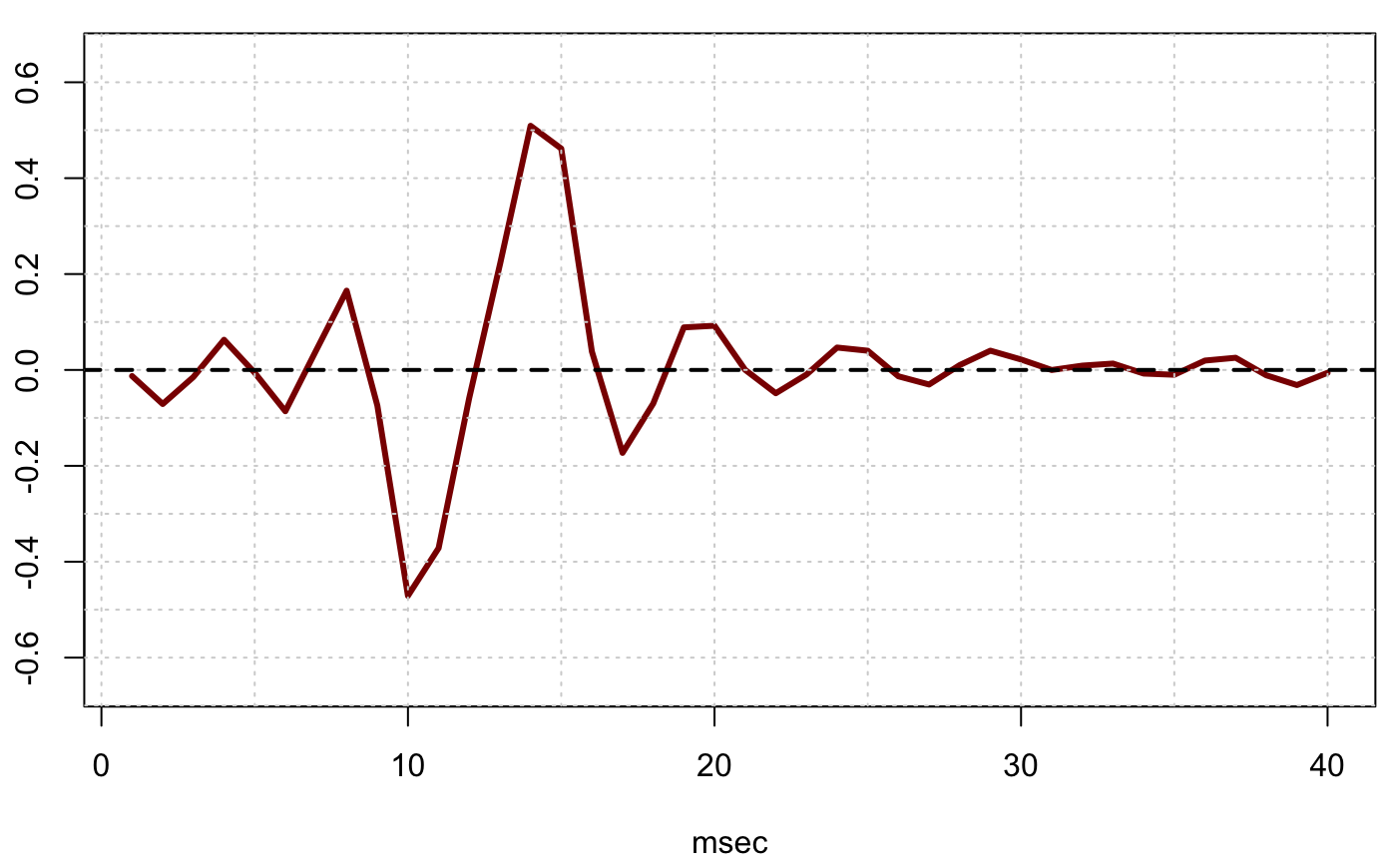}&
\includegraphics[height=4cm, width=4cm]{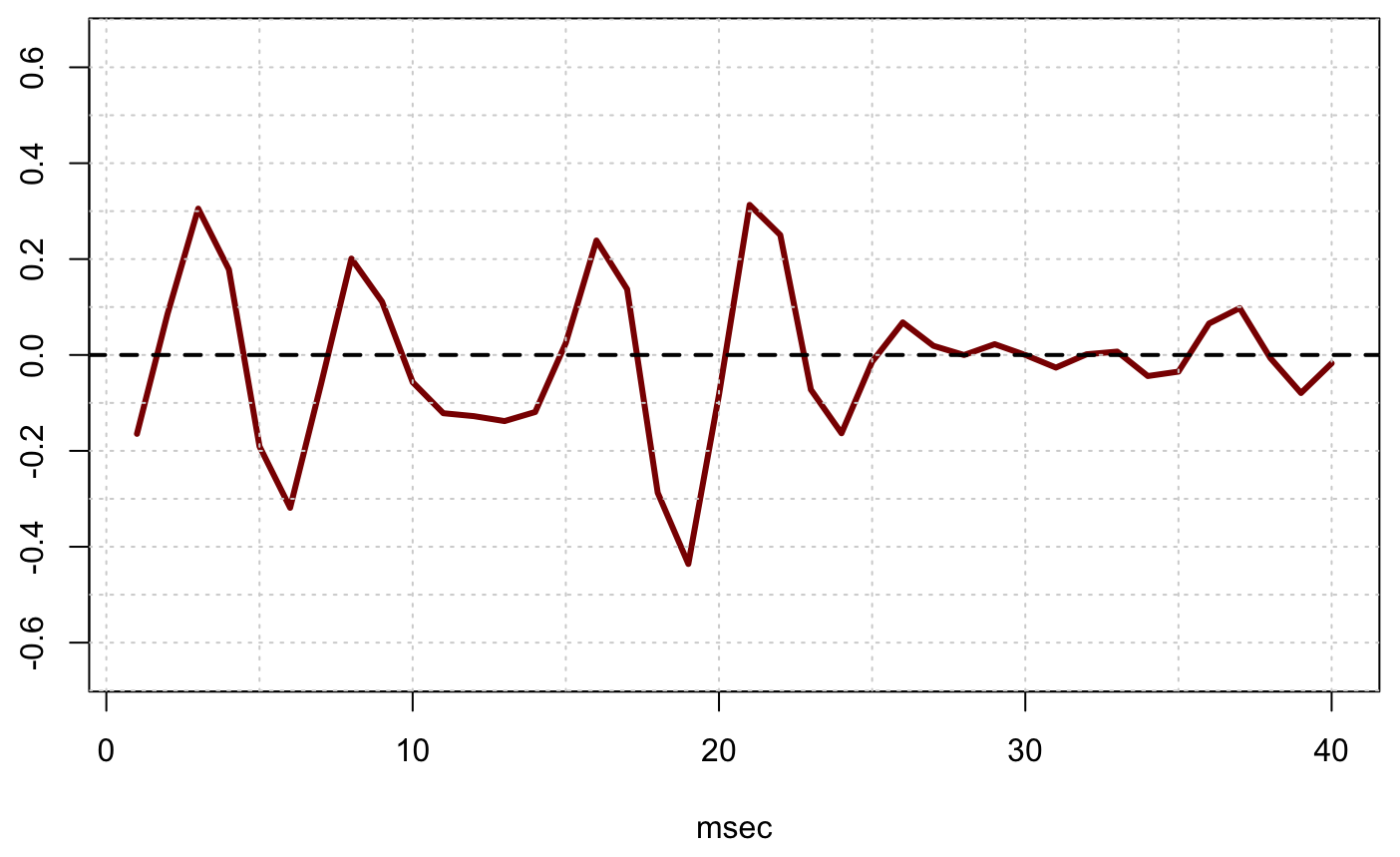}\\
\scriptsize{65.72\%}& \scriptsize{9.79\%}& \scriptsize{4.75\%}& \scriptsize{3.52\%}\\      
\multicolumn{4}{c}{\textbf{\scriptsize{Position level}}}\\
\includegraphics[height=4cm, width=4cm]{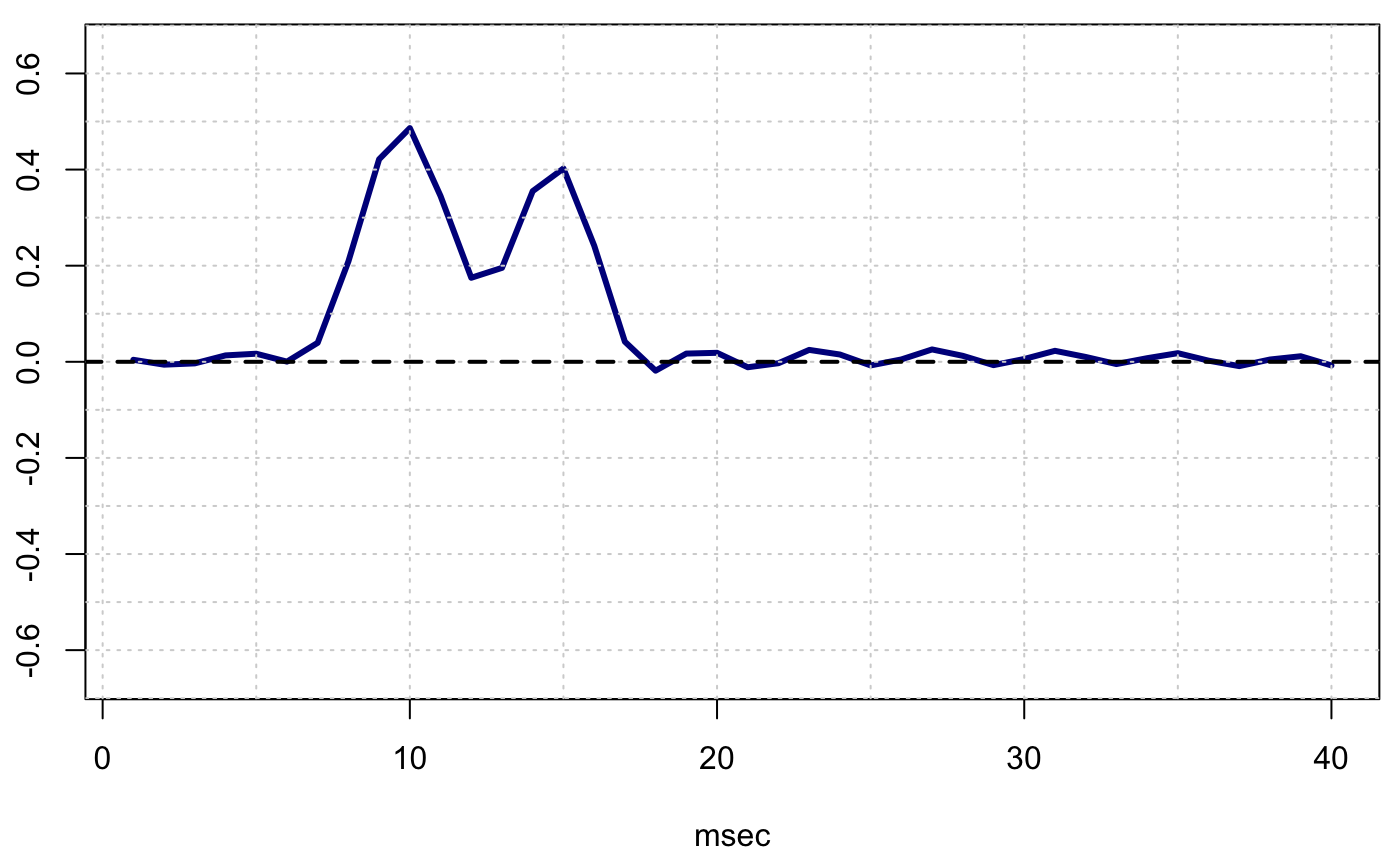}&\includegraphics[height=4cm, width=4cm]{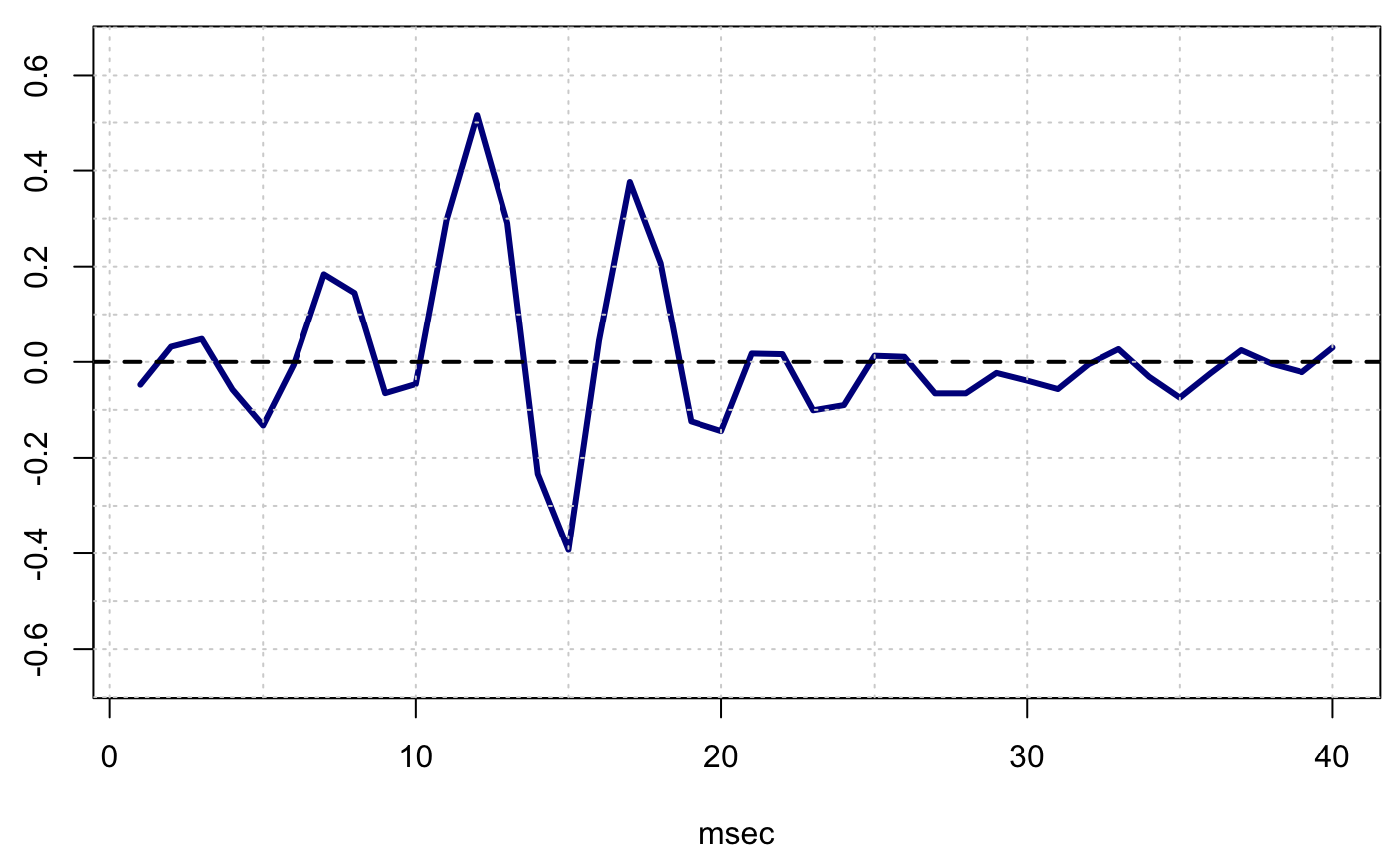}&
\includegraphics[height=4cm, width=4cm]{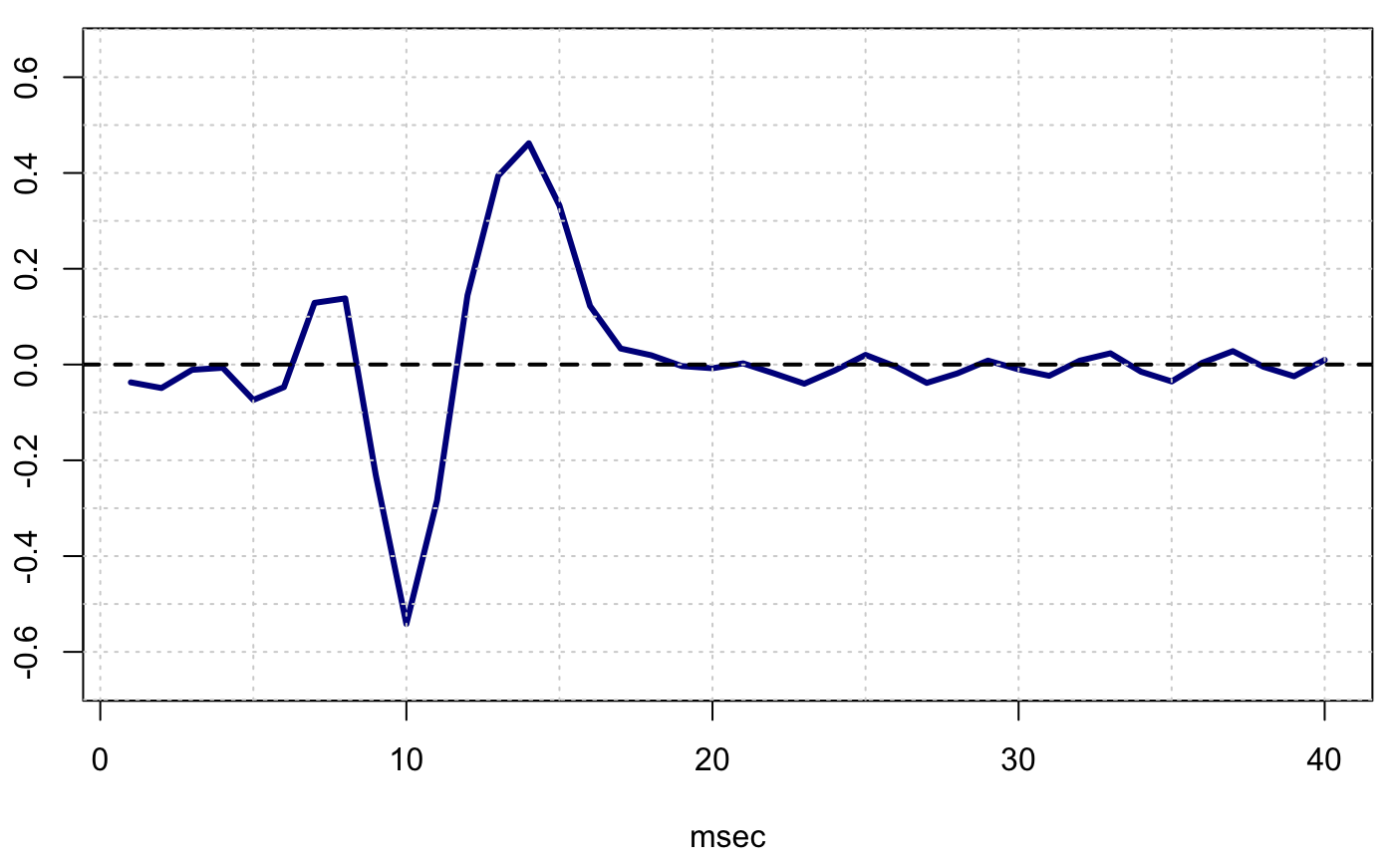}&
\includegraphics[height=4cm, width=4cm]{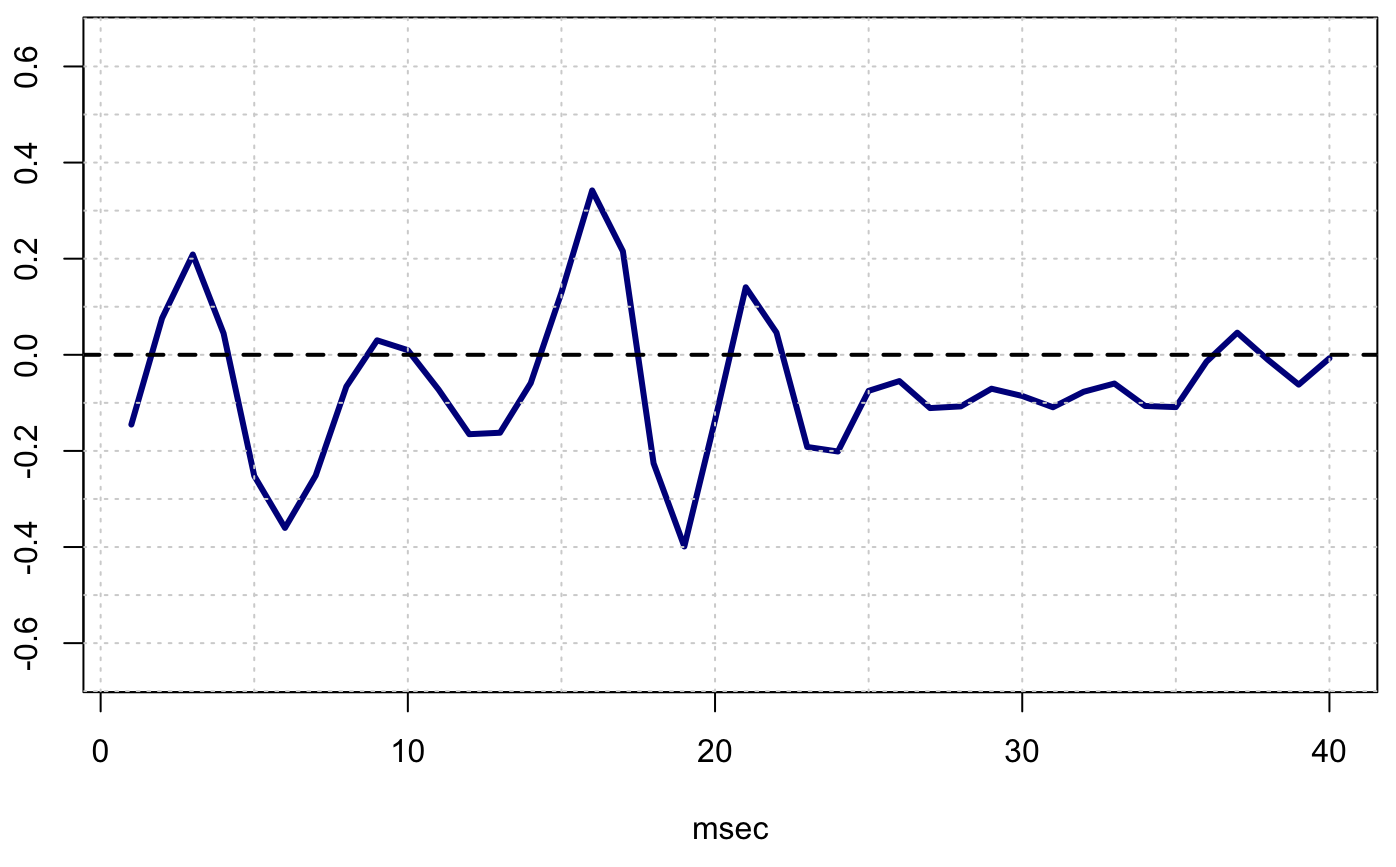}\\
\scriptsize{64.31\%}& \scriptsize{7.87 \%}& \scriptsize{7.09\%}& \scriptsize{3.99\%}\\         
\multicolumn{4}{c}{\textbf{\scriptsize{Player level}}}\\
\includegraphics[height=4cm, width=4cm]{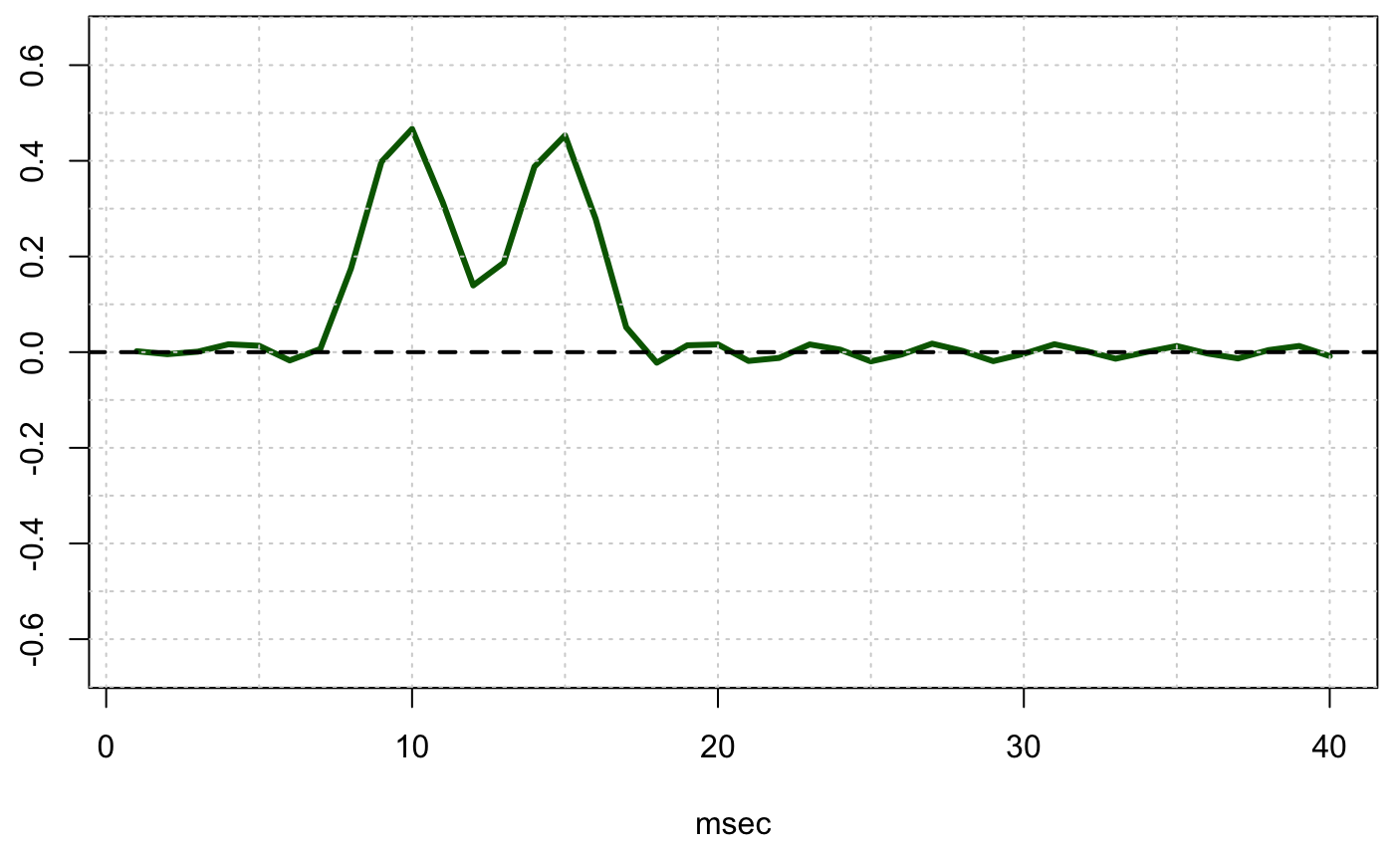}&\includegraphics[height=4cm, width=4cm]{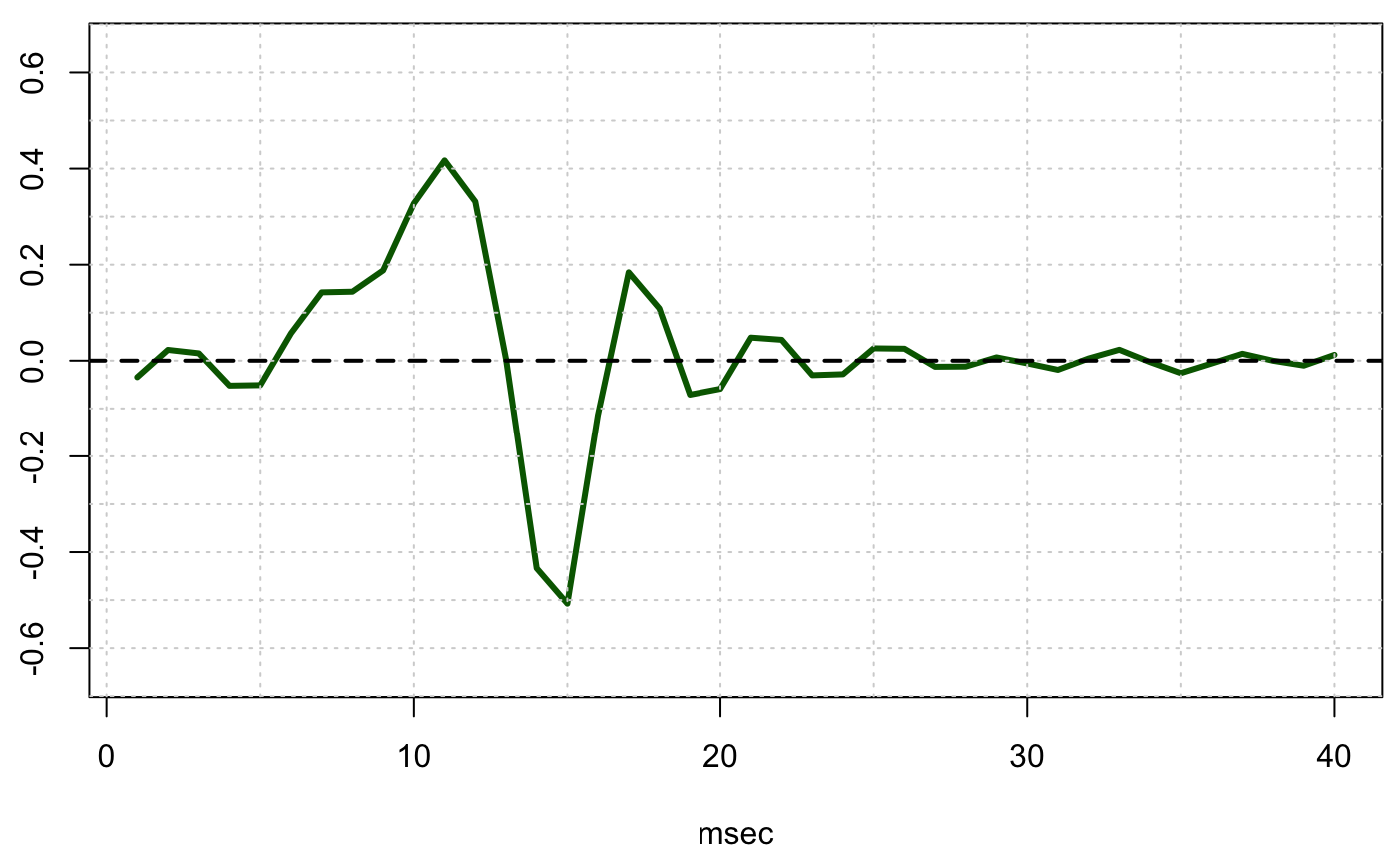}&
\includegraphics[height=4cm, width=4cm]{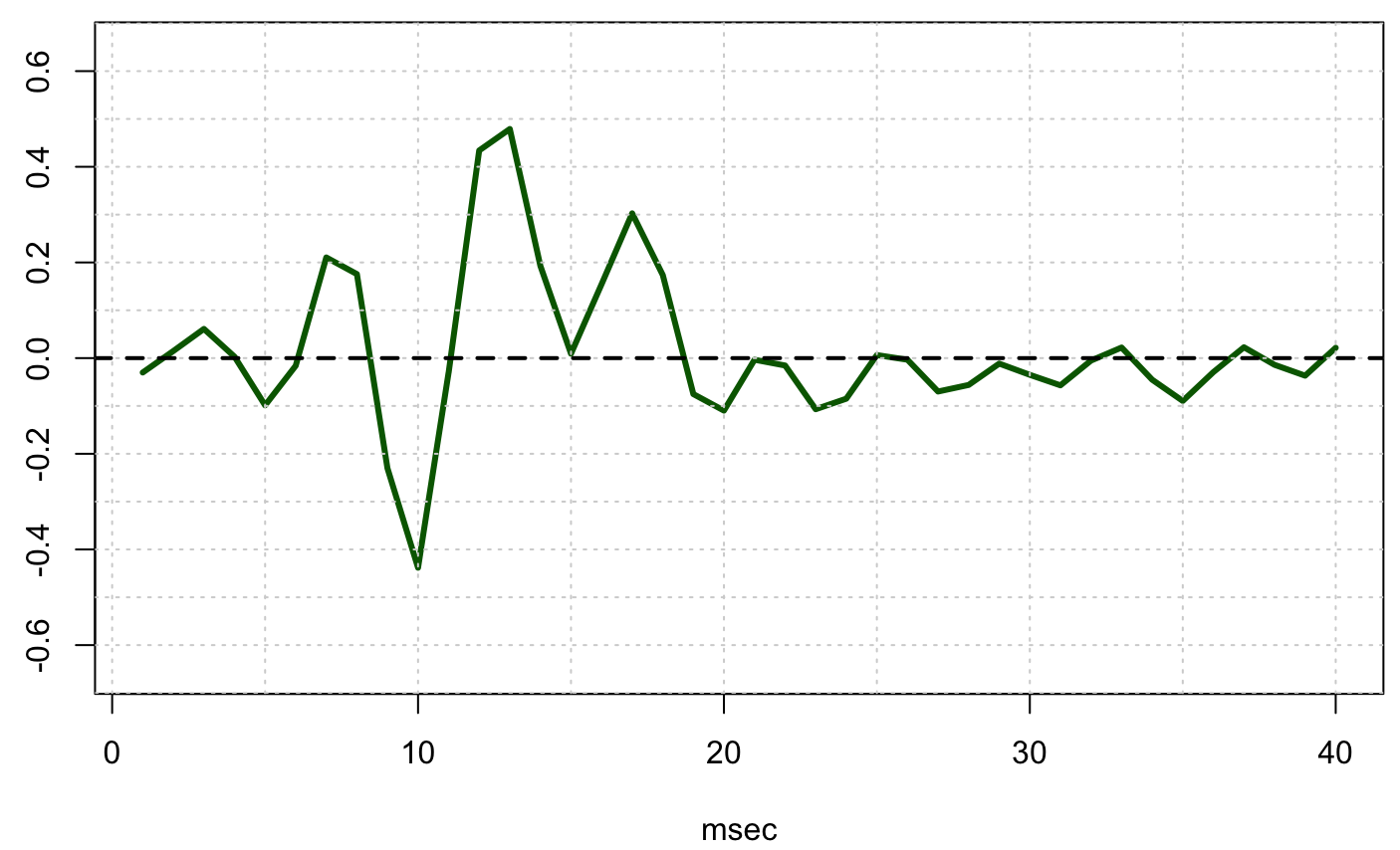}&
\includegraphics[height=4cm, width=4cm]{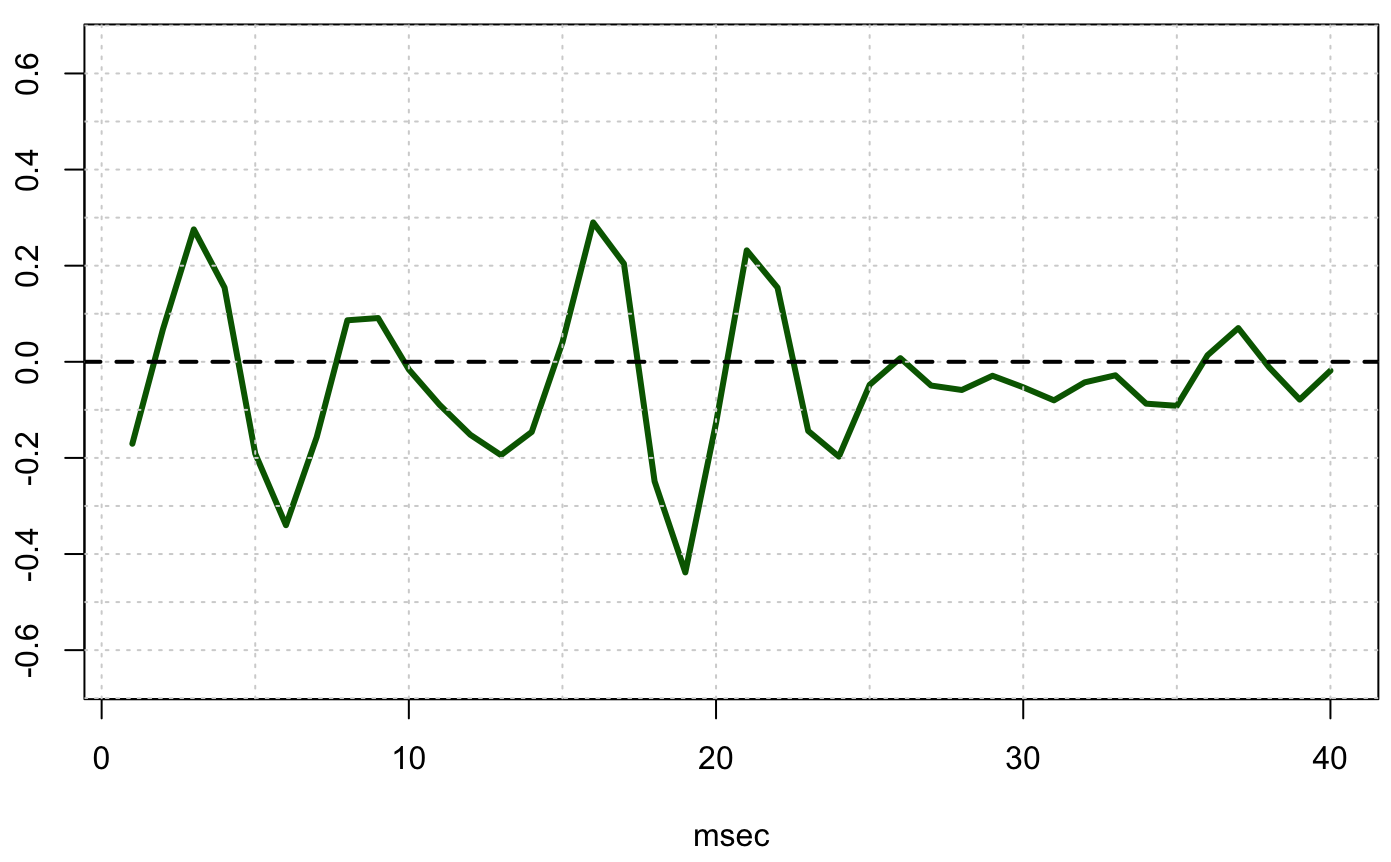}\\
\scriptsize{59.22\%}& \scriptsize{11.83\%}& \scriptsize{8.44\%}& \scriptsize{3.92\%}\\        
\multicolumn{4}{c}{\textbf{\scriptsize{Curve level}}}\\
\includegraphics[height=4cm, width=4cm]{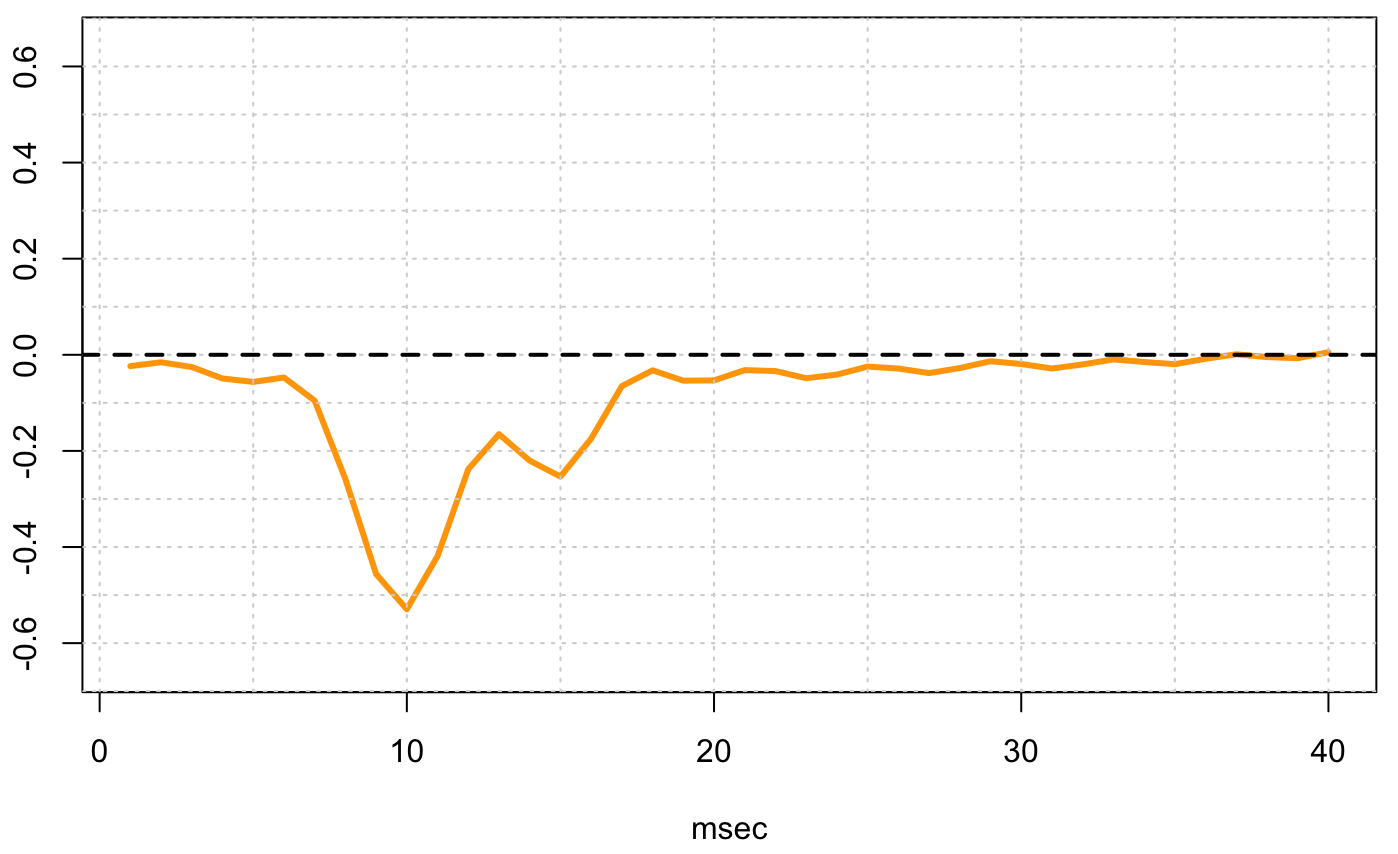}&\includegraphics[height=4cm, width=4cm]{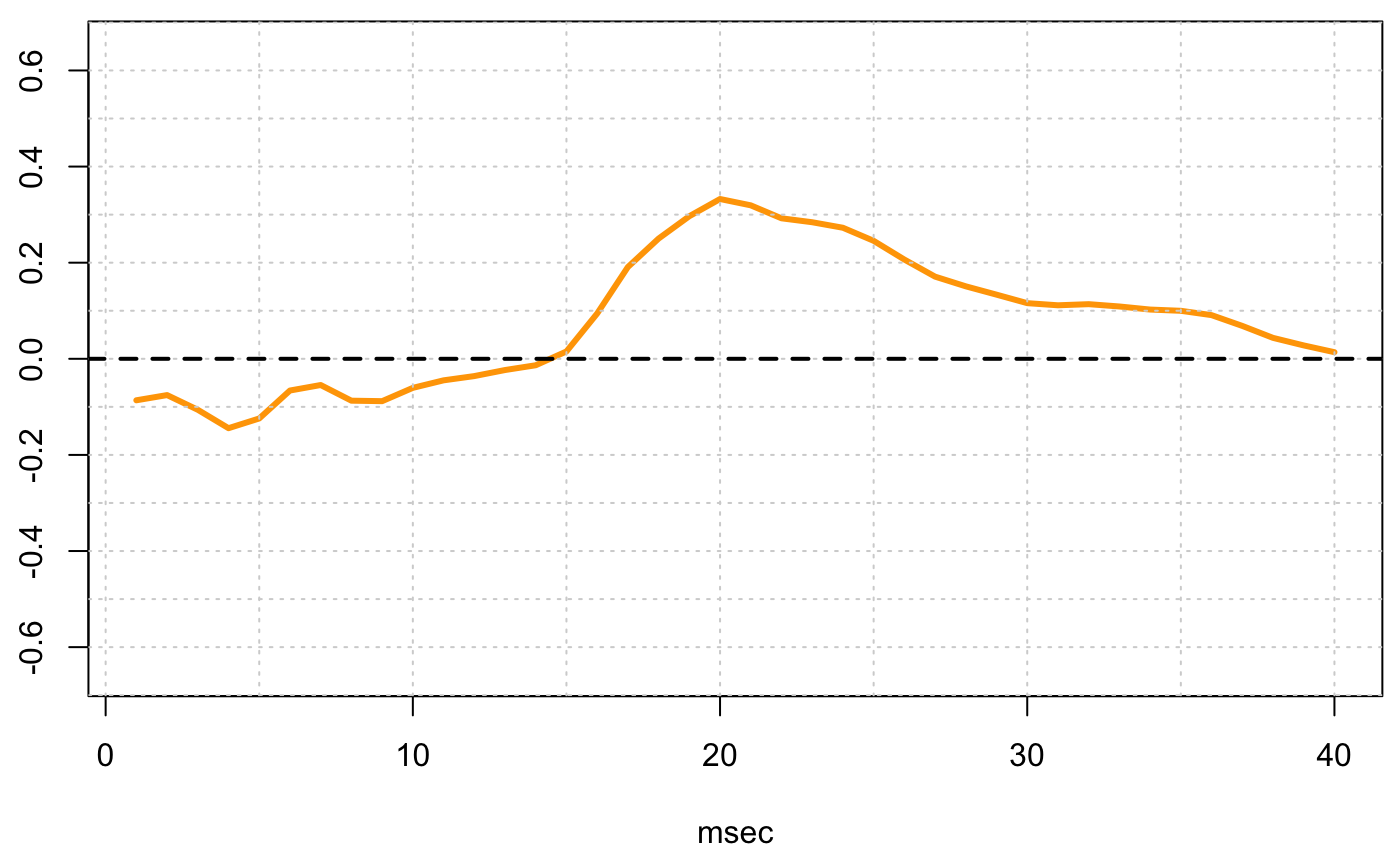}&
\includegraphics[height=4cm, width=4cm]{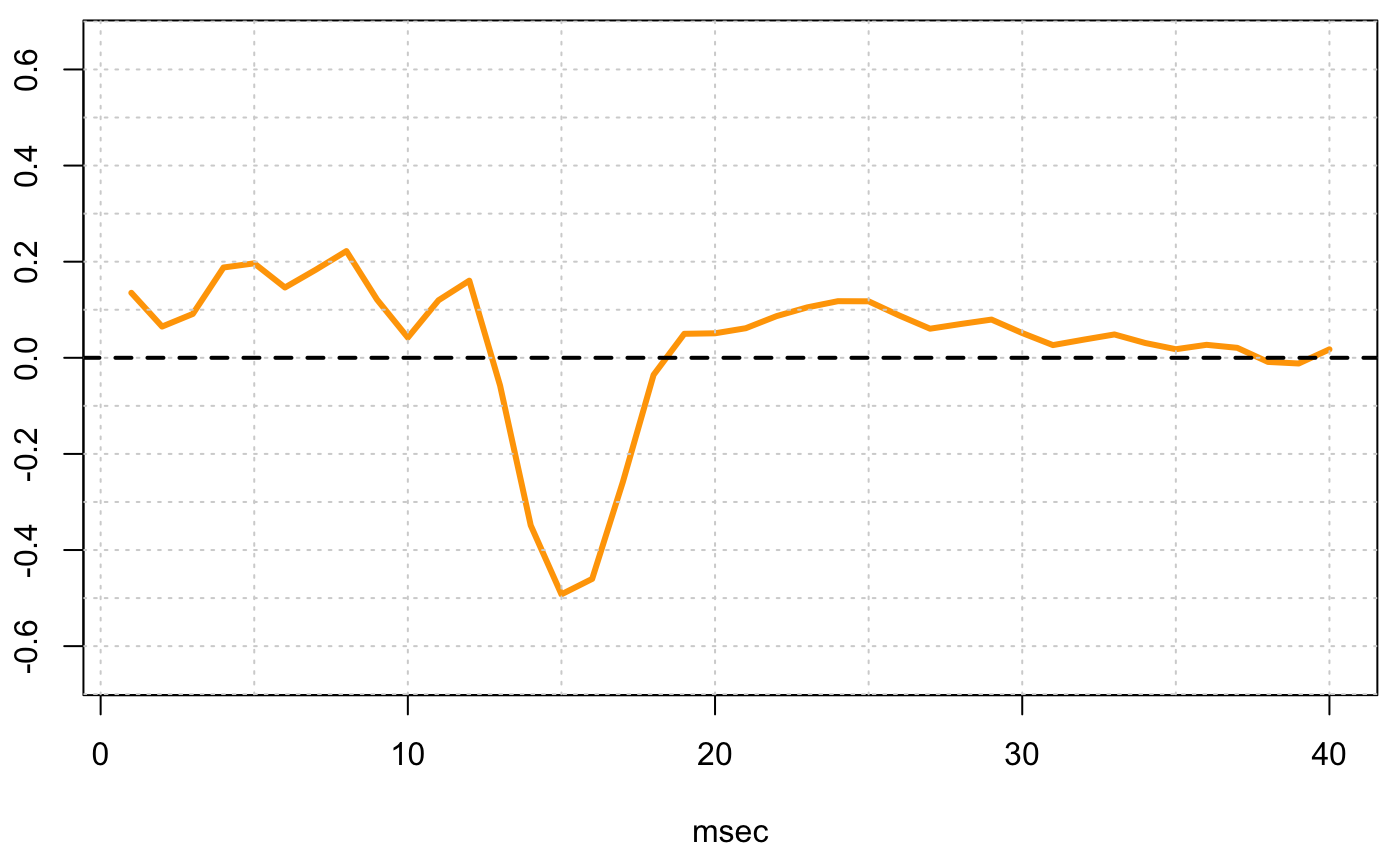}&
\includegraphics[height=4cm, width=4cm]{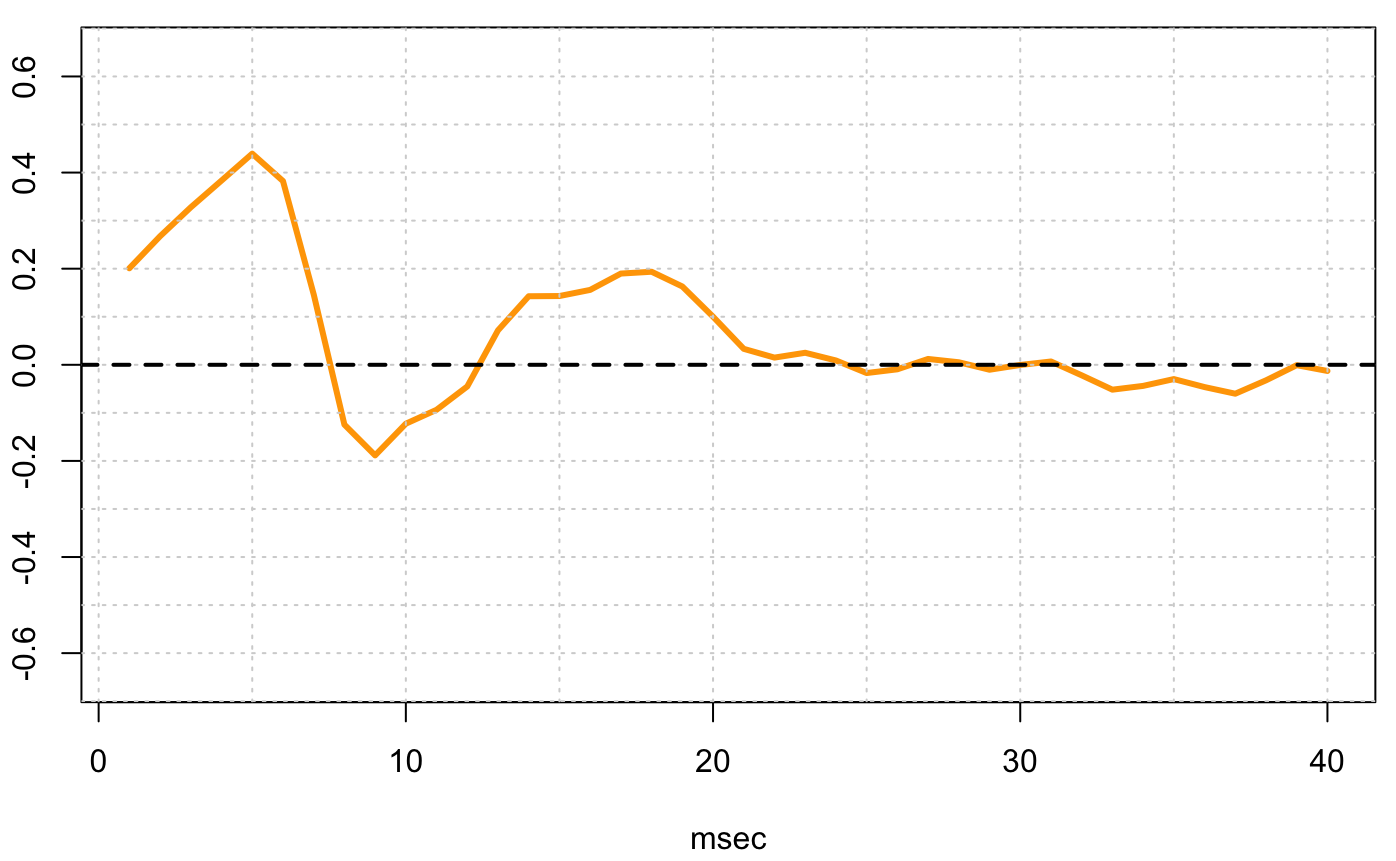}\\
\scriptsize{62.20 \%}& \scriptsize{6.82\%}& \scriptsize{6.11\%}& \scriptsize{5.08\%}\\         
\end{tabular}
\caption{\textbf{Eigenfunctions/principal components: Subset 3 "Preseason Practice".} Four strongest components out of (9, 9, 9, 8) = (team level, position level, player level, curve level) estimated components. Percentage of level-specific variation explained below graphs.}\label{fig:eigfcts_preseason}
\end{figure}

\pagebreak

\appendix

\setcounter{table}{0}
\setcounter{figure}{0}

\renewcommand{\thetable}{\Alph{section}.\arabic{table}}
\renewcommand{\thefigure}{\Alph{section}.\arabic{figure}}


\section{Additional Results}

\begin{table}[H]
  \centering
  {\tablesize
    \begin{tabular}{lcccccccccccc}
    \hline
    \hline
    &\multicolumn{12}{c}{Principal components (Median MSE)}\\
    \hline
    $\boldsymbol\sigma$&$\boldsymbol\phi_1^X$&$\boldsymbol\phi_2^X$&$\boldsymbol\phi_3^X$&$\boldsymbol\phi_1^U$&$\boldsymbol\phi_2^U$&$\boldsymbol\phi_3^U$&$\boldsymbol\phi_1^W$&$\boldsymbol\phi_2^W$&$\boldsymbol\phi_3^W$&$\boldsymbol\phi_1^V$&$\boldsymbol\phi_2^V$&$\boldsymbol\phi_3^V$\\
    \hline
    \textbf{0} & 0.0031 & 0.0078 & 0.0100 & 0.0009 & 0.0013 & 0.0010 & 0.0000 & 0.0001 & 0.0000 & 0.0000 & 0.0000 & 0.0000 \\
    \textbf{0.5} & 0.0032 & 0.0081 & 0.0101 & 0.0009 & 0.0014 & 0.0011 & 0.0001 & 0.0001 & 0.0001 & 0.0000 & 0.0000 & 0.0001 \\
    \textbf{1} & 0.0032 & 0.0082 & 0.0104 & 0.0008 & 0.0014 & 0.0011 & 0.0001 & 0.0001 & 0.0001 & 0.0000 & 0.0001 & 0.0002 \\
    \hline
    &\multicolumn{12}{c}{Principal scores (Median MSE)}\\
    \hline
 $\boldsymbol\sigma$&$\boldsymbol\xi_1$&$\boldsymbol\xi_2$&$\boldsymbol\xi_3$&$\boldsymbol\theta_1$&$\boldsymbol\theta_2$&$\boldsymbol\theta_3$&$\boldsymbol\zeta_1$&$\boldsymbol\zeta_2$&$\boldsymbol\zeta_3$&$\boldsymbol\eta_1$&$\boldsymbol\eta_2$&$\boldsymbol\eta_3$\\
 \hline
    \textbf{0} & 0.0048 & 0.0081 & 0.0189 & 0.0175 & 0.0054 & 0.0014 & 0.0004 & 0.0002 & 0.0001 & 0.0081 & 0.0041 & 0.0020 \\
    \textbf{0.5} & 0.0020 & 0.0047 & 0.0061 & 0.0067 & 0.0031 & 0.0011 & 0.0004 & 0.0002 & 0.0001 & 0.0106 & 0.0066 & 0.0045 \\
    \textbf{1} & 0.0017 & 0.0043 & 0.0048 & 0.0056 & 0.0033 & 0.0011 & 0.0005 & 0.0004 & 0.0003 & 0.0182 & 0.0141 & 0.0121 \\
 \hline
   &\multicolumn{12}{c}{Eigenvalues (MSE)}\\
    \hline
 $\boldsymbol\sigma$&$\boldsymbol\lambda_1^X$&$\boldsymbol\lambda_2^X$&$\boldsymbol\lambda_3^X$&$\boldsymbol\lambda_1^U$&$\boldsymbol\lambda_2^U$&$\boldsymbol\lambda_3^U$&$\boldsymbol\lambda_1^W$&$\boldsymbol\lambda_2^W$&$\boldsymbol\lambda_3^W$&$\boldsymbol\lambda_1^V$&$\boldsymbol\lambda_2^V$&$\boldsymbol\lambda_3^V$\\ 
 \hline
     \textbf{0} & 0.5134 & 0.0463 & 0.0124 & 0.1350 & 0.0167 & 0.0057 & 0.0068 & 0.0018 & 0.0004 & 0.0001 & 0.0000 & 0.0000 \\
    \textbf{0.5} & 0.5129 & 0.0461 & 0.0123 & 0.1358 & 0.0167 & 0.0057 & 0.0066 & 0.0018 & 0.0004 & 0.0002 & 0.0001 & 0.0001 \\
    \textbf{1} & 0.5128 & 0.0460 & 0.0121 & 0.1369 & 0.0167 & 0.0057 & 0.0064 & 0.0018 & 0.0004 & 0.0005 & 0.0002 & 0.0002 \\
 \hline
 \hline
    \end{tabular}
    }
  \caption{\textbf{Comparison true and estimated principal components, principal scores and eigenvalues in unbalanced case ($\boldsymbol{n=23,898}$), increased grid dimension ($\boldsymbol{p=100}$).} 100 simulation repetitions, different signal-to-noise ratios. MSE for estimating $\sigma$:  $2.8476\times 10^{-7}$ for $\sigma=0.5$, $3.2213\times 10^{-7}$ for $\sigma=1$. Simulation for $\sigma = 0$ performed without smoothing}\label{tab:comparison_true_vs_est_unbalanced_p100}%
\end{table}

\begin{table}[H]
  \centering
  {\tablesize
    \begin{tabular}{lcccccccccccc}
    \hline
    \hline
    &\multicolumn{12}{c}{Principal components (Median MSE)}\\
    \hline
    $\boldsymbol\sigma$&$\boldsymbol\phi_1^X$&$\boldsymbol\phi_2^X$&$\boldsymbol\phi_3^X$&$\boldsymbol\phi_1^U$&$\boldsymbol\phi_2^U$&$\boldsymbol\phi_3^U$&$\boldsymbol\phi_1^W$&$\boldsymbol\phi_2^W$&$\boldsymbol\phi_3^W$&$\boldsymbol\phi_1^V$&$\boldsymbol\phi_2^V$&$\boldsymbol\phi_3^V$\\
    \hline
    \textbf{0} & 0.0070 & 0.0184 & 0.0177 & 0.0010 & 0.0020 & 0.0009 & 0.0001 & 0.0001 & 0.0001 & 0.0000 & 0.0000 & 0.0000 \\
    \textbf{0.5} & 0.0069 & 0.0186 & 0.0177 & 0.0011 & 0.0020 & 0.0009 & 0.0001 & 0.0001 & 0.0001 & 0.0000 & 0.0000 & 0.0001 \\
    \textbf{1} & 0.0069 & 0.0187 & 0.0177 & 0.0011 & 0.0021 & 0.0010 & 0.0001 & 0.0002 & 0.0001 & 0.0001 & 0.0001 & 0.0004 \\
    \hline
    &\multicolumn{12}{c}{Principal scores (Median MSE)}\\
    \hline
 $\boldsymbol\sigma$&$\boldsymbol\xi_1$&$\boldsymbol\xi_2$&$\boldsymbol\xi_3$&$\boldsymbol\theta_1$&$\boldsymbol\theta_2$&$\boldsymbol\theta_3$&$\boldsymbol\zeta_1$&$\boldsymbol\zeta_2$&$\boldsymbol\zeta_3$&$\boldsymbol\eta_1$&$\boldsymbol\eta_2$&$\boldsymbol\eta_3$\\
 \hline
    \textbf{0} & 0.0043 & 0.0120 & 0.0096 & 0.0057 & 0.0036 & 0.0012 & 0.0004 & 0.0002 & 0.0001 & 0.0178 & 0.0089 & 0.0044 \\
    \textbf{0.5} & 0.0028 & 0.0079 & 0.0059 & 0.0034 & 0.0033 & 0.0010 & 0.0005 & 0.0004 & 0.0003 & 0.0241 & 0.0152 & 0.0107 \\
    \textbf{1} & 0.0033 & 0.0082 & 0.0073 & 0.0043 & 0.0046 & 0.0011 & 0.0012 & 0.0010 & 0.0009 & 0.0429 & 0.0344 & 0.0297 \\
 \hline
   &\multicolumn{12}{c}{Eigenvalues (MSE)}\\
    \hline
 $\boldsymbol\sigma$&$\boldsymbol\lambda_1^X$&$\boldsymbol\lambda_2^X$&$\boldsymbol\lambda_3^X$&$\boldsymbol\lambda_1^U$&$\boldsymbol\lambda_2^U$&$\boldsymbol\lambda_3^U$&$\boldsymbol\lambda_1^W$&$\boldsymbol\lambda_2^W$&$\boldsymbol\lambda_3^W$&$\boldsymbol\lambda_1^V$&$\boldsymbol\lambda_2^V$&$\boldsymbol\lambda_3^V$\\ 
 \hline
    \textbf{0} & 0.3778 & 0.0435 & 0.0270 & 0.0351 & 0.0106 & 0.0024 & 0.0042 & 0.0007 & 0.0002 & 0.0001 & 0.0000 & 0.0000 \\
    \textbf{0.5} & 0.3788 & 0.0433 & 0.0270 & 0.0356 & 0.0106 & 0.0024 & 0.0042 & 0.0007 & 0.0002 & 0.0001 & 0.0001 & 0.0001 \\
    \textbf{1} & 0.3803 & 0.0431 & 0.0269 & 0.0362 & 0.0108 & 0.0023 & 0.0043 & 0.0007 & 0.0003 & 0.0005 & 0.0002 & 0.0002 \\
\hline
 \hline
    \end{tabular}
    }
  \caption{\textbf{Comparison true and estimated principal components, principal scores and eigenvalues in balanced case ($\boldsymbol{n = 23,868}$), grid dimension $\boldsymbol{p=40}$.} 100 simulation repetitions, different signal-to-noise ratios. MSE for estimating $\sigma$: $1.0940\times10^{-5}$ for $\sigma=0.5$, $4.6980\times10^{-6}$ for $\sigma=1$. Simulation for $\sigma = 0$ performed without smoothing}\label{tab:comparison_true_vs_est_balanced}%
\end{table}

\begin{table}[H]
  \centering
  {\tablesize
    \begin{tabular}{lcccccccccccc}
    \hline
    \hline
    &\multicolumn{12}{c}{Principal components (Median MSE)}\\
    \hline
    $\boldsymbol\sigma$&$\boldsymbol\phi_1^X$&$\boldsymbol\phi_2^X$&$\boldsymbol\phi_3^X$&$\boldsymbol\phi_1^U$&$\boldsymbol\phi_2^U$&$\boldsymbol\phi_3^U$&$\boldsymbol\phi_1^W$&$\boldsymbol\phi_2^W$&$\boldsymbol\phi_3^W$&$\boldsymbol\phi_1^V$&$\boldsymbol\phi_2^V$&$\boldsymbol\phi_3^V$\\
    \hline
    \textbf{0} & 0.0026 & 0.0075 & 0.0071 & 0.0004 & 0.0008 & 0.0003 & 0.0000 & 0.0000 & 0.0000 & 0.0000 & 0.0000 & 0.0000 \\
    \textbf{0.5} & 0.0025 & 0.0073 & 0.0072 & 0.0004 & 0.0008 & 0.0004 & 0.0000 & 0.0000 & 0.0000 & 0.0000 & 0.0000 & 0.0001 \\
    \textbf{1} & 0.0025 & 0.0071 & 0.0073 & 0.0004 & 0.0008 & 0.0005 & 0.0000 & 0.0001 & 0.0001 & 0.0000 & 0.0000 & 0.0002 \\
    \hline
    &\multicolumn{12}{c}{Principal scores (Median MSE)}\\
    \hline
 $\boldsymbol\sigma$&$\boldsymbol\xi_1$&$\boldsymbol\xi_2$&$\boldsymbol\xi_3$&$\boldsymbol\theta_1$&$\boldsymbol\theta_2$&$\boldsymbol\theta_3$&$\boldsymbol\zeta_1$&$\boldsymbol\zeta_2$&$\boldsymbol\zeta_3$&$\boldsymbol\eta_1$&$\boldsymbol\eta_2$&$\boldsymbol\eta_3$\\
 \hline
    \textbf{0} & 0.0018 & 0.0047 & 0.0041 & 0.0022 & 0.0014 & 0.0005 & 0.0002 & 0.0001 & 0.0000 & 0.0081 & 0.0040 & 0.0020 \\
    \textbf{0.5} & 0.0011 & 0.0030 & 0.0025 & 0.0014 & 0.0014 & 0.0004 & 0.0002 & 0.0002 & 0.0001 & 0.0106 & 0.0066 & 0.0045 \\
    \textbf{1} & 0.0011 & 0.0033 & 0.0030 & 0.0017 & 0.0018 & 0.0005 & 0.0004 & 0.0004 & 0.0003 & 0.0182 & 0.0141 & 0.0121 \\
  \hline
   &\multicolumn{12}{c}{Eigenvalues (MSE)}\\
    \hline
 $\boldsymbol\sigma$&$\boldsymbol\lambda_1^X$&$\boldsymbol\lambda_2^X$&$\boldsymbol\lambda_3^X$&$\boldsymbol\lambda_1^U$&$\boldsymbol\lambda_2^U$&$\boldsymbol\lambda_3^U$&$\boldsymbol\lambda_1^W$&$\boldsymbol\lambda_2^W$&$\boldsymbol\lambda_3^W$&$\boldsymbol\lambda_1^V$&$\boldsymbol\lambda_2^V$&$\boldsymbol\lambda_3^V$\\ 
 \hline
    \textbf{0} & 0.3780 & 0.0430 & 0.0268 & 0.0351 & 0.0106 & 0.0024 & 0.0042 & 0.0007 & 0.0002 & 0.0001 & 0.0000 & 0.0000 \\
    \textbf{0.5} & 0.3796 & 0.0430 & 0.0267 & 0.0350 & 0.0106 & 0.0025 & 0.0043 & 0.0007 & 0.0002 & 0.0002 & 0.0001 & 0.0000 \\
    \textbf{1} & 0.3816 & 0.0430 & 0.0264 & 0.0350 & 0.0107 & 0.0026 & 0.0045 & 0.0008 & 0.0002 & 0.0005 & 0.0002 & 0.0002 \\
 \hline
 \hline
    \end{tabular}
    }
  \caption{\textbf{Comparison true and estimated principal components, principal scores and eigenvalues in balanced case ($\boldsymbol{n = 23,868}$) increased grid dimension ($\boldsymbol{p=100}$).} 100 simulation repetitions, different signal-to-noise ratios. MSE for estimating $\sigma$: $2.8099\times 10^{-7}$ for $\sigma=0.5$, $3.1050\times 10^{-7}$ for $\sigma=1$. Simulation for $\sigma = 0$ performed without smoothing}\label{tab:comparison_true_vs_est_balanced_p100}%
\end{table}

\end{document}